\documentclass[preprint]{aastex62}
\usepackage{amsmath,amssymb}
\usepackage{textcomp}
\usepackage{booktabs}
\usepackage{bm}
\def\VEV#1{\left\langle #1\right\rangle}

\newcommand\ltsima{$\; \buildrel <\over\sim \;$}
\newcommand\simlt{\lower.5ex\hbox{\ltsima}}
\newcommand\gtsima{$\; \buildrel >\over\sim \;$}
\newcommand\simgt{\lower.5ex\hbox{\gtsima}}

\shorttitle{Development of a Galactic Model}
\shortauthors{Koshimoto, Baba \& Bennett}

\begin{document}


%\title{A Parametric Galactic Model toward the Galactic Bulge: \\ A New Insight into the Initial Mass Function and Applications to Microlens Analysis}
\title{A Parametric Galactic Model toward the Galactic Bulge Based on Gaia and Microlensing Data}

\author{Naoki Koshimoto}
\affiliation{Laboratory for Exoplanets and Stellar Astrophysics, NASA/Goddard Space Flight Center, Greenbelt, MD 20771, USA}
\affiliation{Department of Astronomy, University of Maryland, College Park, MD 20742, USA}

\author{Junichi Baba}
\affiliation{National Astronomical Observatory of Japan, Mitaka, Tokyo 181-8588, Japan}

\author{David P. Bennett}
\affiliation{Laboratory for Exoplanets and Stellar Astrophysics, NASA/Goddard Space Flight Center, Greenbelt, MD 20771, USA}
\affiliation{Department of Astronomy, University of Maryland, College Park, MD 20742, USA}

\begin{abstract}
We developed a parametric Galactic model toward the Galactic bulge by fitting to spatial distributions of the Gaia DR2 disk velocity, VVV proper motion, BRAVA radial velocity, 
OGLE-III red clump star count, and OGLE-IV star count and microlens rate, optimized for use in microlensing studies. 
We include the asymmetric drift of Galactic disk stars and the dependence of velocity dispersion on Galactic location in the kinematic model, 
which has been ignored in most previous models used for microlensing studies. 
We show that our model predicts a microlensing parameter distribution significantly different from those typically used in previous studies. 
We estimate various fundamental model parameters for our Galaxy through our modeling, including the initial mass function (IMF) in the inner Galaxy. 
Combined constraints from star counts and the microlensing event timescale distribution from the OGLE-IV survey, in addition to a prior on the bulge stellar mass, 
enable us to successfully measure IMF slopes using a broken power-law form over a broad mass range,  
$\alpha_{\rm bd}=0.22^{+0.20}_{-0.55}$ for $M<0.08\,M_{\odot}$, $\alpha_{\rm ms} = 1.16^{+0.08}_{-0.15}$ for $0.08\,M_{\odot}\leq\,M<M_{\rm br}$,
and $\alpha_{\rm hm} = 2.32^{+0.14}_{-0.10}$ for $M\geq\,M_{\rm br}$, as well as a break mass at $M_{\rm br} = 0.90^{+0.05}_{-0.14} \, M_{\odot}$.
This is significantly different from the Kroupa IMF for local stars, but similar to the Zoccali IMF measured from a bulge luminosity function. 
We also estimate the dark matter mass fraction in the bulge region of $28\pm7$\% which could be larger than a previous estimate. 
Because our model is purely parametric, it can be universally applied using the parameters provided in this paper.\footnote{
{ A tool for microlensing simulation using our Galactic model has been published \citep{kos21}, and 
can be downloaded at \dataset[https://github.com/nkoshimoto/genulens]{https://github.com/nkoshimoto/genulens}.}
}
\end{abstract}

\keywords{Milky Way Galaxy (1054), Galactic bulge (2041),  Initial mass function (796), Gravitational microlensing (672)}

\section{Introduction}
Gravitational microlensing is a unique tool that can study the population of planetary to black hole mass objects in our Galaxy \citep{gau19}.
It has been applied to study the population of exoplanets since the proposal by \citet{mao91}, along with more than 100 exoplanets, has been discovered as of 2020\footnote{\dataset[NASA Exoplanet Archive]{https://exoplanetarchive.ipac.caltech.edu/} \citep{ake13}}.
\citet{suz16} used a statistical sample of 30 microlensing planets to reveal a likely peak at $q \sim 10^{-4}$ in 
the mass-ratio function of planets beyond the H$_2$O snow line for the first time, challenging 
the core accretion theory \citep{suz18}.
Microlensing also enables us to study the population of free-floating planets \citep{sum11, mro17}. 
Further, some candidates of isolated black holes have also been discovered \citep{ben02, poi05, wyr16, wyr20}.

Difficulty in measuring the lens mass and distance makes microlensing studies complex.
Although four physical quantities, namely the lens mass $M_{\rm L}$, distance from lens $D_{\rm L}$, distance from source $D_{\rm S}$, 
and lens-source relative proper motion $\mu_{\rm rel}$, are involved in each microlensing event, the Einstein radius crossing time
\begin{align}
t_{\rm E} = \frac{\theta_{\rm E}}{\mu_{\rm rel}}, \label{eq-tE}
\end{align}
is the only quantity that is  measurable for all events, where $\theta_{\rm E}$ is the angular Einstein radius given by
\begin{equation}
\theta_{\rm E} = \sqrt{\kappa M_{\rm L} \pi_{\rm rel}}  \label{eq-thetaE}
\end{equation}
with $\kappa = 8.144 ~{\rm mas} ~M_\odot ^{-1}$ and $\pi_{\rm rel} = {\rm 1\,AU}(D_{\rm L}^{-1} - D_{\rm S}^{-1} )$.
For microlensing events toward the Galactic bulge, the source star is assumed to be a bulge star, i.e., $D_{\rm S} \sim 8~$kpc, but the three remaining quantities 
still degenerate in a $t_{\rm E}$ value.

There are three observable quantities each of which provides a mass-distance relation: the angular Einstein radius, $\theta_{\rm E}$, 
microlens parallax, $\pi_{\rm E}$, and lens star flux.
The degeneracy in $t_{\rm E}$ could be disentangled when any two of the three are observed.
The angular Einstein radius is given by Eq. (\ref{eq-thetaE}) while the microlens parallax is given by 
\begin{equation}
\pi_{\rm E} = \frac{\pi_{\rm rel}}{\theta_{\rm E}}, \label{eq-piE}
\end{equation}
and these two quantities can be measured through { higher} order effects in the light curve, finite source effect \citep{yoo04}
and annual parallax effect caused by the Earth's orbital motion \citep{alc95, an-eros2000blg5}, respectively.
However, these effects are rarely measured, particularly for single-lens events that account for $\sim$90 \% of all microlensing events.
We can measure the lens flux with high-angular resolution follow-up imaging using adaptive optics (AO) or the {\it Hubble Space Telescope} ({\it HST}) \citep{bat15, ben15, bha18}.
However, such measurements typically require us to wait for $\simgt 5$ years until the lens-source separation can be measured, depending on the relative proper motion and contrast between lens and source stars.
These limitations associated with the lens mass and distance measurements have resulted in many microlensing studies suffering from a chronic lack of information.

From this perspective, a Galactic model, which refers to a combination of a stellar mass function and stellar density and velocity distributions 
in our Galaxy, has played a crucial role in providing prior probability distributions for microlensing events toward the Galactic bulge.
For example, a large number of studies on individual event analysis use a Galactic model to calculate a posterior distribution of the lens mass and distance \citep[e.g.,][]{alc95, bea06, kos14, ben14}.
Further, it has also been used for statistical studies to estimate fundamental parameters in our Galaxy, such as the slope of 
initial mass function (IMF) \citep{sum11, mro17, weg17} or dark matter fraction in the bulge region \citep{weg16}.
Galactic models are occasionally used to diagnose whether measurements of microlensing parameters are contaminated by systematic errors \citep{pen16, kos20}.

However, there are concerns regarding the use of Galactic models because the results depend on the choice of models, with 
most Galactic models used in microlensing studies employing several simplified features \citep{yan20}.
This is a concern specifically for statistical studies in which small effects for many individual events can be combined.
For example, \citet{kos20} compared three different Galactic models by \citet{sum11}, \citet{ben14}, and \citet{zhu17} with 50 microlens parallax measurements 
by the 2015 {\it Spitzer} microlensing campaign \citep{zhu17} and concluded that the parallax measurements are significantly contaminated due to systematic errors 
based on diagnoses using the Anderson-Darling (AD) tests.
Here, all of the three models use simplified models for disk kinematics with flat rotation curves, constant velocity dispersions, and no asymmetric drift, which contradict the observed velocity distributions \citep{kat18}.
Although the conclusions seem robust because they are based on the discovered correlations between $p$-values from the AD tests and some characteristics of 
events, which are likely related to vulnerability to systematic errors (e.g., source brightness or peak coverage), the absolute $p$-values might be misestimated due to the simplified features in the Galactic models.\footnote{In Section \ref{sec-stat}, we show that our improved Galactic model does not have a significant effect on the results of \citet{kos20} that were calculated with simplified Galactic models.}

Another case motivating this concern is the IMF in brown dwarf mass range of $M < 0.08 \, M_{\odot}$, parameterized by a slope, $\alpha_{\rm bd}$, with $dN/dM \propto M^{-\alpha_{\rm bd}}$, and inferred by the $t_{\rm E}$ distribution from the OGLE-IV survey \citep{mro17}.
Although \citet{mro17} shows the likelihood peak at $\alpha_{\rm bd} \sim 0.8$ using the Galactic model based on \citet{han95, han03} with simplified 
disk kinematics, \citet{spe20} shows the likelihood peak at $\alpha_{\rm bd} = -0.1$, and 0.8 (-0.8 in their definition) is significantly disfavored according to their Figure 4.
\citet{spe20} used the Besan\c{c}on model \citep{rob03,rob12,rob17} for their analysis using a more realistic model for disk kinematics.
However, the Besan\c{c}on model also has some unrealistic features, as discussed by \citet{pen19}, such as
the small Galactic bar angle of 13$^{\circ}$ compared to $25^{\circ}$--$30^{\circ}$ implied by other observations \citep[see a review by][]{bla16}.
Further, the IMF for bulge stars in the public version has a minimum mass at $0.15 \, M_{\odot}$, a too massive cut-off for applying microlensing analysis, 
which { is} sensitive down to planetary-mass objects.

Several non-parametric dynamical models of our Galaxy are developed with the aid of $N$-body simulations \citep{weg13, por15, por17}.
These models are better than parametric models with respect to consistency with the dynamics, but they have limited resolution.
Numerical models use particles of thousands of solar masses and it is not easy to employ these models for simulations of microlensing events along a specific line of sight.
Also, they are not open to communities outside the group, and thus not useful in terms of accessibility or difficulty of reproduction.
A recent version of the Besan\c{c}on model uses results from an $N$-body simulation \citep{gar14} for the kinematics of bulge stars, and 
it is difficult for other people to reproduce the model for their own use.

In this study, we develop a parametric Galactic model by fitting to spatial distributions of the median velocity and velocity dispersion from Gaia DR2 \citep{kat18}, 
OGLE-III red clump (RC) star count \citep{nat13}, VIRAC proper motion data \citep{smi18, cla19}, BRAVA radial velocity data \citep{ric07, kun12}, and the OGLE-IV star count 
and microlensing event data \citep{mro17, mro19}.
We parameterize the model with 40--48 parameters depending on the selected options for the bulge density profile, and we present the best-fit values for all the
parameters to enable ease of reproduction or make further updates.
Our model aims to fulfill the gap in the market between the simplified parametric models used for microlensing studies
and dynamical models based on $N$-body simulations that are difficult to use for microlensing work.

This model enables, for the first time, simultaneous measurements of the three slopes over entire mass range in a broken power law IMF of $dN/dM \propto M^{-\alpha}$, 
$\alpha_{\rm bd} = 0.22^{+0.20}_{-0.55}$ in $M < 0.08 \, M_{\odot}$, 
$\alpha_{\rm ms} = 1.16^{+0.08}_{-0.15}$ in $0.08 \, M_{\odot} \leq M < M_{\rm br}$, and $\alpha_{\rm hm} = 2.32^{+0.14}_{-0.10}$ in $M \geq M_{\rm br}$, as well as 
a break mass at $M_{\rm br} = 0.90^{+0.05}_{-0.14} \, M_{\odot}$. 
The IMF corresponds to the stellar mass-to-light ratio in $K$-band of  $\Upsilon_K  = 0.72^{+0.05}_{-0.02} \, M_{\odot}/L_{K_{\odot}}$, which is significantly different from 
$\Upsilon_K  = 1.04 \, M_{\odot}/L_{K_{\odot}}$ with the \citet{kro01} IMF for local stars but consistent with $\Upsilon_K  = 0.75 \, M_{\odot}/L_{K_{\odot}}$  with the \citet{zoc00} IMF for bulge stars.
We estimated the dark matter mass inside the VVV bulge box \citep[$\pm 2.2 \times \pm 1.4 \times \pm 1.2$ kpc,][]{weg13} of $M_{\cal DM} = 0.45^{+0.02}_{-0.06} \times 10^{10} \, M_{\odot}$ by 
comparing our $\Upsilon_K$ value with those of the five versions of dynamical Galactic models in \citet{por15}.
This is consistent with another independent estimate of $M_{\cal DM}  = 0.51^{+0.12}_{-0.11} \times 10^{10} \, M_{\odot}$ derived by combining our estimate on the stellar 
mass inside the VVV bulge box of $1.34^{+0.10}_{-0.11} \times 10^{10} \, M_{\odot}$ and the dynamical mass in 
the same region of $(1.85 \pm 0.05) \times 10^{10} \, M_{\odot}$ constrained by \citet{por17}.

Using the developed models, we repeated a part of the analysis by \citet{kos20} and confirmed their finding of a significant discrepancy between the predicted and observed $\pi_{\rm E}$ distributions with the new model,
although the predicted $\pi_{\rm E}$ distribution significantly differed from the previous one.
Further, we applied the model to calculate the lens-source proper motion $\mu_{\rm rel}$ prior for OGLE-2011-BLG-0950, the only ambiguous 
event suffering from degeneracy between the planetary and binary solutions out of the 29 events in the \citet{suz16} combined sample.
Our prior probability distribution for $\mu_{\rm rel}$ indicates that the binary solution is more probable than the planetary solution.

This paper is organized as follows. 
In Section \ref{sec-model}, we describe our parametric Galactic model and introduce all of the 40--48 fit parameters.
Fitting to the Gaia DR2 data is conducted to determine the 10 parameters for the disk velocity model in Section \ref{sec-disk}.
Fitting to the other data, which is toward the bulge sky, is separately conducted in Section \ref{sec-bar} to determine the 26--34 parameters for the bulge density and velocity models,
as well as the 4 parameters for the IMF model, where the determined disk velocity model is used in the fit.
In Section \ref{sec-comp}, we discuss the determined parameters by comparing them with previous studies.
We apply the developed Galactic models to microlensing analysis in Section \ref{sec-app}.
Section \ref{sec-con} presents the summary and conclusion.

\section{Model Parameterization} \label{sec-model}

In this section, we describe our parametric Galactic model with a barred bulge and a multi-component of the stellar thin and thick disks. We disregard a stellar halo in this study because it rarely contributes to the considered data, which are mostly toward the Galactic bulge.
We refer to a bar with a rigid-body rotation as the `bulge' while an inner part of our Galaxy as the `bulge region,' which means stars in the bulge region refer to both the bulge and disk stars present there.
These Galactic structures consist of stars with mass functions (Section \ref{sec-MF}) and have the density and velocity distributions as functions of the Galactocentric coordinate ($x, y, z$) or Galactocentric cylindrical coordinate ($R, \phi, z$). 
We use different density and velocity distributions for each component, where we use $\rho_{\rm d}$ and $v_{\rm d}$ to refer to the disk density and velocity, respectively (Section \ref{sec-disk_model}), and $\rho_{\rm B}$ and $v_{\rm B}$ to refer to the bulge density and velocity, respectively (Section \ref{sec-bar_model}).

Here we introduce 4 fit parameters to model the stellar mass function, 10 fit parameters to model the $v_{\rm d}$ distribution, 7--15 fit parameters to 
model the $\rho_{\rm B}$ distribution, and 19 fit parameters to model the $v_{\rm B}$ distribution, respectively.
Whereas the $\rho_{\rm d}$ model does not contain any fit parameters, two options of flat-scale and linear-scale height models are introduced for the thin disk.
The 10 parameters for the $v_{\rm d}$ distribution are determined by fitting to the Gaia DR2 data \citep{kat18} in Section \ref{sec-disk}, while 
the other 30--38 parameters are determined in Section \ref{sec-bar} by fitting to the other data toward the bulge sky, namely the OGLE-III red clump star count \citep{nat13}, 
VIRAC proper motion data \citep{smi18, cla19}, BRAVA radial velocity data \citep{ric07, kun12}, and OGLE-IV star count and microlensing event data \citep{mro17, mro19}.
%We keep general forms for the parametric models in this section and the actual values are given in the relevant sections.

The OGLE-III red clump star count data from \citet{nat13} is sensitive to the Sun's Galactic position and the bar angle.
In fact, \citet{cao13} used the same dataset to measure the distance to the Galactic center of 8.1--8.2 kpc and the bar angle of 27--32 deg.
However, because the number of fit parameters is already large, 40 to 48, we fix these two parameters to mitigate degeneracies among fit parameters.
For the solar position in our Galaxy, we use $R_{\odot} = 8160~{\rm pc}$, which is consistent with a recent precise measurement of $8178 \pm 13_{\rm stat.} \pm 22_{\rm sys.}~{\rm pc}$
on the distance to Sgr A* \citep{gra19}, and $z_{\odot} = 25~{\rm pc}$ \citep{bla16}.
We use $\alpha_{\rm bar} = 27^\circ$ for the bar angle \citep{weg13, bla16}.
The $R_{\odot}$ and $\alpha_{\rm bar}$ values are consistent with the OGLE-III red clump star count data because these two values are both within 
the above respective ranges derived by \citet{cao13}. Thus, this choice generates no tension.

We also fix the solar velocity to be $(v_{\odot, x}, v_{\odot, y}, v_{\odot, z}) = (-10, 243, 7)$~km/s \citep{bla16}.
The tangential velocity $v_{\odot, y} = 243$ km/s is marginally lower than $248 \pm 3$ km/s suggested by \citet{bla16} (a recent thorough review paper on our Galaxy) because 
we found a better agreement to the VIRAC proper motion data with 243 km/s.
This is consistent with \citet{cla19}, who found that the dynamical model of \citet{por17}, with a slower tangential velocity of $v_{\odot, y} = 245$ km/s, gave 
an improved match to the VIRAC proper motion data, compared to the original value of $v_{\odot, y} = 250.24$ km/s used by \citet{por17}.

\subsection{Stellar Mass Function} \label{sec-MF}
The stellar mass function refers to a present-day mass function, which can be calculated by a combination of IMF, star formation rate, and
initial--final mass relationships for remnants, i.e., for white dwarfs, neutron stars, and black holes.
Here, we introduce the IMF, star formation rate, and initial--final mass relationships used in this paper.

We use a broken-power law form IMF given by
\begin{equation}
\frac{dN}{dM} \propto
\begin{cases}
M^{-\alpha_{\rm hm}}  & \text{ when $M_{\rm br} < M < 120 \, M_{\odot}$} \\
M^{-\alpha_{\rm ms}}  & \text{ when $0.08 \, M_{\odot} \leq M < M_{\rm br}$} \\
M^{-\alpha_{\rm bd}}  & \text{ when $10^{-3} \, M_{\odot} \leq M < 0.08 \, M_{\odot}$} \ . \label{eq-MF}
\end{cases}
\end{equation}
We use four parameters, the three slopes $(\alpha_{\rm hm}, \alpha_{\rm ms}, \alpha_{\rm bd})$ and a break mass $M_{\rm br}$, as fit parameters in Section \ref{sec-bar},
and fit them to the data including the OGLE-IV microlensing $t_{\rm E}$ distribution, which has a sensitivity to objects with brown dwarf mass to black hole mass.
By contrast, we use the local IMF by \citet{kro01}, $(\alpha_{\rm hm}, \alpha_{\rm ms}, \alpha_{\rm bd})$ = (2.30, 1.30, 0.30) with $M_{\rm br} = 0.50\,M_{\odot}$, in the 
fit to the Gaia DR2 data, in Section \ref{sec-disk}, to derive the values of the local density for red giants, which are listed in Table \ref{tab-disks}.
We selected different IMF values because the Gaia data are dominated by local stars, whereas the data used in Section \ref{sec-bar} are dominated by bulge stars.

Further, a star formation rate $\propto \exp [\frac{T} {7 {\rm Gyr}}]$ \citep{bov17} is applied to the thin disk star, where $T$ is stellar age, while 
we assume a mono-age thick disk at $T = 12~$Gyr \citep{rob14} and 
the Gaussian distribution of $9 \pm 1~$Gyr \citep{kos20b} for the bulge stars.
The PARSEC isochrone models \citep{bre12, che14, tan14} are used for the stellar lifetime of a given initial mass.

For initial--final mass relationships of remnants, we apply the model used by \citet{lam20}, where a linear relation of \citet{kal08}, $M_{\rm WD} = 0.109 M_{\rm ini} + 0.394 M_{\odot}$, is 
used for white dwarfs and a modified probabilistic relation based on \citet{rai18} is used for neutron stars and black holes.
See Appendix C of \citet{lam20} and references therein for more details.
Further, we add a birth kick velocity in a random direction of 350 km/s or 100 km/s to the velocity of the progenitor star for a neutron star or a black hole, respectively \citep{lam20}.

\subsection{Galactic Disk Model}  \label{sec-disk_model}
In this subsection, we describe the density and velocity models of a thin and a thick disk, where the thin disk is further divided into seven components depending 
on ages following the classification by \citet{rob03}.
These seven plus one components have different scale height values and velocity dispersions. 
The velocity dispersions differ from each other depending on the 10 fit parameters introduced in the $v_{\rm d}$ model and determined later in Section \ref{sec-disk}.
The disk model contributes to both fits in Sections \ref{sec-disk} and \ref{sec-bar}.

\subsubsection{Disk density ($\rho_{\rm d}$ model)} \label{sec-disk_rho}
A thin and a thick disk are considered with density profiles of 
\begin{align}
\rho_{\rm d}^{\rm thin} (R, z) &= 
\begin{cases}
\rho^{\rm thin}_{\rm d, \odot} \, \frac{z_{\rm d, \odot}^{\rm thin} }{z_{\rm d}^{\rm thin} (R)} \, \exp{\left( - \frac{R - R_{\odot}}{R_{\rm d}^{\rm thin}} \right)} \, {\rm sech}^2 {\left( - \frac{|z|}{z_{\rm d}^{\rm thin} (R)} \right)}   & \text{ if $R > R_{\rm d, break}$} \\
\rho^{\rm thin}_{\rm d, \odot} \, \frac{z_{\rm d, \odot}^{\rm thin} }{z_{\rm d}^{\rm thin} (R)} \, \exp{\left( - \frac{R_{\rm d, break} - R_{\odot}}{R_{\rm d}^{\rm thin}} \right)} \, {\rm sech}^2 {\left( - \frac{|z|}{z_{\rm d}^{\rm thin} (R)} \right)}   & \text{ if $R \leq R_{\rm d, break}$}
\end{cases} \label{eq-rho_thin}\\
\rho_{\rm d}^{\rm thick} (R, z) &= 
\begin{cases}
\rho^{\rm thick}_{\rm d, \odot} \exp{\left( - \frac{R - R_{\odot}}{R_{\rm d}^{\rm thick}} \right)} \, {\rm exp} {\left( - \frac{|z|}{z_{\rm d, \odot}^{\rm thick}} \right)} & \text{ if $R > R_{\rm d, break}$} \\
\rho^{\rm thick}_{\rm d, \odot} \exp{\left( - \frac{R_{\rm d, break} - R_{\odot}}{R_{\rm d}^{\rm thick}} \right)} \, {\rm exp} {\left( - \frac{|z|}{z_{\rm d, \odot}^{\rm thick}} \right)} & \text{ if $R \leq R_{\rm d, break}$,} 
\end{cases}  \label{eq-rho_thick}
\end{align}
where $\rho_{\rm d, \odot}$ is the stellar density in the solar neighborhood, hereafter referred to as the local (stellar) density, 
$R_{\rm d}$ is the disk scale length, and $z_{\rm d}$ is the disk scale height.

The disk profile in the inner Galaxy region is still uncertain. A constant surface density model is adopted within $R < R_{\rm d, break}$ following the findings of \citet{por17} who
developed an $N$-body dynamical model reproducing extensive photometric and kinematic data across our Galaxy. 
In their model, the disk has an exponential feature in the outer part of our Galaxy ($\simgt 5$ kpc) while it has a flat surface density feature in the inner part ($\simlt 5$ kpc).
We use $R_{\rm d, break} = 5.3~{\rm kpc}$ for our model.

For the thin disk scale height $z_{\rm d}^{\rm thin}$, two options are considered: a flat- and a linear-scale height model.
We adopt 
\begin{align}
z_{\rm d}^{\rm thin} (R) &= 
\begin{cases}
z_{\rm d, \odot}^{\rm thin} - (z_{\rm d, \odot}^{\rm thin} - z_{\rm d, 4.5}^{\rm thin}) \frac{R_{\odot} - R}{R_{\odot} - 4.5 {\rm kpc}} & \text{ if $R > 4.5$ kpc} \\
z_{\rm d, 4.5}^{\rm thin} & \text{ if $R \leq 4.5$ kpc}
\end{cases}
\end{align}
for the linear-scale height model with $z_{\rm d, 4.5}^{\rm thin} = 0.6 \, z_{\rm d, \odot}^{\rm thin}$ \citep{weg16} while for the flat-scale height model $z_{\rm d}^{\rm thin} (R) = z_{\rm d, \odot}^{\rm thin}$ is used.
Two options have been considered as we were motivated by a similar attempt of \citet{weg16} who considered the linear-scale height model to smoothly connect the disk to the inner thin long bar 
with a scale height of $\sim 200$ pc suggested by \citet{weg15}.
{ As described in Section \ref{sec-bar_rho}, adding the long bar component to our bulge model did not result in a significant improvement.
However, whether we used the flat- or linear-scale height model in the disk model did affect $\chi^2$ or $\tilde{\chi}^2$ values in the fit.}

% Table 1
\begin{deluxetable}{lcrrrrrr}
\tablecaption{Scale lengths, scale heights and local densities for thin and thick disks. \label{tab-disks}}
\tablehead{
\colhead{}     & \colhead{Age $T$} & \colhead{$R_{\rm d}$} & \colhead{$z_{\rm d, \odot}$} & \colhead{$z_{\rm d, 4.5}$ \tablenotemark{a}} & \colhead{$\rho_{\rm d, \odot}^{\rm MS}$ \tablenotemark{b}} & \colhead{$\rho_{\rm d, \odot}^{\rm WD}$ \tablenotemark{b}} & \colhead{$n_{\rm d, \odot}^{\rm RG}$ \tablenotemark{b}} \\
\colhead{}     & \colhead{[Gyr]}   & \colhead{[pc]}    & \colhead{[pc]}           & \colhead{[pc]}                          & \colhead{[$M_{\odot}$ pc$^{-3}$]}    & \colhead{[$M_{\odot}$ pc$^{-3}$]}    & \colhead{[pc$^{-3}$]}
}
\startdata
Thin disk      & 0 - 0.15         & 5000              &  61                      &   36                                    & $5.1 \times 10^{-3}$              & $5.5 \times 10^{-5}$              & $6.9 \times 10^{-6}$      \\
               & 0.15 - 1         & 2600              & 141                      &   85                                    & $5.0 \times 10^{-3}$              & $2.2 \times 10^{-4}$              & $3.3 \times 10^{-5}$      \\
               &    1 - 2         & 2600              & 224                      &  134                                    & $3.8 \times 10^{-3}$              & $2.9 \times 10^{-4}$              & $4.2 \times 10^{-5}$      \\
               &    2 - 3         & 2600              & 292                      &  175                                    & $3.2 \times 10^{-3}$              & $3.3 \times 10^{-4}$              & $2.1 \times 10^{-5}$      \\
               &    3 - 5         & 2600              & 372                      &  223                                    & $5.9 \times 10^{-3}$              & $7.8 \times 10^{-4}$              & $6.5 \times 10^{-5}$      \\
               &    5 - 7         & 2600              & 440                      &  264                                    & $6.3 \times 10^{-3}$              & $1.0 \times 10^{-3}$              & $6.1 \times 10^{-5}$      \\
               &    7 - 10        & 2600              & 445                      &  267                                    & $1.3 \times 10^{-2}$              & $2.4 \times 10^{-3}$              & $1.3 \times 10^{-4}$      \\
Sum/Mean       &                  &                   & 329                      &  197                                    & $4.2 \times 10^{-2}$              & $5.1 \times 10^{-3}$              & $3.6 \times 10^{-4}$      \\\hline
Thick disk     &    12            & 2200              & 903                      &                                         & $1.7 \times 10^{-3}$              & $4.4 \times 10^{-4}$              & $9.1 \times 10^{-6}$      \\
\enddata
\tablenotetext{a}{For the linear scale height model.}
\tablenotetext{b}{Local densities are given for main sequence stars (MS, including brown dwarfs), white dwarfs (WD), and red giants (RG). In the fit in Section \ref{sec-bar}, the values are updated with a given IMF.}
\tablecomments{$R_{\rm d}$ refers to $R_{\rm d}^{\rm thin}$ in the lines for thin disk, and $R_{\rm d}^{\rm thick}$ in the line for thick disk. The same is true for $z_{\rm d, \odot}$, $z_{\rm d, 4.5}$, $\rho_{\rm d, \odot}$, and $n_{\rm d, \odot}$.}
\end{deluxetable}

The thin disk is divided into 7 components depending on the age ($T$) following the Besan\c{c}on model \citep{rob03,rob12,rob17}.
Table \ref{tab-disks} lists the scale length and height values for each component of the thin and thick disks.
The scale height for each component is calculated using the age-scale height relation for the axis ratio $\epsilon$, that is $\epsilon = {\rm Min} [0.0791, 0.104 \sqrt{(T/{\rm Gyr} + 0.1)/10.1}]$ \citep{sha14}.
Because $\epsilon$ is designed as the ratio of the scale height to the scale length in a disk with Einasto laws \citep{ein79}, we used the surface-to-volume density ratio 
in the Einasto disk calculated with $\epsilon$ to derive the $z_{\rm d, \odot}$ value for each component.

Table \ref{tab-disks} also lists the local stellar density values for main sequence stars, white dwarfs, and red giants for each disk component.
The total thin disk local mass density is normalized by $4.2 \times 10^{-2} \, M_{\odot}$ pc$^{-3}$ for main sequence stars \citep{bov17}, while 
the thick disk local mass density is normalized so that its local contribution to the thin disk becomes $4 \%$ \citep{bla16}.

The local density value for stars at each evolutional stage in each disk component is calculated using the local IMF by \citet{kro01} combined with 
the star formation rate and the initial--final mass relationship described in Section \ref{sec-MF}.
The red giants are selected based on their absolute magnitude ${\cal M}_G < 3.9$ mag and intrinsic color $(G_{\rm BP} - G_{\rm RP})_0 > 0.95$ in the Gaia bands 
using the PARSEC isochrones \citep{bre12, che14, tan14}.
This criteria for red giants is same as that used for the selection of the giant sample by \citet{kat18}, and the $n_{\rm d, \odot}^{\rm RG}$ values listed in the table are 
used in the fit to the Gaia DR2 data in Section \ref{sec-disk}.

An important note here is that the local density values are updated with a given IMF and are different in the fit in Section \ref{sec-bar}.
This apparent inconsistent treatment can be partially justified by considering that our IMF presented in Section \ref{sec-bar} is for the disk stars located in 
the inner Galaxy region ($R \simlt 4$~kpc) contributing to the data used in Section \ref{sec-bar}, and the \citet{kro01} IMF is appropriate for the nearby stars ($R \simgt 4$~kpc) contributing to 
the Gaia data used in Section \ref{sec-disk}.

The local surface density with the \citet{kro01} IMF is $30.01 ~ M_{\odot}$ pc$^{-2}$ for main sequence stars, which is compared with the measurements of 
$23.0 \pm 1.5~ M_{\odot}$ pc$^{-2}$ \citep{bov17} and $28.2 \pm 2.7~ M_{\odot}$ pc$^{-2}$ \citep{mck15}.
The local surface density of white dwarfs is $4.87 ~ M_{\odot}$ pc$^{-2}$, and this is consistent with measurements of $5 \pm 1~ M_{\odot}$ pc$^{-2}$ \citep{bov17} and $4.9 \pm 0.6~ M_{\odot}$ pc$^{-2}$ \citep{mck15}.
The local number density of red giants is $3.6 \times 10^{-4}$~pc$^{-3}$, and this fairly well agrees with the measurement by \citet{bov17} of $(3.9 \pm 0.1) \times 10^{-4}$~pc$^{-3}$.

\subsubsection{Disk kinematics ($v_{\rm d}$ model)} \label{sec-disk_velo}
Our work is highly motivated by Gaia DR2 \citep{kat18} in which skewed distributions for the azimuthal velocity $v_{\phi}$ and
clear dependencies of the velocity dispersion and mean azimuthal velocity of disk stars on their location are shown.
Such disk kinematic structures have not been included in most of the models used for microlensing analysis like the three models \citep{sum11, ben14, zhu17} used in \citet{kos20}.
To include those dependencies as a function of the Galactocentric cylindrical coordinate ($R, \phi, z$) in addition to a skewed $v_{\phi}$ distribution, 
we follow the parameterization of a disk velocity model by \citet{sha14}.

We assume that the Galaxy is in a dynamical equilibrium, and use 
a modified Shu distribution function (DF) model developed by \citet{sch12} and \citet{sha13} to represent the distribution of disk azimuthal velocity $v_{\rm d, \phi}$.
Gaussian velocity models are used for $v_{\rm d, \phi}$ distributions in other Galactic models for microlens analysis \citep{sum11, ben14, zhu17, jun18} and in the Besan\c{c}on model \citep{rob03,rob12,rob17}.
However, a real $v_{\phi}$ distribution is highly skewed to low $v_{\phi}$ \citep[e.g.,][]{nor04, kat18}, and the Shu DF \citep{shu69} provides a much better approximation for it \citep{bin08, sha14, bla16}.

We introduce the guiding-center radius $R_g$ as the radius of a circular orbit with specific angular momentum $L_z$, i.e., $R_g = L_z/v_c$, where $v_c$ is the circular velocity \citep{bin08}.
The modified Shu DF model provides a joint distribution of the Galactocentric radius $R$ and $R_g$,
\begin{align}
P (R, R_g) = \frac{(2 \pi)^2 \Sigma (R_g)}{g\left( \frac{1}{2 a^2} \right)}  \exp {\left[ \frac{2 \ln (R_g/R) + 1 - R_g^2 / R^2}{2a^2} \right]}, \label{eq-ShuDF}
\end{align}
where $a = \sigma_{v_{{\rm d}, R}} (R_g) / v_c$, $g (c) = \frac{e^c \Gamma (c - 1/2)}{2 c^{c- 1/2}}$ with the velocity dispersion along radial direction
$\sigma_{v_{{\rm d}, R}} (R) = \sigma_{R, \odot} \exp[{-\frac{R - R_{\odot}}{R_{\sigma_R}}}]$ and the Gamma function $\Gamma (x)$. 
$R_{\sigma_R}$ is the scale length for the $\sigma_{v_{{\rm d}, R}} (R)$ distribution given by Eqs. (\ref{eq-sigthin})-(\ref{eq-sigthick}) below.
$\Sigma (R_g)$ is a function that controls disk surface density, and we use an empirical formula proposed by \citet{sha13},
\begin{align}
\Sigma (R_g) =  \frac{e^{-R_g/R_{\rm d} }}{2 \pi R_{\rm d} ^2} - \frac{c_3 a_0^{c_4}}{R_{\rm d} ^2} \times s  \left( \frac{R_g}{c_1 R_{\rm d}  (1 + q/c_2)} \right),
\end{align}
where $s(x) = 31.53 e^{-x/0.2743} ((x/0.6719)^2 -1)$ and $q = R_{\rm d}  / R_{\sigma_R}$.
We select $(c_1, c_2, c_3, c_4) = (3.822, 0.524, 0.00567, 2.13)$ from Table 1 of \citet{sha13} that is for the rising rotation curve of $v_c \propto (R/R_{\rm d} )^{0.2}$.
This is because we use a similar rising rotation curve of $v_c (R)$ from \citet{bla16}, which comes from the $N$-body dynamical model of \citet{por15} for $R_{\rm d}  = 2.6$ kpc.

A conditional probability for $R_g$ given $R$, $P (R_g | R)$, which is calculated using Eq. (\ref{eq-ShuDF}), is used to model the disk azimuthal velocity $v_{\rm d, \phi}$ distribution 
through the relation between $v_{\rm d, \phi}$ and $R_g$,
\begin{align}
v_{\rm d, \phi} (R, z) &= v_{c} (R_g, z) R_g/R \notag\\ 
&=  \frac{v_{c} (R_g) R_g/R}{1 + 0.0374 |z/{\rm kpc}|^{1.34}},
\end{align}
where we apply $v_c (R, z) = v_c (R) (1 + 0.0374 |z/{\rm kpc}|^{1.34})^{-1}$ \citep{sha14} for the vertical dependency of $v_c$.
Again, the rotation curve from \citet{bla16} is used for $v_c (R)$.

For the disk velocity along radial ($v_{{\rm d}, R}$) and vertical ($v_{{\rm d}, z}$) directions, we use the Gaussian distribution with mean velocity of 0 (i.e. dynamical equilibrium) with the velocity dispersion given by
\begin{align}
\sigma_{v_{{\rm d}, i}}^{\rm thin} (R) = \sigma_{i, \odot}^{\rm thin} \left ( \frac{T + T_{\rm min}}{T_{\rm max} + T_{\rm min}} \right)^{\beta_i} \exp \left[ - \frac{R - R_{\odot}}{R_{\sigma_i}^{\rm thin}} \right] \label{eq-sigthin}
\end{align}
for the thin disk and 
\begin{align}
\sigma_{v_{{\rm d}, i}}^{\rm thick} (R) = \sigma_{i, \odot}^{\rm thick} \exp \left[ - \frac{R - R_{\odot}}{R_{\sigma_i}^{\rm thick}} \right] \label{eq-sigthick}
\end{align}
for the thick disk, where $i$ takes $R$ or $z$, and we use $T_{\rm min} = 0.01$ Gyr and $T_{\rm max} = 10$ Gyr in this study.
Further, we introduce the dependence on stellar age $T$ for the thin disk to include the age-velocity dispersion relation owing to secular heating in the disk.
With this formula, the local velocity dispersion value calculated for thin disk, $\sigma_{i, \odot}^{\rm thin}$, is for stars with $T = 10$ Gyr.

In Section \ref{sec-disk_fit}, we investigate acceptable combinations of the following 10 fit parameters by comparing with the data from the giant sample by \citet{kat18};
the local velocity dispersion values, $\sigma_{i, \odot}^{\rm thin}$ and $\sigma_{i, \odot}^{\rm thick}$, slope of age-velocity dispersion relation, $\beta_i$, 
and scale lengths of the velocity dispersion distribution, $R_{\sigma_i}^{\rm thin}$ and $R_{\sigma_i}^{\rm thick}$, where $i$ takes $R$ or $z$. 
Note that additional parameters are not needed to represent the distribution of velocity dispersion along the azimuthal direction, such as the above parameters with $i = \phi$, because 
Eq. (\ref{eq-ShuDF}) naturally relates the $v_{{\rm d}, \phi}$ distribution to the $v_{{\rm d}, R}$ distribution.

\subsection{Barred Bulge Model}  \label{sec-bar_model}
Compared to the disk model, an analytical approximated expression for the bulge dynamical model is less developed because of its difficulty in the treatment of a non-axisymmetric property of the bar.
An $N$-body model is dynamically correct; however, fitting the model to the observational data is difficult.
A probable optimal technique is a made-to-measure method \citep{sye96}, where weights of particles are updated such that 
observables of the model match a given dataset during the simulation.
\citet{por17} developed an $N$-body dynamical model matching extensive photometric and kinematic data across our Galaxy using the made-to-measure method.
However, such a dynamical simulation is beyond the scope of this study because we aim to develop a parametric Galactic model, which can be easily implemented, reproduced, and updated { by} anybody.
Although some studies developed parametric models for the bar \citep{dwe95, rat07, rob12, cao13}, they lack constraints from some recent data such as the one by 
\citet{mro19}, who performed the largest statistical study for single-lens microlensing events through the OGLE-IV Galactic bulge survey, which is especially important for microlensing studies.

In this subsection, we describe our parameterization for the bulge density ($\rho_{\rm B}$) and velocity ($v_{\rm B}$) models.
We consider a total of four different shapes for the $\rho_{\rm B}$ model: two `one-component' models and two `two-components' models, 
in which 7 and 15 fit parameters are introduced, respectively.
The one-component model is designed following the parameterization used in the previous studies \citep{dwe95, rat07, rob12, cao13}, while 
the two-components model is designed to express the X-shape structure \citep{nat10}, which is not considered in the previous parametric models.
We consider a bar's rigid-body rotation and a streaming motion in the $v_{\rm B}$ model and introduce 19 fit parameters to model it.
Note that the bulge model contributes to the fits in Section \ref{sec-bar} rather than those in Section \ref{sec-disk}.

\subsubsection{Bulge density ($\rho_{\rm B}$ model)} \label{sec-bar_rho}
Our one-component bulge model follows the parameterization by \citet{rob12}, in which we consider each of E (exponential) and G (Gaussian) models given by
\begin{align}
\rho_{\rm B}^{(i)} = \rho_{\rm 0, B} \, f_i (x', y', z'; \bm p_{r_s}) \, {\rm Cut} \left[ \frac{R - R_{\rm c}}{0.5 \, {\rm kpc}} \right] \hspace{0.6cm} (i = {\rm E, G}), \label{eq-rhob1}
\end{align}
where ${\rm Cut} (x)$ is a cut-off function given by 
\begin{align}
{\rm Cut} (x) = 
\begin{cases}
\exp (-x^2)  & \text{ if $x > 0$} \\
1   & \text{ if $x \leq 0$}
\end{cases} \label{eq-Cut}
\end{align}
and $R_{\rm c}$ is the cut-off radius.
{ The two functions, $f_{\rm E} (x', y', z'; \bm p_{r_s})$ and $f_{\rm G} (x', y', z'; \bm p_{r_s})$, are defined as }
\begin{align}
f_{\rm E} (x', y', z'; \bm p_{r_s}) &= \exp[-r_s(x', y', z' ; \bm p_{r_s})], \notag\\
f_{\rm G} (x', y', z'; \bm p_{r_s}) &= \exp[- 0.5 r_s^2(x', y', z'; \bm p_{r_s})] \label{eq-fEG}
\end{align}
with 
\begin{align}
r_s (x', y', z'; {\bm p_{r_s}}) = \left\{ \left[ \left( \frac{x'}{x_0} \right)^{C_\perp} +  \left( \frac{y'}{y_0} \right)^{C_\perp} \right]^{C_\parallel/C_\perp} + \left( \frac{z'}{z_0} \right)^{C_\parallel} \right\}^{1/C_\parallel}
\end{align}
and $\bm p_{r_s} = (x_0, y_0, z_0, C_\perp, C_\parallel)$.
We use $(x', y', z')$ to refer to a Galactocentric coordinate system rotated around the $z$-axis by an angle $\alpha_{\rm bar}$ 
such that the $x'$ axis is aligned with the major axis of the Galactic bar, where $\alpha_{\rm bar} = 27^\circ$ is applied as the bar angle.
The parameters $(x_0, y_0, z_0)$ are the scale lengths along $(x', y', z')$ axes, and $C_\perp$ and $C_\parallel$ allow the bar to take various shapes \citep{rob12}.

Motivated by the X-shape structure confirmed both observationally and dynamically \citep{mcw10, nat10, weg13}, 
we consider a two-components model, { $\rho_{\rm B} = \rho_{\rm B}^{(i)} + \rho_{\rm X}^{(j)}$, where $\rho_{\rm B}^{(i)}$ is given by Eq. (\ref{eq-rhob1}) and }
\begin{align}
\rho_{\rm X}^{(j)} = \rho_{\rm 0, X} \left[ f_j (x' - b_{\rm X} z', y', z'; \bm p_{r_s, {\rm X}}) + f_j (x' + b_{\rm X} z', y', z'; \bm p_{r_s, {\rm X}})  \right] {\rm Cut} \left[ \frac{R - R_{\rm c, X}}{0.5 \, {\rm kpc}} \right] \ \hspace{0.6cm} (j = {\rm E, G}).  \label{eq-rhoX}
\end{align}
{ The} parameter $b_{\rm X}$ controls the slope of an X-shape, and we use another parameter set,  
$\bm p_{r_s, {\rm X}} = (x_{\rm 0, X}, y_{\rm 0, X}, z_{\rm 0, X}, C_{\rm \perp, X}, C_{\rm \parallel, X})$, which is different from the $\bm p_{r_s}$ for the first component given by Eq. (\ref{eq-rhob1}). 
Note that the X-shape structure with this expression is centered on the Galactic center, although \citet{por15} found a slightly off-centered X-shape structure in their $N$-body dynamical model.
We considered all four combinations of $i = {\rm E, G}$ and $j = {\rm E, G}$, and found no significant difference among those combinations with respect to agreement with the fitted data.
Hence, herein, we present results of two combinations of $(i, j)$ = (E, E) and (G, G) among the four.
Hereafter, we refer to these models as the E+E$_{\rm X}$ and G+G$_{\rm X}$ models, respectively.

Furthermore, other structures known in the bulge region are locally significant but not captured by either of the four $\rho_{\rm B}$ models.
For example, a nuclear stellar disk exists in a central sub-kpc region \citep{lau02, nis13, por17}.
The scale height and the outer edge of the nuclear stellar disk are $\sim 45~{\rm pc}$ \citep{nis13} and $\sim 230~{\rm pc}$ \citep{bla16}, respectively, 
which indicates no influence on the used data in this study ranged in $|b| \simgt 2^{\circ}$.
A long bar component was found to be distributed in the outer bulge region along the major axis by \citet{weg15}; hence, it was added in our model to observe its effect, but no significant improvement was observed regarding the ${ \tilde{\chi}^2_{\rm sum}}$ value defined in Section \ref{sec-bar_fit}.
This is probably because the used data lacks the sky region in $|l| > 10^\circ$, where the long bar component becomes prominent.
Therefore, we consider each of the E, G, E+E$_{\rm X}$ and G+G$_{\rm X}$ models with no additional components.

The 7 fit parameters for the E and G models are: $x_0, y_0, z_0, C_\perp, C_\parallel, R_c$, and $\rho_{\rm 0, B}$.
In the E+E$_{\rm X}$ and G+G$_{\rm X}$ models, the additional 8 fit parameters are: $x_{\rm 0, X}, y_{\rm 0, X}, z_{\rm 0, X}, C_{\rm \perp, X}, C_{\rm \parallel, X}, R_{\rm c, X}$, $b_{\rm X}$, and $f_{\rm 0, X}$, where 
$f_{\rm 0, X} \equiv \rho_{\rm 0, X}/\rho_{\rm 0, B}$.
%Note that the normalization factor $\rho_{\rm 0, B}$ moves under a prior constraint of $M_{\rm VVV} = 1.32 \pm 0.08 \times 10^{10}~M_{\odot}$ \citep{por17}, 
%where $M_{\rm VVV}$ is the model integrated mass within the VVV bulge box which is defined as (\pm 2.2 \times \pm 1.4 \times \pm 1.2 kpc) along the bar axes by \citet{weg13}.
%Details of our prior constraints are described in Section \ref{sec-prior}.

\subsubsection{Bulge kinematics ($v_{\rm B}$ model)}
For the velocity distribution of the bulge star, we use the Gaussian distribution with a mean velocity and velocity dispersion, both varying as a function of $(x', y', z')$.
The mean velocity is calculated by combining the rigid-body rotation of the bar and streaming motion along the bar.
We denote the angular velocity of the rigid-body rotation or the bar pattern speed by $\Omega_{\rm p}$ and consider the streaming motion along the major axis as 
\begin{align}
v_{x'}^{\rm str} (y') = v_0^{\rm str} (1 - \exp [-(y'/y_0^{\rm str})^2] ),
\end{align}
where $v_0^{\rm str}$ is the streaming velocity at $y' \gg y_0^{\rm str}$, and $y_0^{\rm str}$ is the scale length along $y'$ axis.
This form of distribution for the streaming motion is motivated by the bottom panels in Figure 14 of \citet{san19} that show the 
$v_{x'}^{\rm str}$ of their dynamical model, which represents our Galaxy, increasing from $y' = 0$ to $|y'| > 0$ along $y'$ axis.

For the velocity dispersion, we use
\begin{align}
\sigma_{v_{\rm B}, i} (x', y', z')  = \sigma_{v_{\rm B}, i, 0} + \sigma_{v_{\rm B}, i, 1} \, f_{\rm E} (x', y', z'; \bm p_{r_s, \sigma_i})  \hspace{0.8cm} (i = x', y', z'), \label{eq-sigb}
\end{align}
where we apply $\bm p_{r_s, \sigma_{x'}} = \bm p_{r_s, \sigma_{y'}} \ne \bm p_{r_s, \sigma_{z'}}$ and denote the parameter set for $\sigma_{x'}$ and $\sigma_{y'}$ by $\bm p_{r_s, \sigma_R}$, 
i.e., $\bm p_{r_s, \sigma_R} = \bm p_{r_s, \sigma_{x'}} = \bm p_{r_s, \sigma_{y'}} = (x_{\rm 0, \sigma_R}, y_{\rm 0, \sigma_R}, z_{\rm 0, \sigma_R}, C_{\rm \perp, \sigma_R}, C_{\rm \parallel, \sigma_R})$.
The constant $\sigma_{v_{\rm B}, i, 0}$ provides a minimum value of $\sigma_{v_{\rm B}, i}$ while $\sigma_{v_{\rm B}, i, 0} + \sigma_{v_{\rm B}, i, 1}$ provides the maximum $\sigma_{v_{\rm B}, i}$ value at the Galactic center.

This model is not dynamically consistent with the density model described in Section \ref{sec-bar_rho}; 
however, the profile of Eq. (\ref{eq-sigb}) peaking at the Galactic center and gradually decreasing as it goes around is motivated by 
the velocity dispersion field of the dynamical model of our Galaxy by \citet{san19} illustrated in their Figure 15.

There are a total of 19 fit parameters for the velocity model; 
$\Omega_{\rm p},  v_0^{\rm str}, y_0^{\rm str}, \sigma_{v_{\rm B}, i, 0}, \sigma_{v_{\rm B}, i, 1} (i = x', y', z')$,
$x_{\rm 0, \sigma_R}, y_{\rm 0, \sigma_R}, z_{\rm 0, \sigma_R}, C_{\rm \perp, \sigma_R}, C_{\rm \parallel, \sigma_R}$, 
$x_{\rm 0, \sigma_{z'}}, y_{\rm 0, \sigma_{z'}}, z_{\rm 0, \sigma_{z'}}, C_{\rm \perp, \sigma_{z'}},$ and $C_{\rm \parallel, \sigma_{z'}}$.
%Although our choice of $\bm p_{r_s, \sigma_{y'}} \ne \bm p_{r_s, \sigma_{z'}}$ adds five parameters to describe the velocity distribution, 
%we found that the improvement with this addition was not that great with respect to the agreement (or $\chi^2$) with the data.
%However, this 

\section{Fitting for the Disk Velocity Parameters}  \label{sec-disk}
In this section, we determine the 10 fit parameters for the $v_{\rm d}$ model by fitting our disk model to the spatial distribution 
of the median velocity and the velocity dispersion for giant stars in the Gaia DR2 \citep{kat18}, where the data are given in grids of 200 pc by 200 pc in $(R, z)$.
Bulge stars rarely contribute to the Gaia data and the barred bulge model is not used in the fit in this section.
The fit is conducted through a grid search. 
The disk model with the determined parameters in this section is used for the other fits in Section \ref{sec-bar}.

\subsection{Gaia DR2 Velocity Data} \label{sec-disk_data}
We use the median velocity and velocity dispersion distributions of the red giant sample consisting of 3,153,160 sources by \citet{kat18} as a function of the Galactocentric radius $R$ and 
the height from the Galactic plane $z$.
The stars in this sample are selected based on their absolute magnitude ${\cal M}_G < 3.9$ mag and intrinsic color $(G_{\rm BP} - G_{\rm RP})_0 > 0.95$ in the Gaia bands.
This is the same data plotted in Figure 11 of \citet{kat18}, and we obtained the data through private communication with the lead author, D. Katz.
The medians and dispersions are given in grids of 200 pc by 200 pc in $(R, z)$ over $3340~{\rm pc} < R < 13340~{\rm pc}$, $-3400~{\rm pc} < z < 3400~{\rm pc}$.
The center of $(i_R, i_z)$th grid is $(R, z) = (200\, i_R + 3440, 200\, i_z - 3300)~{\rm pc}$.
The total grids in the range of $R$ and $z$ are $50 \times 34 = 1700$; however, we do not use the grids with $< 30$ stars contributing to the statistics, which results in 1207 grids being available.

\begin{figure}
\centering
\includegraphics[width=15cm]{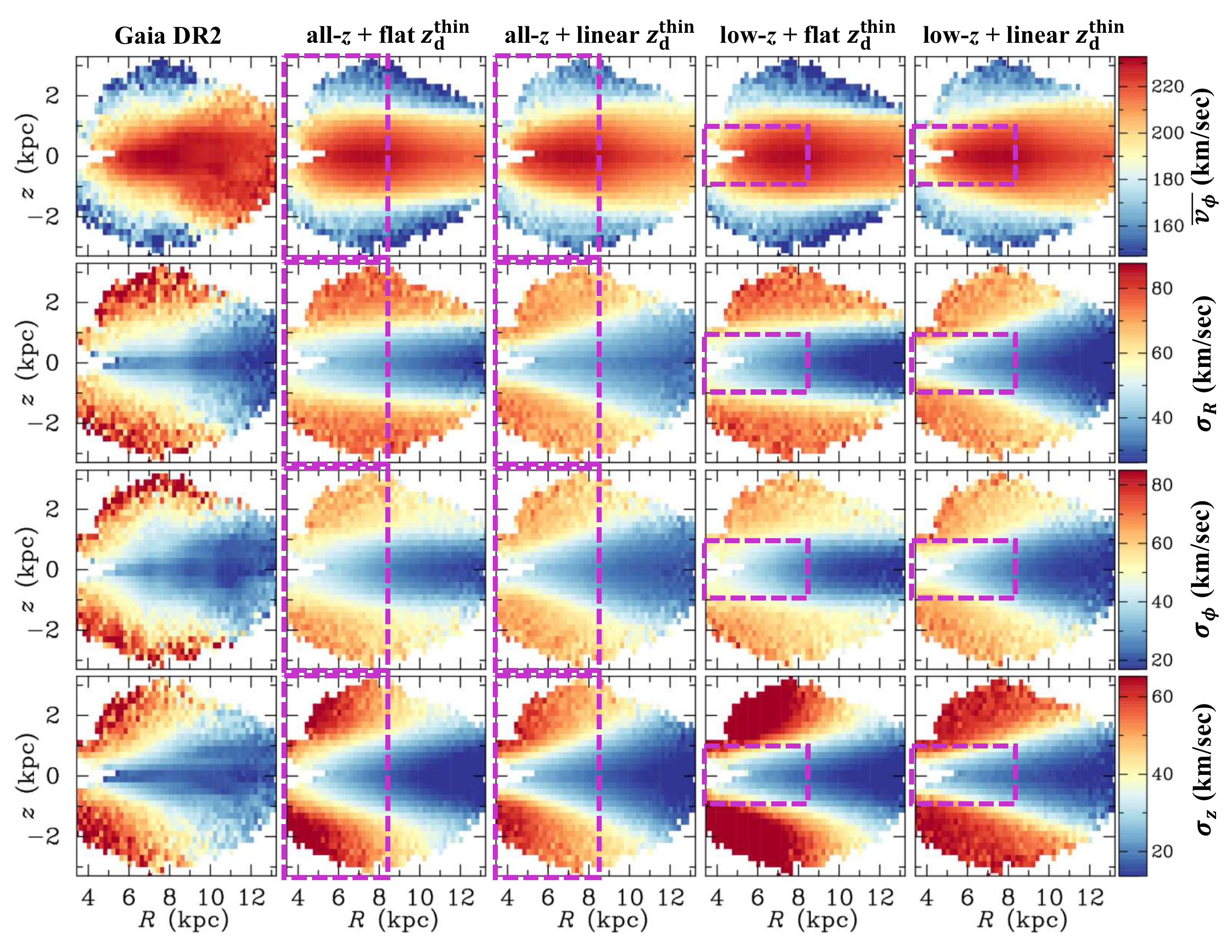}
\caption{Far left: The Gaia DR2 giant sample's $R$--$z$ distributions for median azimuthal velocity ($\overline{v_{\phi}}$) and velocity dispersions along the radial ($\sigma_R$), azimuthal ($\sigma_R$), 
and vertical ($\sigma_z$) directions from top to bottom \citep[reproduction of Figure 11 of ][]{kat18}. The right four columns: The same distributions from each indicated model. 
Only the grids inside the purple boxes are used in the calculation of $\tilde{\chi}^2$ for each model.}
\label{fig-velo-gaia}
\end{figure}

The medians of the $v_R$ and $v_z$ distribution are assumed as 0; hence, only the data of the median of the azimuthal velocity $\overline{v_{\phi}}$ and 
velocity dispersions along the three axes, $\sigma_{R}, \sigma_{\phi}$, and $\sigma_z$, are used.
The data distribution for these four parameters is plotted in the far-left panels in Fig. \ref{fig-velo-gaia}.

\subsection{Definition of Goodness of Fit}
The far-right column of Table \ref{tab-disks} lists the local number density of red giants in our model, which is calculated using the same criteria as for the Gaia DR2 giant sample.
We use these $n_{\rm d, \odot}^{\rm RG}$ values instead of $\rho_{\rm d, \odot}$ in Eqs. (\ref{eq-rho_thin})-(\ref{eq-rho_thick}) to calculate model values for the median $\overline{v_{\phi}}$ and 
the velocity dispersions $\sigma_{R}, \sigma_{\phi}$, and $\sigma_z$.
Thereafter, Monte Carlo random sampling is used to calculate the model values for each grid of the data with a particular combination of the 10 fit parameters, 
i.e., $\sigma_{i, \odot}^{\rm thin}$, $\sigma_{i, \odot}^{\rm thick}$, $\beta_i$, $R_{\sigma_i}^{\rm thin}$, and $R_{\sigma_i}^{\rm thick}$ ($i = R, z$).
We assume that data incompleteness does not affect the kinematic statistics in each grid and do not consider the completeness correction for the comparison between the data and model values.

%Figure 1

%Figure 2
\begin{figure}
\centering
\includegraphics[width=12cm]{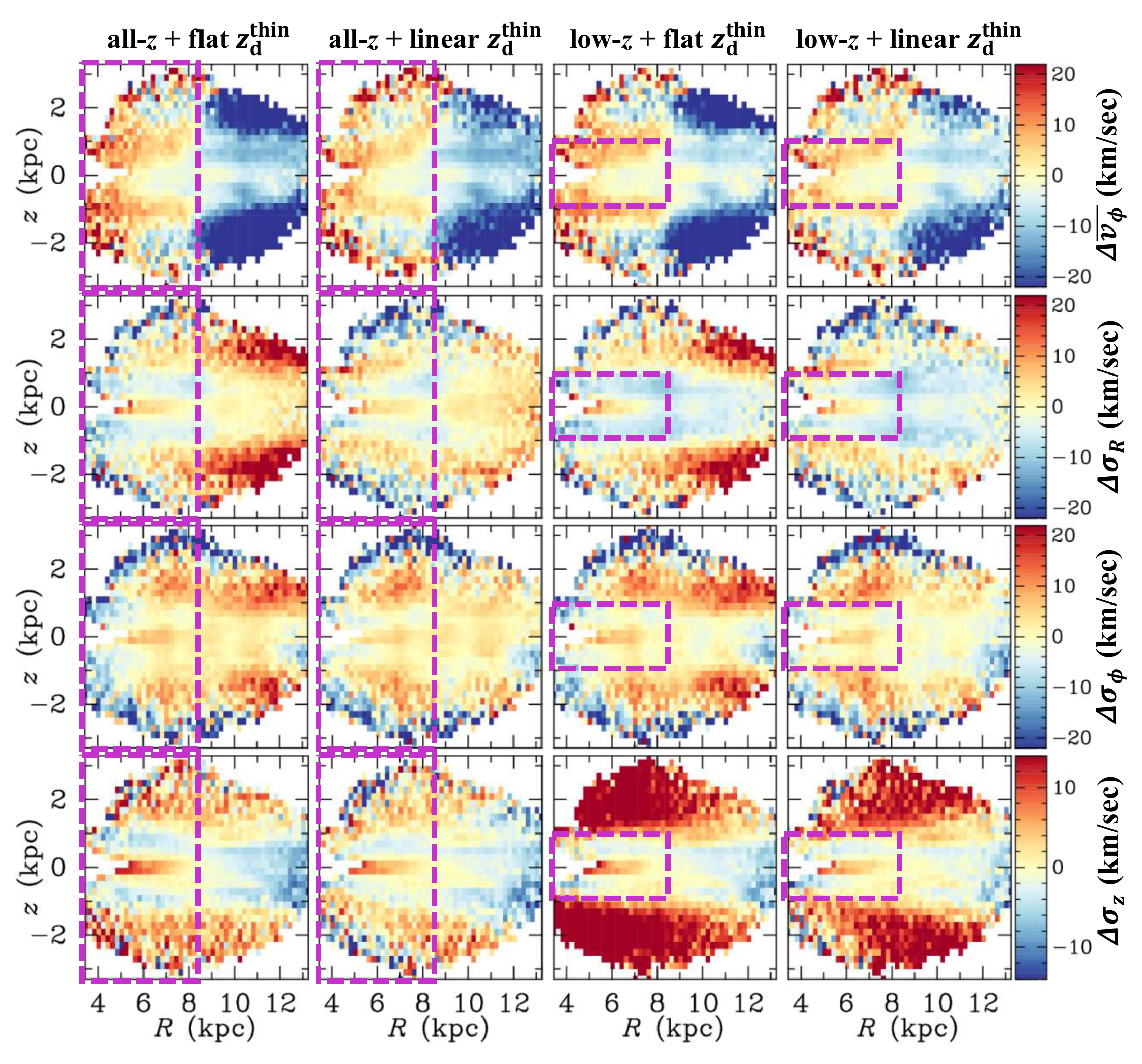}
\caption{Residuals (model $-$ data) corresponding to the four models plotted in Fig. \ref{fig-velo-gaia}.}
\label{fig-disk_resi}
\end{figure}

To evaluate an agreement with the data at $(i_R, i_z)$th grid, we use
\begin{align}
\tilde{\chi}^2_{i_R, i_z} = \sum_{p = \overline{v_{\phi}}, \sigma_{R}, \sigma_{\phi}, \sigma_z} w_{i_R, i_z} (p_{i_R, i_z}^{\rm simu} - p_{i_R, i_z}^{\rm obs} )^2, \label{eq-chi2Rz}
\end{align}
where 
\begin{align}
w_{i_R, i_z} = \frac{\left(1/N_{i_R, i_z}^{\rm obs} + 1/N_{i_R, i_z}^{\rm simu} \right)^{-1}}{\sum_{j_R, j_z} \left(1/N_{j_R, j_z}^{\rm obs} + 1/N_{j_R, j_z}^{\rm simu} \right)^{-1}}, \label{eq-wRz}
\end{align}
and $N_{i_R, i_z}^{\rm obs}$ is the number of observed stars in a grid, while $N_{i_R, i_z}^{\rm simu}$ is the number of simulated stars in the same grid.
$N_{i_R, i_z}^{\rm obs}$ takes 30--34243 depending on grid position, where the median is $\sim 250$ and 67 out of 1207 grids have $N_{i_R, i_z}^{\rm obs} > 10^4$.
The weight for each grid, $w_{i_R, i_z}$, disregards each measurement uncertainty because we were not provided the uncertainties of measurements of the median velocity and 
velocity dispersion through the private communication.
This corresponds to an assumption that the average measurement uncertainty in each grid is the same regardless of the grid; however,
this is not true because velocity measurements in a grid further from the Sun tend to have larger uncertainties than those in a closer grid due to its relative faintness, and hence 
a closer grid should have a larger weight considering this effect.
Nevertheless, the relative weights among different grids are, qualitatively, correctly set by the current form of $w_{i_R, i_z}$ because a closer grid tends to 
have a larger $N_{i_R, i_z}^{\rm obs}$ value, which makes its relative weight larger.
Therefore, although there should be an underestimation of relative weight for a grid closer to the Sun, our result is not expected to change significantly with the missed effect.

Another concern with the $w_{i_R, i_z}$ expression by Eq. (\ref{eq-wRz}) is its dependency on $N_{i_R, i_z}^{\rm simu}$.
To reduce the dependency on the simulated number of stars but not increase the computation time significantly, we adopt
$N_{i_R, i_z}^{\rm simu} = {\rm min} (10^4, 4 \, N_{i_R, i_z}^{\rm obs})$ for each grid.
The maximum simulation number of $10^4$ indicates that the $N_{i_R, i_z}^{\rm simu}$ term becomes dominant in $w_{i_R, i_z}$ when $N_{i_R, i_z}^{\rm obs} > 10^4$.
We set this maximum primarily to reduce the computation time, but also to avoid placing too much weight for grids with large $N_{i_R, i_z}^{\rm obs}$ values, because 
we want to have a model that matches a wide range of grid points instead of being optimized to a small fraction of the grid points that is highly weighted.

A particular set of the 10 fit parameters is evaluated by $\tilde{\chi}^2 = \sum_{i_R, i_z} \tilde{\chi}^2_{i_R, i_z}$. 
The summation for $\tilde{\chi}^2$ only run over 651 grids with $R \leq 8440~{\rm pc}$ out of a total of 1207 grids because 
our primary science interest is in microlensing events toward the Galactic bulge and because the outer disk has a warp and/or a flare feature, which is not considered in our model.
The cut at $8440~{\rm pc}$ is motivated by the starting Galactocentric radius of the warp used in the Besan\c{c}on model \citep{rob03} of $R = 8400~{\rm pc}$.
We refer to a model fitted with this data selection as an all-$z$ model.
Additionally, we consider another option, which is highly optimized for microlensing studies toward the Galactic bulge, where the summation for $\tilde{\chi}^2$ run 
over 234 grids with $R \leq 8440~{\rm pc}$ and $|z| \leq 900~{\rm pc}$.
With this option, we also ignore the agreement in $\sigma_{R}$, i.e., the $\tilde{\chi}^2_{i_R, i_z}$ in 
Eq. (\ref{eq-chi2Rz}) just run over $p = \overline{v_{\phi}}, \sigma_{\phi}, \sigma_z$ and not over $p = \sigma_{R}$.
This is because $\sigma_{R}$ is irrelevant to observables of microlensing events in fields near the Galactic center.
We refer to a model from this data selection as a low-$z$ model.
%Note that the difference between these two models is not only in $z$ but also in weight for the $\sigma_{R}$ data, although the names do not contain information about $\sigma_{R}$.

\subsection{Grid Search}  \label{sec-disk_fit}
Because the Monte Carlo simulation to calculate $\tilde{\chi}^2$ is computationally expensive, we conduct 
a grid search to find a combination of parameters with a better agreement to the data.
We conduct a sparse initial grid search of the 10 fit parameters with a large interval between adjacent grids at first, and then repeat 
it in narrower parameter space with a smaller interval.
The initial grid search run over the following ranges;
$\sigma_{R, \odot}^{\rm thin}= 30$--$50~{\rm km/s}$, 
$\sigma_{z, \odot}^{\rm thin}= 15$--$30~{\rm km/s}$, 
$\sigma_{R, \odot}^{\rm thick}= 40$--$70~{\rm km/s}$, 
$\sigma_{z, \odot}^{\rm thick}= 35$--$65~{\rm km/s}$, 
$\beta_{R} = 0.1$--0.4,
$\beta_{z} = 0.2$--0.8,
$R_{\sigma_R}^{\rm thin} = 8$--$30~{\rm kpc}$,
$R_{\sigma_z}^{\rm thin} = 8$--$30~{\rm kpc}$,
$R_{\sigma_R}^{\rm thick} = 8$--$30~{\rm kpc}$,
and $R_{\sigma_z}^{\rm thick} = 8$--$30~{\rm kpc}$.
Once a best-fit combination was found at an edge of the parameter space searched, we expanded the parameter space in the next iteration.
This procedure was repeated until no significant $\tilde{\chi}^2$ improvement was found with a smaller interval or an expansion of the parameter space.

We conduct the above search for each of the four options, i.e., all combinations of the two options for the data selection (all-$z$ or low-$z$) and 
thin disk scale height (flat or linear), respectively.
Table \ref{tab-disk_result} shows the best-fit parameters given by the calculations. 
Figs. \ref{fig-velo-gaia} and \ref{fig-disk_resi} show color maps of each model values and residuals (model value $-$ data value), respectively.
In each panel of these figures, we indicate the selected grid region with magenta dashed boxes. 
Fig. \ref{fig-disk_resi} shows that our models fail to reproduce the outer disk distributions, in particular, the $\overline{v_{\phi}}$, but this is expected because 
our density model is designed for the inner disk and does not consider the warp or flare structure seen in the outer disk.
Similarly, the $\sigma_{z}$ distribution of the low-$z$ model significantly overestimates the values in $z > 900$ pc which is outside the magenta box.
This is because the best-fit $\sigma_{z, \odot}^{\rm thick}$ value for a low-$z$ model (61.4 or 59.0 km/s) is much higher than that for an all-$z$ model (49.2 or 47.8 km/s), as shown in Table \ref{tab-disk_result}.
However, this difference has little effect in the microlensing region toward the Galactic bulge because the thick disk stars are relatively rare in this region.
Focusing on the region related to the microlensing study (i.e., inside the magenta box of a low-$z$ model), all the four models show moderate agreements with the Gaia data.

The $\tilde{\chi}^2$ value is smallest for the low-$z$ $+$ linear $z_{\rm d}^{\rm thin}$ model; however, comparison of the $\tilde{\chi}^2$ values between low-$z$ and all-$z$ models is unreasonable because 
the grids contributing to the $\tilde{\chi}^2$ are different.
The $\tilde{\chi}^2$ values, defined by Eq (\ref{eq-chi2Rz}), is the weighted root mean square of deviation of the model from the data, that is, $\sqrt{\tilde{\chi}^2}$ provides a weighted average of the deviations in km/s.
Notably, the linear scale height models are preferred for both the all-$z$ and low-$z$ models. 
Although this might indicate that the scale height is not constant inside the solar radius, the flat scale height models are favored in the fits to the data toward bulge regions as described in Section \ref{sec-modsel}.

We keep all the four models as options for the disk model, and use each of them combined with a bulge model in the fits conducted in Section \ref{sec-bar}, and 
the four models are compared in Section \ref{sec-modsel} with respect to the best-fit $\chi^2$ values to the bulge data.
We discuss the determined values in Table \ref{tab-disk_result} by comparing with the previous studies in Section \ref{sec-disk_comp}.

% Table 2
\begin{deluxetable}{lccccccccccccccccc}
\tabletypesize{\scriptsize}
\tablecaption{Best-fit parameters for the disk kinematic model. \label{tab-disk_result}}
\tablehead{
\colhead{Model}  & \colhead{$N_{\rm grid}$} & \colhead{$\sigma_{R, \odot}^{\rm thin}$} & \colhead{$\sigma_{z, \odot}^{\rm thin}$} & \colhead{$\sigma_{R, \odot}^{\rm thick}$} & \colhead{$\sigma_{z, \odot}^{\rm thick}$}
                 & \colhead{$\beta_{R}$} & \colhead{$\beta_{z}$} & \colhead{$R_{\sigma_R}^{\rm thin}$} & \colhead{$R_{\sigma_z}^{\rm thin}$} & \colhead{$R_{\sigma_R}^{\rm thick}$} & \colhead{$R_{\sigma_z}^{\rm thick}$}
                 & \colhead{$\tilde{\chi}^2_{\overline{v_{\phi}}}$} & \colhead{$\tilde{\chi}^2_{\sigma_{R}}$}   & \colhead{$\tilde{\chi}^2_{\sigma_{\phi}}$}& \colhead{$\tilde{\chi}^2_{\sigma_z}$}  \\
\colhead{}       &\colhead{} & \colhead{[km/s]} & \colhead{[km/s]} & \colhead{[km/s]} & \colhead{[km/s]}  &      &      & \colhead{[kpc]} & \colhead{[kpc]} & \colhead{[kpc]} & \colhead{[kpc]}  &      &        &      &       
}
\startdata
all-$z$ $+$ flat   $z_{\rm d}^{\rm thin}$ & 651 &   42.0             & 24.4               & 75                 &   49.2              & 0.32 & 0.77 & 14.3            &       5.9       &   180           & 9.4              & 19.5 & 16.3   & 15.9 & 9.6 \\
all-$z$ $+$ linear $z_{\rm d}^{\rm thin}$ & 651 &   44.0             & 25.4               & 68                 &   47.8              & 0.34 & 0.81 & 21.4            &       8.1       &   57.6          & 15.6             & 14.9 & 13.7   & 14.1 & 6.7 \\
low-$z$ $+$ flat   $z_{\rm d}^{\rm thin}$ & 234 &   35.2             & 22.2               & 75                 &   61.4              & 0.22 & 0.77 &  9.5            &      10.4       &   90.0          & 6.9              & 12.6 &  7.2   &  --  & 4.2 \\
low-$z$ $+$ linear $z_{\rm d}^{\rm thin}$ & 234 &   37.6             & 23.4               & 68                 &   59.0              & 0.30 & 0.82 & 11.1            &       7.8       &   47.0          & 52.0             &  8.7 &  5.8   &  --  & 3.5 \\
\enddata
\end{deluxetable}

\section{Fitting for the Bulge and IMF parameters}   \label{sec-bar}
In this section, we determine the 4 fit parameters for the IMF model, 7--15 fit parameters for the $\rho_{\rm B}$ model, and 19 fit parameters for the $v_{\rm B}$ model, through 
the Markov Chain Monte Carlo (MCMC) fitting to the observed distributions toward the bulge sky of the OGLE-III RC star count \citep{nat13}, VIRAC proper motion \citep{smi18, cla19}, BRAVA radial velocity \citep{ric07, kun12}, and star and microlensing event count by OGLE-IV \citep{mro17, mro19}.

We use the bulge model combined with the disk model in the fit, where we fix and use the 10 fit parameters for the $v_{\rm d}$ model determined in Section \ref{sec-disk}.
However, the local density values are recalculated in the fit for every given set of the 4 fit parameters in the IMF.

\subsection{Data and Corresponding $\chi^2$ Values} \label{sec-bar_data}
This section describes details of each dataset used to constrain the 30--38 fit parameters.
For each dataset, we also describe model values compared with the observed values and introduce the corresponding $\chi^2$ values.
All the used data are plotted in Figs. \ref{fig-mod_rhob}, \ref{fig-mod_vb}, and \ref{fig-mod_tE}.

\subsubsection{OGLE-III red clump star count} \label{sec-RCdata}
\citet{nat13} divided the OGLE-III Galactic bulge fields over 90.25 deg$^2$ in $-10^\circ < l < 10^\circ$ and $2^\circ < |b| < 7^\circ$ into 9019 areas and 
measured the red clump (RC) star count in each area.
They used a luminosity function model that primarily consists of a Gaussian RC component on an exponential red giant branch continuum for fit to the observed $I$-mag distribution in each area.
As a result of the fits, they provided a catalog of 9019 sets of the number count $N_{\rm RC}$, mean distance modulus ${\rm DM}_{\rm RC}$, variance of distance modulus $\sigma_{\rm DM}^2$, and error matrix for them, as well as the extinctions $A_I$ and $A_V$, where $N_{\rm RC}$ and ${\rm DM}_{\rm RC}$ are equivalent to the area and the peak of the Gaussian, respectively.
The variance of distance modulus, $\sigma_{\rm DM}^2$, was determined by subtracting the sum of variances of the extinction 
in the area ($= \sigma_{A_I}^2$) and of the intrinsic brightness of RC ($= \sigma_{I, {\rm RC}, 0}^2$) from the variance of the Gaussian ($= \sigma_{I, {\rm RC}}^2$).

\citet{nat13} used $\sigma_{I, {\rm RC}, 0} = 0.09$ with no uncertainty to derive the $\sigma_{\rm DM}^2$ values.
To be conservative, we apply $\sigma_{I, {\rm RC}, 0} = 0.15 \pm 0.06$, i.e., we modify the data by $\sigma_{\rm DM}^2 = \sigma_{\rm DM, org}^2 + 0.09^2 - 0.15^2$ and 
increase the uncertainty accordingly.
This choice for $\sigma_{I, {\rm RC}, 0}$ is motivated by \citet{haw17} who measured the RC mag dispersions of $0.20 \pm 0.02$ in both $G$- and $J$-bands.
Because the wavelength of $I$-band is between these bands, we conservatively take $0.15 \pm 0.06$ so that its 1-$\sigma$ range includes both 0.09 used by \citet{nat13} and 0.20 measured for $G$- and $J$-bands.

Moreover, because the measurements of $N_{\rm RC}, {\rm DM}_{\rm RC}$, and $\sigma_{\rm DM}^2$ are all from the Gaussian fit, it could 
overestimate/underestimate the RC population when it is not distributed following a Gaussian shape.
In particular, because our disk model has a constant surface density at $R < 5.3$ kpc (see Eqs. \ref{eq-rho_thin}--\ref{eq-rho_thick}), which is continuously distributed in the bulge region mildly,
most or part of the disk RC population is expected to be absorbed into the exponential red giant branch continuum component in the fit by \citet{nat13}, although the fraction of absorption probably depends on the line of sight.
To account for this uncertainty, we considered only the bulge population in the model observables given below in Eqs. (\ref{eq-NRC})--(\ref{eq-sig2DM}), in addition to increasing the uncertainties of $N_{\rm RC}$ by 10\% and 
adding 0.04 mag error to the uncertainties of ${\rm DM}_{\rm RC}$ in quadrature. 

\citet{cao13} modeled a bulge density distribution by fitting to the \citet{nat13} data, and we similarly follow their parameterization but with a modification on the integration range.
For a particular $i$th line of sight toward $(l_i, b_i)$, the model RC number count is expressed as
\begin{align}
N_{{\rm RC}, i}^{\rm mod}  = \Omega_i N_{{\rm B}, i} \times \frac{\sum_j N_{{\rm RC}, j}^{\rm obs}}{\sum_j \Omega_j N_{{\rm B}, j}}, \label{eq-NRC}
\end{align}
where $\Omega_i$ is the sky area of the $i$th field, and $N_{{\rm B}, i}$ is the model number of bulge stars integrated along $i$th line of sight, which is defined as
\begin{align}
N_{{\rm B}, i} \equiv \int_{s_{{\rm min}, i}}^{s_{{\rm max}, i}} n_{\rm B} (l_i, b_i, s) s^2 ds, \notag
\end{align}
and $n_{\rm B} (l_i, b_i, s)$ is the number density of bulge stars at the distance $s$ from the Sun toward $(l_i, b_i)$.
The summation for $j$ in the second factor in Eq. (\ref{eq-NRC}) runs over all the 9019 lines of sight, and the factor is for a normalization to let the total $N_{{\rm RC}, i}^{\rm mod}$ be same as the 
observed one, i.e., to make $\sum_i N_{{\rm RC}, i}^{\rm mod}  = \sum_i N_{{\rm RC}, i}^{\rm obs}$.

The mean and variance of distance modulus are expressed as
\begin{align}
{\rm DM}_{{\rm RC}, i}^{\rm mod}  &= \frac{1}{N_{{\rm B}, i}} \int_{s_{{\rm min}, i}}^{s_{{\rm max}, i}} n_{\rm B} (l_i, b_i, s) s^2 {\rm DM} (s) ds  \label{eq-DMRC}
\end{align}
and
\begin{align}
(\sigma_{{\rm DM},i}^2)^{\rm mod}  &= \frac{1}{N_{{\rm B}, i}} \int_{s_{{\rm min}, i}}^{s_{{\rm max}, i}} n_{\rm B} (l_i, b_i, s) s^2 [{\rm DM} (s)]^2 ds - ({\rm DM}_{{\rm RC}, i}^{\rm mod})^2,   \label{eq-sig2DM}
\end{align}
respectively, where 
${\rm DM} (s)$ denotes the distance modulus at distance $s$, which is given by ${\rm DM} (s) = 5 \log_{10} [s/(10 {\rm pc})]$.

For the integration range, we use a pair of $s_{{\rm min}, i}$ and $s_{{\rm max}, i}$ that satisfies
\begin{align}
{\rm DM} (s_{{\rm min}, i}) &= {\rm DM}_{{\rm RC}, i}^{\rm obs} - {\rm min} ( 3 \sigma_{I, {\rm RC}, i}^{\rm obs} \, , 1.5 ) \\
{\rm DM} (s_{{\rm max}, i}) &= {\rm DM}_{{\rm RC}, i}^{\rm obs} + {\rm min} ( 3 \sigma_{I, {\rm RC}, i}^{\rm obs} \, , 1.5), 
\end{align} 
where $\sigma_{I, {\rm RC}, i}^{\rm obs}$ is the standard deviation of the Gaussian fit for the RC component in the $I$-band luminosity function of $i$th field.
We chose these values rather than $s_{{\rm min}, i} = 3~{\rm kpc}$ and $s_{{\rm max}, i} = 13~{\rm kpc}$ used by \citet{cao13} because 
the RC stars outside of $\pm 3$-$\sigma$ from the mean have a negligible contribution to the measurements of the three observables, which are equivalent to the area, 
peak position, and variance of the Gaussian distribution.
The value of 1.5 for the maximum range of integration originates from the limit on the fitting range of magnitude, $-1.5 < I - I_{\rm RC} < 1.5$, set by \citet{nat13}.

Following \citet{cao13}, we calculate the $\chi^2$ for this dataset by
\begin{align}
\chi^2_{\rm RC} = \sum_i ({\bm X}_i^{\rm mod} -  {\bm X}_i^{\rm obs})^T {\bm S}^{-1}_i ({\bm X}_i^{\rm mod} -  {\bm X}_i^{\rm obs}),
\end{align}
where ${\bm X}_i = (N_{{\rm RC}, i}, {\rm DM}_{{\rm RC}, i}, \sigma_{{\rm DM},i}^2)$ and ${\bm S}^{-1}_i$ is the covariance matrix of the uncertainties.
We further define the following four $\chi^2$ values to quantify a contribution to $\chi^2_{\rm RC}$ from each observable; 
\[ \chi^2_p = \sum_i \left( \frac{p_i^{\rm mod} - p_i^{\rm obs}}{p_i^{\rm err}} \right)^2  \ \ \ \ (p = N_{\rm RC}, {\rm DM}_{\rm RC},  \sigma_{{\rm DM}}^2), \] 
and $\chi^2_{\rm cov} = \chi^2_{\rm RC} - \chi^2_{N_{\rm RC}} - \chi^2_{{\rm DM}_{\rm RC}} - \chi^2_{\sigma_{{\rm DM}}^2}$, where $p_i^{\rm err}$ denotes the error-bar of the $i$th data of parameter $p$.

\subsubsection{VIRAC red giants' proper motions} \label{sec-mudata}
\citet{smi18} provided the VVV infrared astrometric catalogue (VIRAC), which is a near-infrared proper motion catalog of the five years
VISTA Variables in the Via Lactea (VVV) survey \citep{min10}, which includes 312,587,642 sources over 560 deg$^2$ of the bulge and southern disk.
\citet{cla19} calibrated the VIRAC proper motions by comparing the VIRAC values to the corresponding Gaia values.
They carefully selected red giants with $11.8 < K_{s, 0} < 13.6$ from the calibrated catalog, where $K_{s, 0}$ indicates the extinction-corrected $K_s$ magnitude.
Then they split each of the VVV tiles located in the bulge region into 4 sub-tiles and presented means $\VEV{\mu_i}$ and dispersions $\sigma_{\mu_i}$ ($i = l, b$) of proper motions
 of the selected giant stars in each sub-tile.
We use the data for $\VEV{\mu_l}$,  $\sigma_{\mu_l}$, and  $\sigma_{\mu_b}$ in 676 sub-tiles distributed roughly over $-9^\circ < l < 9.4^\circ$ and $-9.4^\circ < b < 4.2^\circ$.
Out of the 676 sub-tiles, we do not use 90 sub-tiles with $|b| \simlt 1^\circ$ because of high extinction values. Thus, we use the data in 586 sub-tiles for the fit.
We do not use the $\VEV{\mu_b}$ data because they are little sensitive to the fit parameters and not very useful to constrain them.

%Figure 3
\begin{figure}
\centering
\includegraphics[width=6cm]{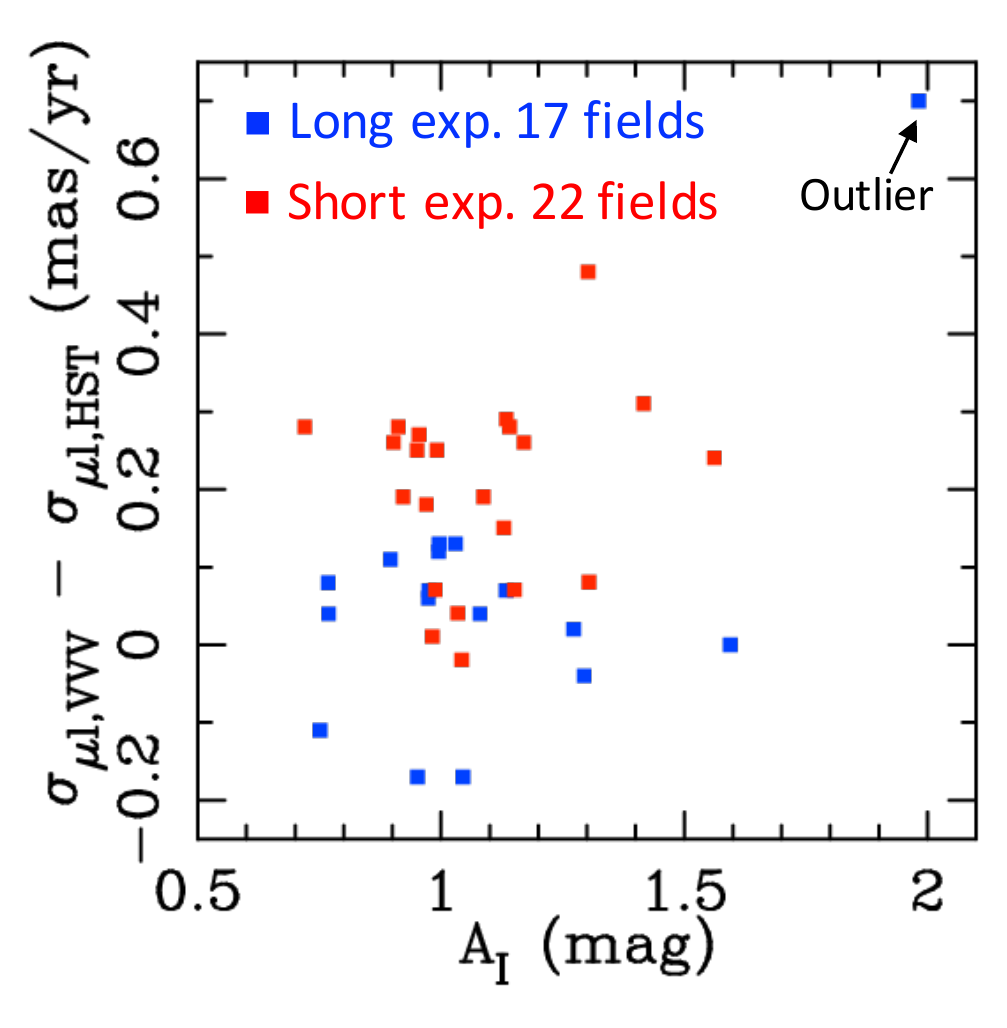}
\includegraphics[width=6cm]{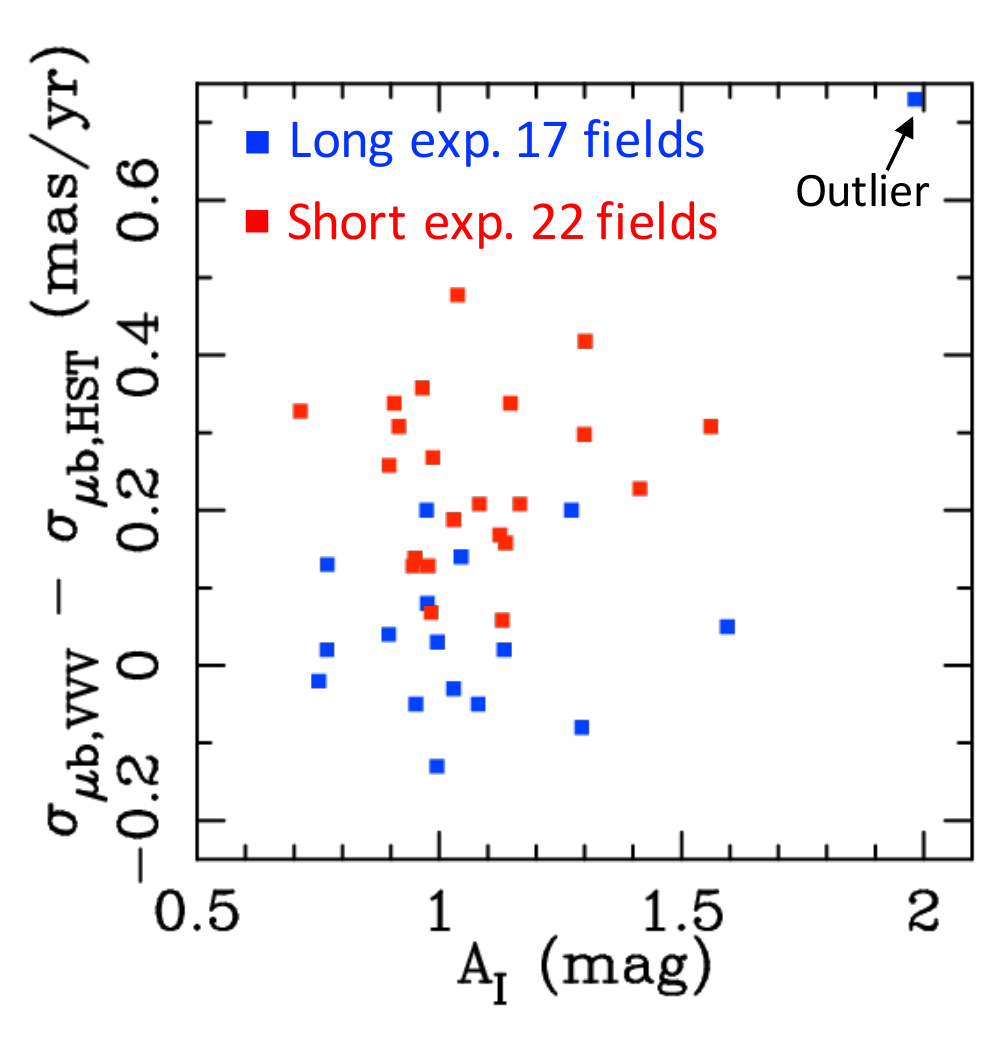}
\caption{Comparison of proper motion dispersions (left $\sigma_{\mu_l}$, right $\sigma_{\mu_b}$) between VVV and {\it HST}. Differences of measured 
values are plotted as a function of extinction for 35 fields from \citet{koz06} and 4 fields from Table 2 of \citet{ter20}.
22 fields with $\leq 40$ secs exposures from \citet{koz06} are classified as the short exp. fields while the other 17 fields are classified as the long exp. fields. An outlier pointed by \citet{ter20} is indicated.}
\label{fig-VVV2HST}
\end{figure}

Seeing-limited observations toward the Galactic bulge often suffer from systematic errors due to blended stars.
Because the data are derived from bright stars with $11.8 < K_{s, 0} < 13.6$, they are not expected to be affected by such systematics compared to fainter stars.
To verify this, we compared the proper motion dispersion measurements from VIRAC with those from the {\it Hubble Space Telescope} ({\it HST}) observations.
Fig. \ref{fig-VVV2HST} shows the comparison as a function of extinction $A_I$. 
We used 35 fields observed by \citet{koz06} and 4 fields summarized in Table 2 of \citet{ter20} for the comparison, and the {\it HST} values are taken from the two papers.
The 4 {\it HST} values from \citet{ter20} were originally measured by \citet{cal14}, \citet{kui02}, and \citet{ter20}.
If there are no systematic errors, the mean value of $\sigma_{\mu_i, {\rm VVV}} - \sigma_{\mu_i, {\rm HST}}$ should be consistent with 0.
However, we found the mean and standard deviation values of 
$\VEV{\sigma_{\mu_l, {\rm VVV}} - \sigma_{\mu_l, {\rm HST}}} = 0.124 \pm 0.135$~mas/yr and 
$\VEV{\sigma_{\mu_b, {\rm VVV}} - \sigma_{\mu_b, {\rm HST}}} = 0.153 \pm 0.144$~mas/yr
when we used the 39 fields without the outlier indicated in the figure that is specified by \citet{ter20}.
This implies the existence of systematic offset between the VVV and {\it HST} measurements.
We, at first, checked whether there was any correlation between the offset and the extinction, but Fig. \ref{fig-VVV2HST} shows no such a correlation except for the outlier.

Thereafter, we divide the sample of 39 fields into two subsamples depending on the exposure time of the {\it HST} observations because 
certain fields have significantly shorter exposure times than others.
One of the two subsamples consists of 22 short exposure fields from \citet{koz06} with each exposure less than 40~s, and 
the other consists of 17 long exposure fields with each exposure longer than 100~s.
Note that \citet{koz06} used 1st epoch data taken by the WFPC2/PC camera, which is much less sensitive than the ACS camera used to take their 2nd epoch data.
The mean and standard deviation values are $\VEV{\sigma_{\mu_l, {\rm VVV}} - \sigma_{\mu_l, {\rm HST}}} = 0.205 \pm 0.116$~mas/yr and 
$\VEV{\sigma_{\mu_b, {\rm VVV}} - \sigma_{\mu_b, {\rm HST}}} = 0.246 \pm 0.108$~mas/yr for the 22 short exposure fields and 
$\VEV{\sigma_{\mu_l, {\rm VVV}} - \sigma_{\mu_l, {\rm HST}}} = 0.023 \pm 0.096$~mas/yr and 
$\VEV{\sigma_{\mu_b, {\rm VVV}} - \sigma_{\mu_b, {\rm HST}}} = 0.034 \pm 0.093$~mas/yr for the 17 long exposure fields without the outlier.
The long exposure subsample shows no clear difference between VVV and {\it HST} measurements while the short exposure subsample does.

Thus, we conclude that the implied offset is due to the systematic error in the {\it HST} measurements for the 22 short exposure fields.
Given the standard deviation of $\sim 0.1$~mas/yr for the long exposure subsample, we use 0.1 mas/yr for 
the uncertainty of $\sigma_{\mu_l}$ and $\sigma_{\mu_b}$ of each sub-tile because formal statistic errors are comparatively small.
For the uncertainty of $\VEV{\mu_l}$, we use 0.14 mas/yr by additionally considering the error of calibration to the Gaia scale of $\sim 0.1$~mas/yr \citep{cla19}.

For a particular $i$th subtile's coordinate of $(l_i, b_i)$, the model $\VEV{\mu_l}$ value is calculated by 
\begin{align}
\VEV{\mu_l}_i^{\rm mod} &= \frac{\int_{3 {\rm kpc}}^{16 {\rm kpc}} \left[ n_{\rm d}^{\rm RG} (l_i, b_i, s) \VEV{\mu_{l, {\rm d}} (l_i, b_i, s)} + n_{\rm B}^{\rm RG} (l_i, b_i, s) \VEV{\mu_{l, {\rm B}} (l_i, b_i, s)} \right]  w(s) ds}{\int_{3 {\rm kpc}}^{16 {\rm kpc}}
 \left[n_{\rm d}^{\rm RG} (l_i, b_i, s) + n_{\rm B}^{\rm RG} (l_i, b_i, s) \right] w(s) ds}, 
\end{align}
where $n_{\rm d}^{\rm RG} (l_i, b_i, s)$ and $n_{\rm B}^{\rm RG} (l_i, b_i, s)$ are the number densities of red giants of the disk and bulge components at $(l_i, b_i, s)$, respectively.
We used the same definition for red giants in Section \ref{sec-disk_rho}, i.e., stars with ${\cal M}_G < 3.9$ mag and $(G_{\rm BP} - G_{\rm RP})_0 > 0.95$. 
$\VEV{\mu_{l, {\rm d}} (l_i, b_i, s)}$ and $\VEV{\mu_{l, {\rm B}} (l_i, b_i, s)}$ are mean proper motions calculated 
using our disk and bulge velocity model at $(l_i, b_i, s)$, respectively. 
Calculations of proper motion require the solar velocity, and again we use $(v_{\odot, x}, v_{\odot, y}, v_{\odot, z}) = (-10, 243, 7)$~km/s in this study.
The weight $w(s)$ is given by
\begin{align}
 w(s) = s^2 \int_{11.8}^{13.6} L_{{\cal M}_{K_s}} \left( K_{s,0} - {\rm DM} (s) \right) dK_{s, 0},
\end{align}
where $L_{{\cal M}_{K_s}} ({\cal M}_{K_s})$ is luminosity function for the red giant's absolute magnitude ${\cal M}_{K_s}$, given by Eqs. (2)--(5) of \citet{cla19}.
Further, the model $\sigma_{\mu_j}$ $(j = l, b)$ values are expressed as
\begin{align}
\sigma_{\mu_j, i}^{\rm mod} &= \sqrt{\VEV{\mu_j^2}_i^{\rm mod} - \left( \VEV{\mu_j}_i^{\rm mod} \right)^2},
\end{align}
where 
\[ \VEV{\mu_j^2}_i^{\rm mod} = \frac{\int_{3 {\rm kpc}}^{16 {\rm kpc}} \left[ n_{\rm d}^{\rm RG} (l_i, b_i, s) \VEV{\mu_{j, {\rm d}}^2 (l_i, b_i, s)} + n_{\rm B}^{\rm RG} (l_i, b_i, s) \VEV{\mu_{j, {\rm B}}^2 (l_i, b_i, s)} \right]  w(s) ds}{\int_{3 {\rm kpc}}^{16 {\rm kpc}}  \left[n_{\rm d}^{\rm RG} (l_i, b_i, s) + n_{\rm B}^{\rm RG} (l_i, b_i, s) \right] w(s) ds}. \]

$\chi^2$ for this dataset is denoted by 
\[ \chi^2_{\mu} =  \chi^2_{\mu_l} + \chi^2_{\sigma_{\mu_l}} + \chi^2_{\sigma_{\mu_b}}, \]
where
\[ \chi^2_p = \sum_i \left( \frac{p_i^{\rm mod} - p_i^{\rm obs}}{p_i^{\rm err}} \right)^2 \ \ \ \ (p = \mu_l, \sigma_{\mu_l}, \sigma_{\mu_b}). \]

\subsubsection{BRAVA radial velocity data} \label{sec-RVdata}
The Bulge Radial Velocity Assay \citep[BRAVA,][]{ric07, kun12} is a large spectroscopic survey of M giant stars in the Galactic bulge to constrain 
the bulge dynamics by measuring their radial velocities (RVs).
We use the mean and dispersion values of the RV measurements in 82 fields where 80 fields are located at  
$-10^\circ \leq l \leq 10^\circ$ and $3^\circ \leq |b| \leq 8^\circ$, while the other 2 are located at $(l, b) = (0^\circ, -1^\circ)$ and $(l, b) = (0^\circ, -2^\circ)$.
Although RV is not very relevant to our interest in microlensing observables, we use these data as a constraint because RV provides direct information of velocity compared to
proper motion, which is a combination of distance and velocity.

The BRAVA mean RV values are given in the Galactocentric frame by a conversion from the originally observed values in the heliocentric frame.
The conversion was done using the Sun velocity of $(-9.0, 231.9, 7.0)$ km/s along $(x, y, z)$ axes \citep{how08}, which is slightly different from our values of (-10, 243, 7) km/s. 
Thus, we reconverted the mean RV values using our values and then use them as data for our fits.

Similar to $\VEV{\mu_l}_i^{\rm mod}$, the model value for mean RV is given by
\begin{align}
\VEV{\rm RV}_i^{\rm mod} &= \frac{\int_{3 {\rm kpc}}^{16 {\rm kpc}} \left[ n_{\rm d}^{\rm RG} (l_i, b_i, s) \VEV{{\rm RV}_{\rm d} (l_i, b_i, s)} + n_{\rm B}^{\rm RG} (l_i, b_i, s) \VEV{{\rm RV}_{\rm B} (l_i, b_i, s)} \right]  s^{0.6} ds}{\int_{3 {\rm kpc}}^{16 {\rm kpc}}
 \left[n_{\rm d}^{\rm RG} (l_i, b_i, s) + n_{\rm B}^{\rm RG} (l_i, b_i, s) \right] s^{0.6} ds}, 
\end{align}
where the weight $s^{0.6}$ follows \citet{por15} who combined the volume effect ($\propto s^2$) with an approximate luminosity function of 
giant stars of $\propto 10^{0.28 {\cal M}_K}$ \citep{weg13}, which resulted in a dependency on the distance of $\propto s^{-1.4}$.
Further, the model RV dispersion value is calculated using the following
\begin{align}
\sigma_{{\rm RV}, i}^{\rm mod} &= \sqrt{\VEV{\rm RV^2}_i^{\rm mod} - \left( \VEV{\rm RV}_i^{\rm mod} \right)^2}.
\end{align}

$\chi^2$ for this dataset is denoted by 
\[ \chi^2_{\rm BRA} =   \chi^2_{\rm RV} + \chi^2_{\sigma_{\rm RV}}, \]
where
\[ \chi^2_p = \sum_i \left( \frac{p_i^{\rm mod} - p_i^{\rm obs}}{p_i^{\rm err}} \right)^2 \ \ \ \ (p = \VEV{\rm RV}, \sigma_{\rm RV}). \]

\subsubsection{OGLE-IV star count data} \label{sec-N1821}

\citet{mro19} analyzed long-term photometric observations of the Galactic bulge by the OGLE group using their fourth-generation wide-field camera OGLE-IV.
The analyzed fields consist of 9 high-cadence fields and 112 low-cadence fields, with each field further divided into 32 subfields having an individual area of $\sim 0.044~{\rm deg}^2$.
They measured the number of stars with $I < 21~{\rm mag}$ and $I < 18~{\rm mag}$, denoted by $N_{I < 21}$ and $N_{I < 18}$ respectively, 
for each subfield in addition to the microlens-related observables described in Section \ref{sec-Neve}.
Note that $N_{I < 21}$ is used as the number of candidate source stars in the microlensing events.
Although an extinction value $A_I$ is needed for each line of sight to model $N_{I < I_{\rm c}}$ ($I_{\rm c} = 18, 21$), there is no extinction catalog in $I$-band covering all the OGLE-IV fields.
Thus, we only use 1456 subfields covered by the $A_I$ map of \citet{nat13} for our analysis.

The model $N_{I < I_{\rm c}}$  for a particular $i$th subfield is calculated from 
\begin{align}
N_{I < I_{\rm c}, i}^{\rm mod} &= \Omega_i  \int_{3 {\rm kpc}}^{16 {\rm kpc}} n_{I < I_{\rm c}} (l_i, b_i, s)  s^2 ds   \hspace{0.7cm} (I_{\rm c} = 18, 21), \label{eq-NIc}
\end{align}
where $\Omega_i$ is the sky area, and $n_{I < I_{\rm c}} (l_i, b_i, s)$ is the number density of stars with $I < I_{\rm c}$ at a distance $s$ from the Sun toward $(l_i, b_i)$, which is expressed as
\begin{align}
n_{I < I_{\rm c}} (l_i, b_i, s) = [ n_{\rm d} (l_i, b_i, s) + n_{\rm B} (l_i, b_i, s) ] \int_{I < I_{\rm c}} L_{{\cal M}_I} \left( I - {\rm DM} (s) - A_I (l_i, b_i, s) \right) dI 
\end{align}
with the stellar number density for disk and bulge stars, $n_{\rm d} (l_i, b_i, s)$ and $n_{\rm B} (l_i, b_i, s)$, respectively, 
and the luminosity function for $I$-band absolute magnitude $L_{{\cal M}_I} ({\cal M}_I)$ calculated using the PARSEC isochrone models \citep{bre12, che14, tan14} for a given IMF.
We use a 3D extinction distribution of 
\begin{align}
A_I (l_i, b_i, s) = A_{I, {\rm RC}} (l_i, b_i) \frac{1 - \exp [- s/(h_{\rm dust}/\sin |b_i|)]}{1 - \exp [- s_{\rm RC} (l_i, b_i)/(h_{\rm dust}/\sin |b_i|)]},
\end{align}
where $A_{I, {\rm RC}} (l_i, b_i)$ and $s_{\rm RC} (l_i, b_i)$ are the mean extinction for the RC and mean distance to RC, respectively, both taken from the \citet{nat13} extinction catalog, and 
we use $h_{\rm dust} = 164~{\rm pc}$ based on the measurement by \citet{nat13}.
Because the extinction distribution is not uniform even inside a subfield of $\sim 0.044~{\rm deg}^2$, we further divide each subfield into 8 sub-subfields and then calculate the 
$N_{I < I_{\rm c}, i}^{\rm mod}$ value for each subfield by summing the values over the 8 sub-subfields with different $A_{I, {\rm RC}}$ values.

$\chi^2$ for the star count data is given by 
\[ \chi^2_{N_{I < I_{\rm c}}} =  \sum_i \left( \frac{N_{I < I_{\rm c}, i}^{\rm mod} - N_{I < I_{\rm c}, i}^{\rm obs}}{N_{I < I_{\rm c}, i}^{\rm err}} \right)^2   \hspace{0.7cm} (I_{\rm c} = 18, 21), \]
where we apply $N_{I < 21, i}^{\rm err} = 0.12 \, N_{I < 21, i}^{\rm obs}$ because \citet{mro19} estimated the errors as 10 to 15 \% by comparing to star count in {\it HST} images.
We use a larger uncertainty of $N_{I < 18, i}^{\rm err} = 0.14 \, N_{I < 18, i}^{\rm obs}$ because evolved stars like red giants mainly contribute to $N_{I < 18}$ and 
the isochrone model uncertainty for such evolved stars is likely larger than that for main-sequence stars contributing to $N_{I < 21}$.

The model value $N_{I < I_{\rm c}}^{\rm mod}$ depends on the density model, as well as the IMF, because the luminosity function $L_{{\cal M}_I} ({\cal M}_I)$ depends on the IMF.
As a quantity that is less dependent on the density model but more on the IMF, we defined $f_{18/21} \equiv N_{I < 18}/N_{I < 21}$.
$\chi^2$ for this quantity is calculated by
\[ \chi^2_{f_{18/21}} =  \sum_i \left( \frac{f_{18/21, i}^{\rm mod} - f_{18/21, i}^{\rm obs}}{f_{18/21, i}^{\rm err}} \right)^2, \]
where we used $f_{18/21, i}^{\rm err} = 0.16 \, f_{18/21, i}^{\rm obs}$. 
This uncertainty is taken slightly smaller than just a square root of the sum of the $N_{I < 21, i}^{\rm err}$ and $N_{I < 18, i}^{\rm err}$ considering 
a positive correlation between $N_{I < 21, i}^{\rm obs}$ and $N_{I < 18, i}^{\rm obs}$.

\subsubsection{Number of microlensing events and $t_{\rm E}$ distribution by OGLE-IV} \label{sec-Neve}

%Figure 4
\begin{figure}
\centering
\includegraphics[height=16.8cm]{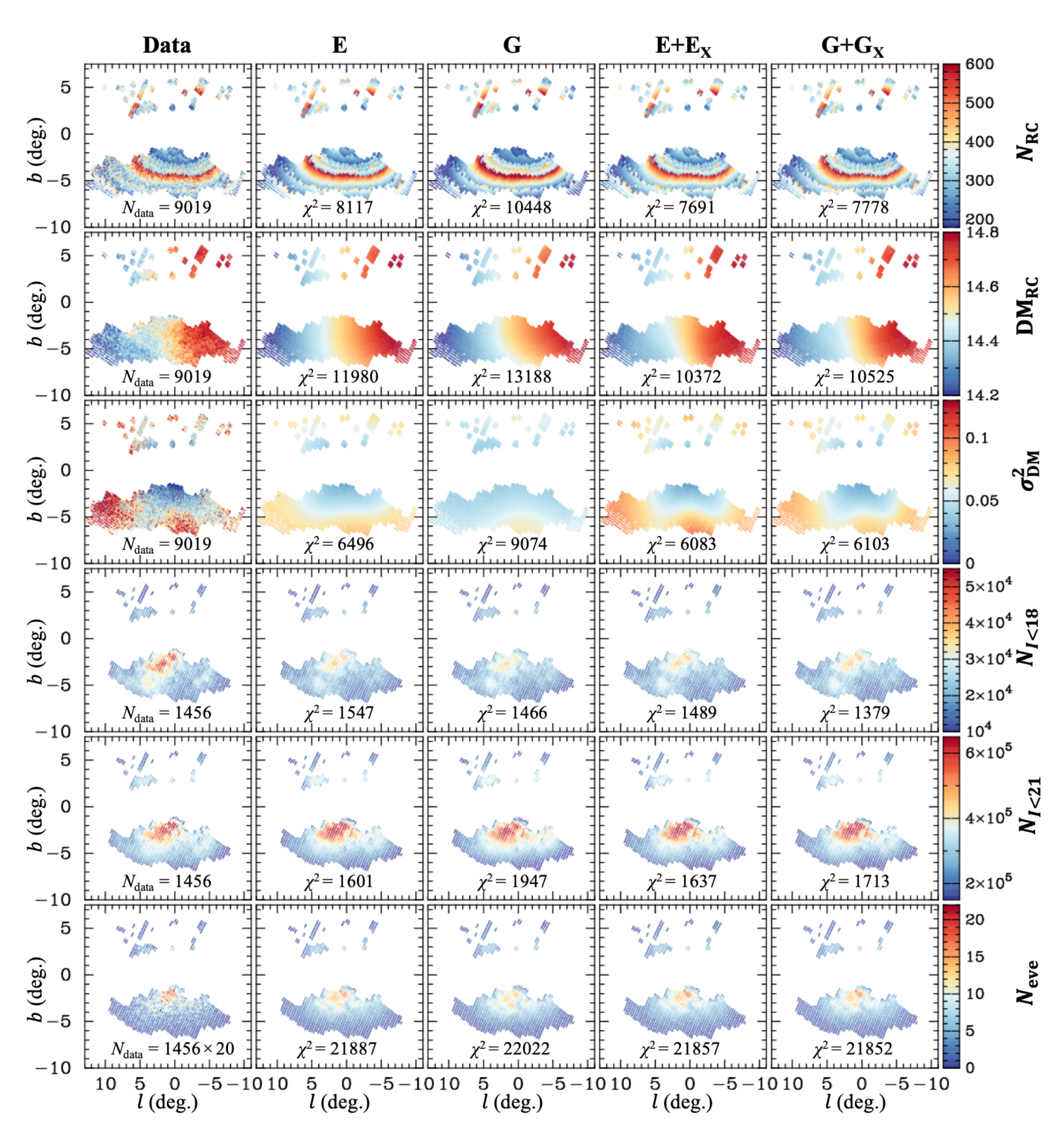}
\caption{Far left: Data distribution of six quantities used to constrain the bulge density ($\rho_{\rm B}$) model \citep{nat13, mro19}.
The right six rows: The same distributions from each indicated model. $\chi^2$ values defined for each quantity are shown.
In the bottom panels indicated by $N_{\rm eve}$, total number of (expected) event detections in $i$th subfield, $\sum_{j= 1}^{20} N_{{\rm eve}, i} (t_{{\rm E}, j})$, are plotted.}
\label{fig-mod_rhob}
\end{figure}

%Figure 5
\begin{figure}
\centering
\includegraphics[height=14cm]{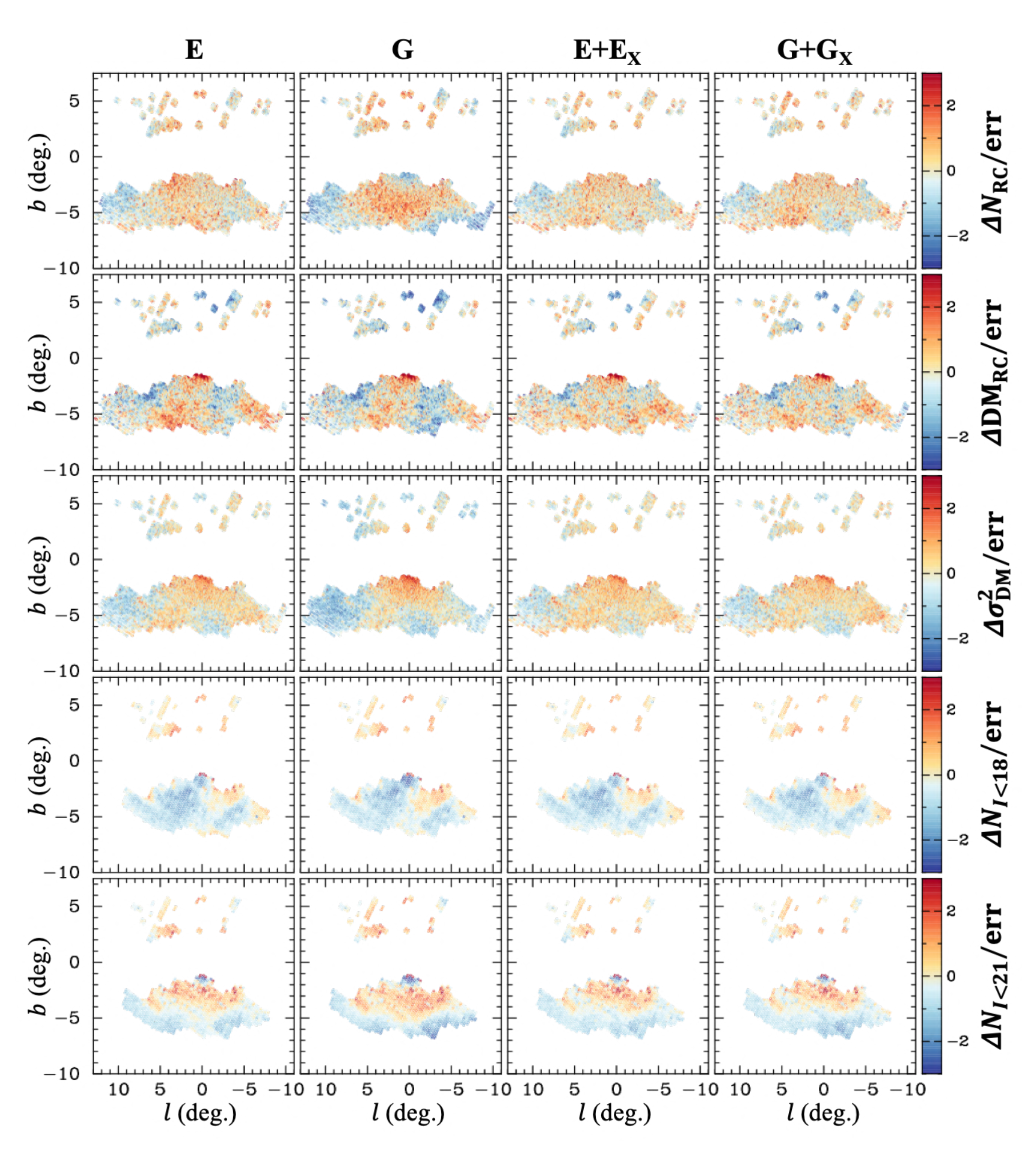}
\caption{Residuals corresponding to Fig. \ref{fig-mod_rhob}. 
In $i$th subfield, $(p_i^{\rm mod} - p_i^{\rm obs})/p_i^{\rm err}$, is plotted where $p$ is each indicated parameter on the far right.}
\label{fig-resi_rhob}
\end{figure}

%Figure 6
\begin{figure}
\centering
\includegraphics[height=14cm]{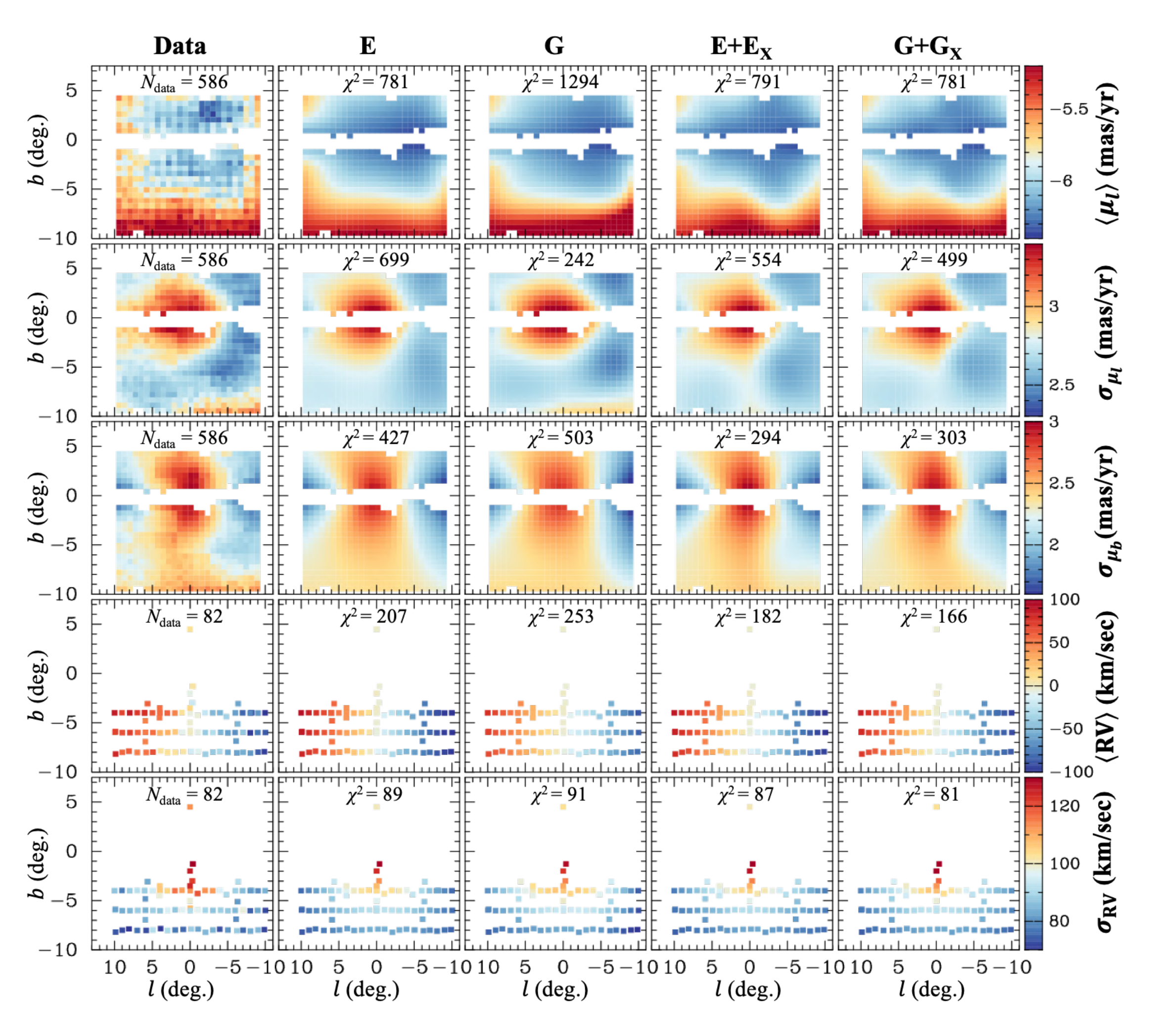}
\caption{Far left: Data distribution of five quantities used to constrain the bulge velocity ($v_{\rm B}$) model \citep{cla19, ric07, kun12}.
The right six rows: The same distributions from each indicated model. $\chi^2$ values defined for each quantity are shown.}
\label{fig-mod_vb}
\end{figure}

%Figure 7
\begin{figure}
\centering
\includegraphics[height=14cm]{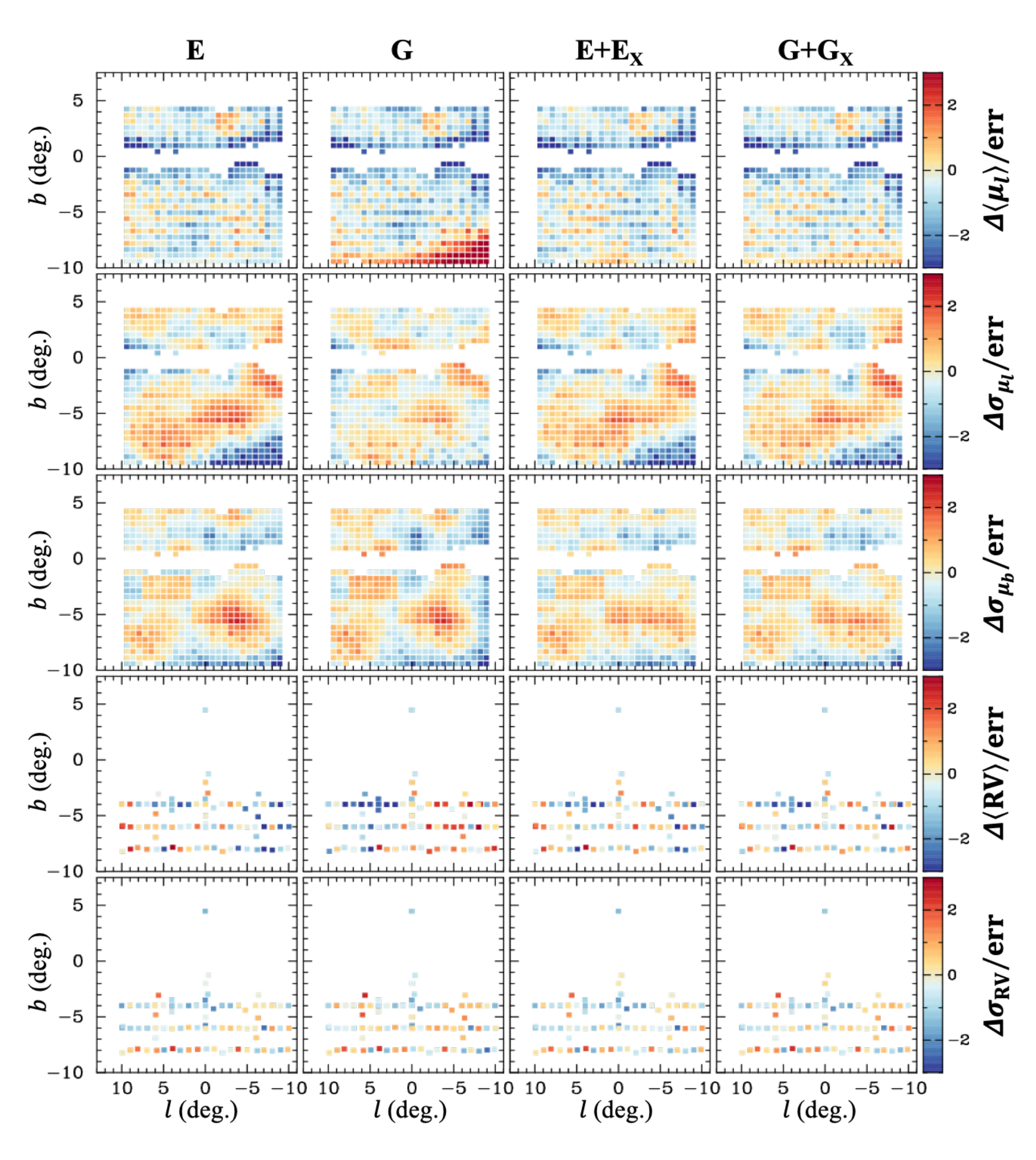}
\caption{Residuals corresponding to Fig. \ref{fig-mod_vb}. 
In $i$th subfield, $(p_i^{\rm mod} - p_i^{\rm obs})/p_i^{\rm err}$, is plotted where $p$ is each indicated parameter.}
\label{fig-resi_vb}
\end{figure}

%Figure 8
\begin{figure}
\centering
\includegraphics[height=10cm]{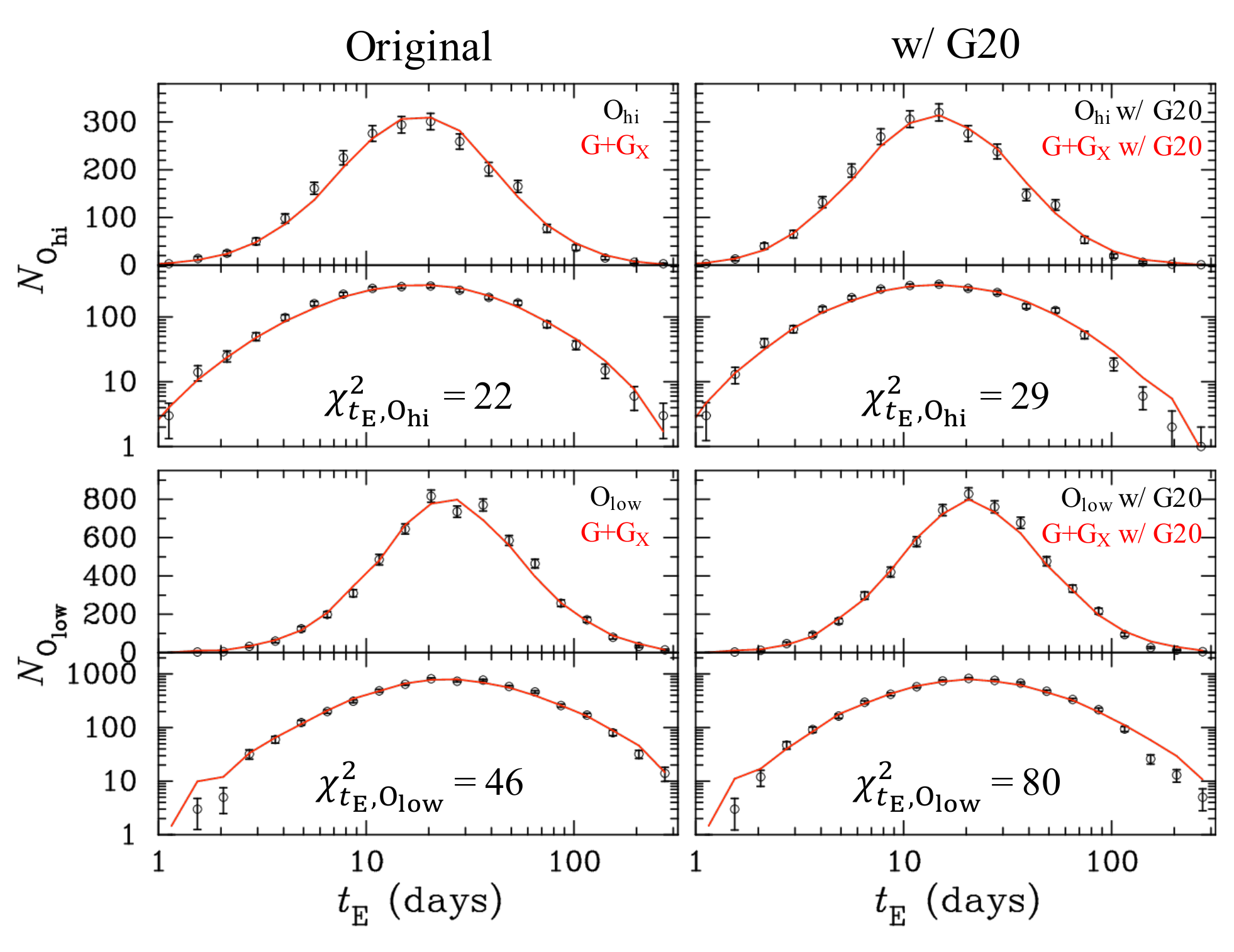}
\caption{Left: The original OGLE-IV $t_{\rm E}$ distributions of 2212 events in the high-cadence fields by \citet{mro17} (top) and 5788 events in the low-cadence fields by \citet{mro19} (bottom). 
Right: Modified $t_{\rm E}$ distributions by applying the \citet{gol20} correction with the Gaussian Process model (G20 option, considered in Appendix \ref{sec-G20}).
The black open circles and the red curves are indicated data and models, respectively.}
\label{fig-mod_tE}
\end{figure}

\citet{mro19} also provided the microlensing optical depth and the event rate for each OGLE-IV field.
However, it is statistically easier to deal with another equivalent quantity, $N_{{\rm eve}, i} (t_{\rm E})$, which is the number of 
detected events as a function of $t_{\rm E}$ for each $i$th subfield, because $N_{{\rm eve}, i} (t_{\rm E})$ simply follows a Poisson distribution.
For a particular $i$th subfield and a $j$th $t_{\rm E}$ bin, the expected number of event detections is given by
\begin{align}
N_{{\rm eve}, i}^{\rm mod} (t_{{\rm E}, j}) &= \frac{2 N_{I < 21, i}^{\rm mod} T_{\rm o}}{\pi \VEV{t_{\rm E}}_i} \tau_{I < 21, i} \, f_i (t_{{\rm E}, j})  \epsilon_i (t_{{\rm E}, j}) \Delta t_{\rm E}, \label{eq-Neve}
\end{align}
where $T_{\rm o}$ is the survey duration that takes 2741 or 2011 days depending on the field, $\tau_{I < 21, i}$ is the average optical depth over the monitored $N_{I < 21, i}^{\rm mod}$ stars, which is given by
\begin{align}
\tau_{I < 21, i} =  \frac{1}{N_{I < 21, i}^{\rm mod}} \int_{3 {\rm kpc}}^{16 {\rm kpc}}  \tau (l_i, b_i, s)  \, n_{I < 21} (l_i, b_i, s)  s^2 ds 
\end{align}
with 
\begin{align}
\tau (l_i, b_i, s) = \frac{4 \pi G}{c^2} \int_{0}^{s} \rho (l_i, b_i, s_{\rm L}) \frac{s_{\rm L} (s - s_{\rm L})}{s} ds_{\rm L},
\end{align}
$f_i (t_{\rm E})$ is the probability density function of $t_{\rm E}$, $\VEV{t_{\rm E}}_i$ is the mean $t_{\rm E}$ value given by $\VEV{t_{\rm E}}_i = \int f_i (t_{\rm E}) \, t_{\rm E} \, dt_{\rm E}$, 
$\epsilon_i (t_{\rm E})$ is the detection efficiency for an event with $t_{\rm E}$ in the subfield, and $\Delta t_{\rm E}$ is the $t_{\rm E}$ bin size.
We have been provided with the detection efficiency values in addition to the number of detected events as a function of 
$t_{\rm E}$ for each subfield through a private communication with P. Mr{\'o}z.

We determine $\chi^2$ for the number of microlensing event detections with
\begin{align}
\chi^2_{N_{\rm eve}} = \sum_{i, j} -2 \ln P[N_{{\rm eve}, i}^{\rm obs} (t_{{\rm E}, j}) ; N_{{\rm eve}, i}^{\rm mod} (t_{{\rm E}, j})], \label{eq-chi2Neve}
\end{align}
where $P (x ; \lambda) = \lambda^x e^{-\lambda}/x!$ is the Poisson probability of $x$ with the mean value $\lambda$.
A $j$th $t_{\rm E}$ bin includes events with $0.125 (j -1) \leq \log [t_{\rm E}/ {\rm days}] < 0.125 j$ ($j=$ 1, ..., 20) for $i \in {\rm O_{\rm low}}$, while 
it includes events with $-0.30 + 0.14 (j -1) \leq \log [t_{\rm E}/ {\rm days}] < -0.30 + 0.14 j$  ($j=$ 1, ..., 20) for $i \in {\rm O_{\rm hi}}$, where 
$i \in {\rm O_{\rm low}}$ and $i \in {\rm O_{\rm hi}}$ indicate when $i$th subfield is located in the low- and high-cadence fields, respectively.
Note that 4,057 bins have $N_{{\rm eve}, i}^{\rm obs} (t_{{\rm E}, j}) > 0$ out of the total 29,120 bins (1456 subfields $\times$ 20 bins) while the remaining have $N_{{\rm eve}, i}^{\rm obs} (t_{{\rm E}, j}) = 0$.

The probability density function of $t_{\rm E}$, $f_i (t_{\rm E})$, is calculated through a Monte Carlo simulation of $\sim 10^8$ microlensing events for each $i$th subfield with a given model.
This is computationally expensive and time-consuming when $f_i (t_{\rm E})$ is updated for every 
proposed model in the MCMC fitting procedure described in Section \ref{sec-bar_fit}.
Thus, in a fitting run, we instead use a representative $f_i (t_{\rm E})$ calculated with a tentative best-fit model because we found that 
the $\chi^2_{N_{\rm eve}}$ value had negligible variance when the $f_i (t_{\rm E})$ used was calculated with a model showing similar $\chi^2_{t_{\rm E}}$ introduced below.
 
\citet{mro17} and \citet{mro19} separately provided $t_{\rm E}$ distributions consisting of 2212 events in the high-cadence fields 
and 5788 events in the low-cadence fields, respectively, which are denoted by $N_{\rm O_{\rm hi}}^{\rm obs} (t_{{\rm E}, j})$ and $N_{\rm O_{\rm low}}^{\rm obs} (t_{{\rm E}, j})$, respectively. These are given by
\[ N_o^{\rm obs} (t_{{\rm E}, j}) = \sum_{i \in o} N_{{\rm eve}, i}^{\rm obs} (t_{{\rm E}, j}) \hspace{0.7cm} (o = {\rm O_{\rm hi}}, {\rm O_{\rm low}}), \]
where the summation is taken over all subfields located in $i \in o$ ($o = {\rm O_{\rm hi}}$ or ${\rm O_{\rm low}}$).
The range of summation is different from the case for $\chi^2_{N_{\rm eve}}$ where only the subfields covered by \citet{nat13} are considered because a shape of $t_{\rm E}$ distribution is 
not sensitive to variation in extinction values.
A compared model value for a particular $j$th $t_{\rm E}$ bin is
\begin{align}
N_o^{\rm mod} (t_{{\rm E}, j}) = \eta_o \VEV{\epsilon_o}_j \sum_{i \in o}  \omega_i f_i (t_{{\rm E}, j}) \Delta t_{\rm E}  \hspace{0.7cm} (o = {\rm O_{\rm hi}}, {\rm O_{\rm low}}),
\end{align}
where the mean detection efficiency $\VEV{\epsilon_o}_j$ and weight for each subfield $\omega_i$  are
\[ \VEV{\epsilon_o}_j  \equiv  \frac{\sum_{i \in o} N_{{\rm eve}, i}^{\rm obs} (t_{{\rm E}, j})}{\sum_{i \in o} N_{{\rm eve}, i}^{\rm obs} (t_{{\rm E}, j})/\epsilon_i (t_{{\rm E}, j})}, \]
and
\[ \omega_i = \sum_j  \frac{N_{{\rm eve}, i}^{\rm obs} (t_{{\rm E}, j})}{\epsilon_i (t_{{\rm E}, j})}, \] 
respectively, with $\eta_o$ being an arbitrary constant expected to be approximately 1. 
Although $\eta_o = 1$ provides $\sum_j N_o^{\rm mod} (t_{{\rm E}, j}) = \sum_j  N_o^{\rm obs} (t_{{\rm E}, j})$, $\eta_o$ can take a random value 
because $\sum_j N_o^{\rm mod} (t_{{\rm E}, j})$ is simply an expected number of the total event detections, and $\sum_j  N_o^{\rm obs} (t_{{\rm E}, j})$ can differ from it.
Thus, $\eta_o$ is chosen such that it minimizes $\chi^2_{t_{\rm E}, o}$ introduced below in every step of the MCMC fitting.

We consider $\chi^2$ for the two $t_{\rm E}$ distributions by
\begin{align}
\chi^2_{t_{\rm E}} &= \chi^2_{t_{\rm E}, {\rm O_{\rm hi}}} + \chi^2_{t_{\rm E}, {\rm O_{\rm low}}} \notag\\
& = \sum_{j} \left( \chi^2_{t_{{\rm E}, j}, {\rm O_{\rm hi}}} + \chi^2_{t_{{\rm E}, j}, {\rm O_{\rm low}}} \right),
\end{align}
where 
\begin{equation}
\chi^2_{t_{{\rm E}, j}, o} =
\begin{cases}
-2 \ln P[N_o^{\rm obs} (t_{{\rm E}, j}) ; N_o^{\rm mod} (t_{{\rm E}, j})]  & \text{ when $N_o^{\rm obs} (t_{{\rm E}, j}) \leq 15$} \\
\left[ \frac{N_o^{\rm mod} (t_{{\rm E}, j}) - N_o^{\rm obs} (t_{{\rm E}, j})}{N_o^{\rm err} (t_{{\rm E}, j})} \right]^2  & \text{ when $N_o^{\rm obs} (t_{{\rm E}, j}) > 15$} 
\end{cases}
\end{equation}
with 
\[N_o^{\rm err} (t_{{\rm E}, j}) = \VEV{\epsilon_o}_j \sqrt{\sum_{i \in o}  \frac{N_{{\rm eve}, i}^{\rm obs} (t_{{\rm E}, j})}{\epsilon_i^2 (t_{{\rm E}, j})} }.  \]
Recall that $P (x ; \lambda) = \lambda^x e^{-\lambda}/x!$ is the Poisson probability of $x$ with the mean value $\lambda$.
The uncertainty $N_o^{\rm err} (t_{{\rm E}, j})$ is not simply a square root of the sum of the Poisson errors of each $N_{{\rm eve}, i}^{\rm obs} (t_{{\rm E}, j})$, but instead, it is 
weighted by $1/\epsilon_i^2 (t_{{\rm E}, j})$.
Therefore, this makes the $N_o (t_{{\rm E}, j})$ distribution different from a simple Poisson probability distribution;
however, we consider $\chi^2_{t_{{\rm E}, j}, o}$ with the Poisson distribution when $N_o^{\rm obs} (t_{{\rm E}, j})$ is small ($\leq 15$) because  
the dependency of $\epsilon_i (t_{{\rm E}, j})$ on $i$ (i.e., subfield) is much smaller than that on $j$ (i.e., $t_{\rm E}$), and
the distribution is expected to remain similar to the Poisson distribution, especially when number of subfields contributing to $N_o^{\rm obs} (t_{{\rm E}, j})$ is small.
Because the Poisson distribution can be approximated to the Gaussian distribution when $N_o^{\rm obs} (t_{{\rm E}, j})$ becomes larger, 
we consider $\chi^2_{t_{{\rm E}, j}, o}$ with the Gaussian distribution with $N_o^{\rm err} (t_{{\rm E}, j})$ to consider the weight of $1/\epsilon_i^2 (t_{{\rm E}, j})$.

\subsection{Prior Constraints and $\chi^2$ Penalty} \label{sec-prior}

% Table 3
\begin{deluxetable}{lrrccccccccccccccccc}
\tabletypesize{\normalsize}
\tablecaption{Priors used in fits for the bulge model and posterior values. \label{tab-bar_prior}}
\tablehead{
\colhead{}  & \colhead{Parameter}                                      & \colhead{Reference}   & \colhead{Value/Prior\tablenotemark{a}}   & \colhead{Posterior\tablenotemark{b}}
}
\startdata
Sun         & $(R_{\odot}, z_{\odot})$ [pc]                            & (1), (2)              & (8160, 25)                               & (8160, 25)    \\
            & $(v_{\odot, x}, v_{\odot, y}, v_{\odot, z})$ [km/s]      & (1)                   & (-10, 243, 7)                            & (-10, 243, 7) \\
Bulge       & $\alpha_{\rm bar}$ [deg.]                                & (1)                   & 27                                       & 27            \\
            & $M_{\rm VVV}$ [$10^{10} M_{\odot}$]                      & (3)                   & $1.32 \pm 0.08$                          & $1.14^{+0.10}_{-0.11}$ \\  % frho = 0.839 for EXE to 0.914 for E
            & $\VEV{\sigma_{v_{\rm B}, x'}}$ [km/s]                    & (1), (4)              & $135 \pm 5$                              & $141.0^{+1.7}_{-4.4}$  \\
            & $\VEV{\sigma_{v_{\rm B}, y'}}$ [km/s]                    & (1), (4)              & $105 \pm 5$                              & $113.6^{+3.4}_{-0.9}$  \\
            & $\VEV{\sigma_{v_{\rm B}, z'}}$ [km/s]                    & (1), (4)              & $96  \pm 5$                              & $108.3^{+0.2}_{-1.7}$  \\
            & $\Omega_{\rm p}$  [km/s/kpc]                             & (3)                   & $39.0 \pm 3.5$                           & $45.9^{+4.0}_{-5.4}$   \\
IMF         & $M_{\rm br}$ [$M_{\odot}$]                               & (5), (6)              & $0.60 \pm 0.10$                          & $0.90^{+0.05}_{-0.14}$ \\  
            & $\alpha_{\rm hm}$                                        & (5), (6)              & $2.30 \pm 0.10$                          & $2.32^{+0.14}_{-0.10}$ \\ 
            & $\alpha_{\rm ms}$                                        & (5), (6)              & $1.30 \pm 0.15$                          & $1.16^{+0.08}_{-0.15}$ \\ 
            & $\alpha_{\rm bd}$                                        & (5)                   & $0.30 \pm 0.70$                          & $0.22^{+0.20}_{-0.55}$ \\ 
\enddata
\tablerefs{(1) \citet{bla16}; (2) \citet{gra19}; (3) \citet{por17}; (4) \citet{por15}; (5) \citet{kro01}; (6) \citet{cal15}.}
\tablenotetext{a}{ $x \pm x_e$ indicates Gaussian prior and $\chi^2_{\rm pena} = \sum_x a_x \times (\frac{p_x -x}{x_e})^2$ is applied in the fit, where $a_x = 5$ for $M_{\rm VVV}$ and $a_x = 1$ for the others, and $p_x$ is the value of parameter $x$ in a given model.}
\tablenotetext{b}{Posterior values are from the best-fit G+G$_{\rm X}$ model in Table \ref{tab-params}, and the uncertainties are combinations of statistic errors and systematic errors due to model choice.}
%\tablecomments{.}
\end{deluxetable}

We have prior information about certain fundamental parameters in our Galaxy from several previous studies, and reasonable prior constraints on such parameters help us efficiently examine 
the huge parameter space or disentangle degeneracies among fit parameters.
Particularly, in this case where our parametric model lacks a dynamical consistency, prior constraints on the bulge mass or kinematic parameters help us 
avoid converging into a completely unphysical model.

Table \ref{tab-bar_prior} lists the fundamental parameters on which we apply prior constraints during the fit.
Other than the solar position, velocity and bar angle that are fixed, there are 9 parameters: 
the model integrated mass within the VVV bulge box, $M_{\rm VVV}$, 
mass-weighted velocity dispersions inside the bulge half mass radius, $\VEV{\sigma_{v_{\rm B}, i}}$ ($i = x', y', z'$), 
bar pattern speed, $\Omega_{\rm p}$,  three IMF slopes, ($\alpha_{\rm hm}$, $\alpha_{\rm ms}$, $\alpha_{\rm bd}$), and break mass, $M_{\rm br}$,
where the VVV bulge box is defined the central region inside $(x', y', z') = (\pm 2.2, \pm 1.4, \pm 1.2)$ kpc \citep{weg13}.
Practically, we add a $\chi^2$ penalty depending on a deviation from the applied prior value for each parameter, as described below.

\citet{por17} found a total dynamical mass inside the VVV bulge box of $(1.85 \pm 0.05) \times 10^{10}~M_{\odot}$, where the mass budget was $(0.32 \pm 0.05) \times 10^{10}~M_{\odot}$ for dark matter,
$0.2 \times 10^{10}~M_{\odot}$ for a nuclear stellar disk in the very center, and 
the remaining $(1.32 \pm 0.08) \times 10^{10}~M_{\odot}$ was for stellar objects traceable by the RC stars.
Because the data introduced in Section \ref{sec-bar_data} is not sensitive to the central nuclear stellar disk with a $\sim 50~{\rm pc}$ scale height, 
we use $(1.32 \pm 0.08)  \times 10^{10}~M_{\odot}$ as a prior for $M_{\rm VVV}$ and apply a $\chi^2$ penalty of $5 \times (\frac{M_{\rm VVV} \, [10^{10} \, M_{\odot}] - 1.32}{0.08})^2$ in the fit.
Our best-fit models presented in Section \ref{sec-bar_fit} all have $M_{\rm VVV} < 1.32 \times 10^{10}~M_{\odot}$, and the large factor of 5 was multiplied by the $\chi^2$ penalty 
to avoid needlessly reducing the bulge stellar mass, because our parametric Galactic model does not ensure dynamical consistency by itself.
Note that the light stellar mass in the bulge region does not violate dynamics in itself; 
rather it indicates an additional stellar mass in a non-sensitive region and/or a larger dark matter mass fraction compared to the \citet{por17} model. 
We discuss the implied dark matter mass in our model later, in Section \ref{sec-M2L}.

For the mass-weighted velocity dispersions inside the bulge half mass radius, $\VEV{\sigma_{v_{\rm B}, i}}$ ($i = x', y', z'$), we apply a 
$\chi^2$ penalty of $\sum_{i = x', y', z'} \left(\frac{\VEV{\sigma_{v_{\rm B}, i}} - \VEV{\sigma_{v_{\rm B}, i}}_{\rm P15}}{5 {\rm km/s}}\right)^2$, 
where $(\VEV{\sigma_{v_{\rm B}, x'}}, \VEV{\sigma_{v_{\rm B}, y'}}, \VEV{\sigma_{v_{\rm B}, z'}})_{\rm P15} = $ (135, 105, 96) km/s that were derived by \citet{bla16} using 
the \citet{por15} dynamical bulge model. Further, a $\chi^2$ penalty of $(\frac{\Omega_{\rm p} - 39.0}{3.5})^2$ is applied for the pattern speed of $\Omega_{\rm p}$ in 
unit of km/s/kpc based on the value derived by \citet{por17}.

For the IMF parameters, the following $\chi^2$ penalties were applied for each slope and break mass; 
$(\frac{M_{\rm br}/M_{\odot} - 0.60}{0.10})^2$, $(\frac{\alpha_{\rm hm} - 2.30}{0.10})^2$, $(\frac{\alpha_{\rm ms} - 1.30}{0.15})^2$, and $(\frac{\alpha_{\rm bd} - 0.30}{0.70})^2$, where the 
prior values were determined based on the local IMF by \citet{kro01} combined with the break mass at $\sim 0.56 \, M_{\odot}$ found by \citet{cal15}.

Hereafter, we denote the sum of the $\chi^2$ penalty values for the above nine parameters by $\chi^2_{\rm pena}$.

\subsection{Fitting} \label{sec-bar_fit}

\subsubsection{Fitting procedure} \label{sec-fit_proc}
We use the Markov Chain Monte Carlo (MCMC) methods \citep{metrop} for our fitting.
The bulge model consists of 39 fit parameters, i.e., 7--16, 19, and 4 parameters for the density ($\rho_{\rm B}$), velocity ($v_{\rm B}$), and IMF models, respectively.
In the previous sections, we defined the following eight $\chi^2$ values, 
$\chi^2_{\rm RC}$ for the OGLE-III RC star count data in 9019 lines of sight (Section \ref{sec-RCdata}), 
$\chi^2_{\mu}$ for the VIRAC proper motion measurements in 586 VVV sub-tiles (Section \ref{sec-mudata}), 
$\chi^2_{\rm BRA}$ for the BRAVA radial velocity data in 82 fields (Section \ref{sec-RVdata}), 
$\chi^2_{N_{I < I_{\rm c}}} (I_{\rm c} = 18, 21)$ and $\chi^2_{f_{18/21}}$ for the OGLE-IV star count data in 1456 subfields (Section \ref{sec-N1821}),
$\chi^2_{N_{\rm eve}}$ for the number of microlensing event detections as a function of $t_{\rm E}$ by the OGLE-IV survey in the 1456 subfields (Section \ref{sec-Neve}),
and $\chi^2_{t_{\rm E}}$ for the two $t_{\rm E}$ distributions in the OGLE-IV high- and low-cadence fields divided into 20 $\log [t_{\rm E} / {\rm days}]$ bins (Section \ref{sec-Neve}), 
in addition to the $\chi^2$ penalty from priors on the nine parameters summarized in Table \ref{tab-bar_prior} (Section \ref{sec-prior}).
Of the eight $\chi^2$ values, $\chi^2_{\mu}$ and $\chi^2_{\rm BRA}$ depend on the $\rho_{\rm B}$ and $v_{\rm B}$ models, $\chi^2_{N_{I < I_{\rm c}}} (I_{\rm c} = 18, 21)$ depends on the $\rho_{\rm B}$ and IMF models, and $\chi^2_{N_{\rm eve}}$ and $\chi^2_{t_{\rm E}}$ depend on the $\rho_{\rm B}$, $v_{\rm B}$, and IMF models.
Thus, ideally, all the 39 fit parameters are simultaneously fitted because of the correlations among the $\rho_{\rm B}$, $v_{\rm B}$, and IMF models in the parameter space.

However, such a simultaneous fit is difficult for the following reasons: 
First, the volume of the parameter space is very large and difficult to examine in a reasonable amount of time.
Second, the probability density function of $t_{\rm E}$, $f_i (t_{\rm E})$, is needed to calculate $\chi^2_{N_{\rm eve}}$, 
which involves time-consuming calculations, for a given combination of $\rho_{\rm B}$, $v_{\rm B}$, and IMF models.
As described in Section \ref{sec-Neve}, a representative $f_i (t_{\rm E})$ can be used for models showing 
similar $\chi^2_{t_{\rm E}}$, but it is difficult to ensure the similarity in such a simultaneous fit.
Third, even if time permitted, the ``best-fit" model depends on what $\chi^2$ value was minimized in the fit.
Because the number of data points in the used datasets ranges from 40 bins for $\chi^2_{t_{\rm E}}$ to $9019 \times 3$ for $\chi^2_{\rm RC}$, agreement to a small dataset could be neglected depending on 
arbitrarily selected weights among the eight $\chi^2$ values. The weights are ideally 1 if every data does not contain any systematic error, and the model selection is perfect; however, 
since neither of these statements is true, the ``best-fit" model would be dominated by an arbitrary choice of weights.

% Table 4
{\tabcolsep = 1.2mm
\begin{deluxetable}{rlrrccccccccccccccccc}
\tabletypesize{\scriptsize}
\tablecaption{Each step in the iterative process to find best-fit models. \label{tab-bar_fit}}
\tablehead{
\colhead{Step}  & \colhead{Fit parameters}                                                                                                                         & \colhead{Minimized $\chi^2$}                    & \colhead{Input\tablenotemark{a}}                            & \colhead{Output}
}
\startdata
1             & $M_{\rm br}, \alpha_{\rm hm}, \alpha_{\rm ms}, \alpha_{\rm bd}, \rho_{\rm 0, B}$                                                                     & ${ \tilde{\chi}^2_{\rm IMF}}\tablenotemark{b} + \chi^2_{\rm pena}$          & Tentative best $\rho_{\rm B}$ and $v_{\rm B}$ models          & $L_{{\cal M}_I} ({\cal M}_I)$, $f_i(t_{{\rm E}, j})$ \\
2            & \ \ \ $\rho_{\rm 0, B}, \bm p_{r_s}, R_c$                                                                                                            & ${ \tilde{\chi}^2_{\rho_{\rm B}}}\tablenotemark{b} + 5\chi^2_{\rm pena}$    & $L_{{\cal M}_I} ({\cal M}_I)$ \& $f_i(t_{{\rm E}, j})$ from 1 & $\rho_{\rm B}$ model         \\
                & + $x_{\rm KM, th}$ for KM option                                                                                                                     &    &  &        \\
                & + $f_{\rm 0, X}, \bm p_{r_s, {\rm X}}, R_{\rm c, X}, b_{\rm X}$ for 2-comp. model                                                                    &    &  &        \\
3           & $\Omega_{\rm p}, v_0^{\rm str}, y_0^{\rm str}, \sigma_{v_{\rm B}, i, 0}, \sigma_{v_{\rm B}, i, 1}, \bm p_{r_s, \sigma_R}, \bm p_{r_s, \sigma_{z'}}$  & $2\chi^2_{v_{\rm B}}\tablenotemark{b} + 5\chi^2_{\rm pena}$       & $\rho_{\rm B}$ model from 2                                & $v_{\rm B}$ model            \\
4            & Same as step 2                                                                                                                                    & ${ \tilde{\chi}^2_{\rho_{\rm B}}} + 2\chi^2_{v_{\rm B}} + 5\chi^2_{\rm pena}$   & $v_{\rm B}$ model from 3, $L_{{\cal M}_I} ({\cal M}_I)$ \& $f_i(t_{{\rm E}, j})$ from 1 & $\rho_{\rm B}$ model \\
5             & Same as step 3                                                                                                                                   & $2\chi^2_{v_{\rm B}} + 5\chi^2_{\rm pena}$     & $\rho_{\rm B}$ model from 4                                & $v_{\rm B}$ model \\
% 6            & \multicolumn{4}{c}{Calculate $\chi^2_{t_{\rm E}}$ and if the difference from that from step 1 is $< 4$, stop iteration. Otherwise, repeat steps 1-5.} \\
\enddata
\tablenotetext{a}{Input models or functions are fixed during each step other than $\rho_{\rm 0, B}$ in step 1.}
\tablenotetext{b}{ ${ \tilde{\chi}^2_{\rm IMF}} \equiv \chi^2_{t_{\rm E}} + 0.2(\chi^2_{N_{I < 21}} + \chi^2_{N_{I < 18}} + \chi^2_{f_{18/21}})/3$, \ \  ${ \tilde{\chi}^2_{\rho_{\rm B}}} \equiv \chi^2_{\rm RC} + \chi^2_{N_{I < 21}} + \chi^2_{N_{I < 18}} + 2\chi^2_{N_{\rm eve}}$, and $\chi^2_{v_{\rm B}} \equiv \chi^2_{\mu} + \chi^2_{\rm BRA}$.}
% \tablecomments{Input models or functions are fixed during each step other than $\rho_{\rm 0, B}$ in step 1.}
\end{deluxetable}
}

Because of these difficulties associated with the simultaneous fit, we perform modeling by iterating the following step-by-step procedure, where steps 1--5 are also summarized in Table \ref{tab-bar_fit}:
\begin{enumerate}
\item[0.] At the beginning, tentative best-fit $\rho_{\rm B}$ and $v_{\rm B}$ models are determined to use in step 1 by performing steps 2--5 without using 
$\chi^2_{t_{\rm E}}$, $\chi^2_{N_{I < 21}}$, $\chi^2_{N_{I < 18}}$, $\chi^2_{f_{18/21}}$, or $\chi^2_{N_{\rm eve}}$. The \citet{kro01} IMF is used when an IMF is needed.
\item[1.] An IMF model is determined by minimizing ${ \tilde{\chi}^2_{\rm IMF}} + \chi^2_{\rm pena}$, where $\chi^2$ to evaluate an IMF model is defined as 
${ \tilde{\chi}^2_{\rm IMF}} \equiv \chi^2_{t_{\rm E}} + 0.2(\chi^2_{N_{I < 21}} + \chi^2_{N_{I < 18}} + \chi^2_{f_{18/21}})/3$,
and the best-fit $\rho_{\rm B}$ and $v_{\rm B}$ models from the previous run or from step 0 are fixed and used.
Subsequently, the probability density function of $t_{\rm E}$ in $i$th subfield, $f_i(t_{\rm E})$, is calculated using the $\rho_{\rm B}$, $v_{\rm B}$, and IMF models.
The luminosity function $L_{{\cal M}_I} ({\cal M}_I)$ is calculated using the IMF model.
A $\chi^2_{t_{\rm E}}$ value, $\chi^2_{t_{\rm E}, {\rm ini}}$, is calculated using the $\rho_{\rm B}$, $v_{\rm B}$, and IMF models.
\item[2.] A $\rho_{\rm B}$ model is determined by minimizing ${ \tilde{\chi}^2_{\rho_{\rm B}}} + 5 \chi^2_{\rm pena}$, where $\chi^2$ to evaluate a $\rho_{\rm B}$ model is defined as
${ \tilde{\chi}^2_{\rho_{\rm B}}} \equiv \chi^2_{\rm RC} + \chi^2_{N_{I < 21}} + \chi^2_{N_{I < 18}} + 2\chi^2_{N_{\rm eve}}$, 
and the $f_i (t_{\rm E})$ and $L_{{\cal M}_I} ({\cal M}_I)$ from step 1 are fixed and used. 
\item[3.] A $v_{\rm B}$ model is determined by minimizing $2 \chi^2_{v_{\rm B}} + 5 \chi^2_{\rm pena}$, where $\chi^2$ to evaluate a $v_{\rm B}$ model is defined as
$\chi^2_{v_{\rm B}} \equiv \chi^2_{\mu} + \chi^2_{\rm BRA}$, and the best-fit $\rho_{\rm B}$ model from step 2 is fixed and used.
\item[4.] The $\rho_{\rm B}$ model from step 2 is updated by minimizing ${ \tilde{\chi}^2_{\rho_{\rm B}}} + 2 \chi^2_{v_{\rm B}} + 5 \chi^2_{\rm pena}$, where the best-fit $v_{\rm B}$ model from step 3 is fixed and used.
\item[5.] The $v_{\rm B}$ model from step 3 is updated by minimizing $2 \chi^2_{v_{\rm B}} + 5 \chi^2_{\rm pena}$, where the best-fit $\rho_{\rm B}$ model from step 4 is fixed and used.
\item[6.] A $\chi^2_{t_{\rm E}}$ value, $\chi^2_{t_{\rm E}, {\rm fin}}$, is calculated using the $\rho_{\rm B}$, $v_{\rm B}$, and IMF models from steps 4, 5, and 1, respectively. 
If $|\chi^2_{t_{\rm E}, {\rm ini}} - \chi^2_{t_{\rm E}, {\rm fin}}| < 4$, the iteration is stopped; otherwise, steps 1--6 are repeated.
\end{enumerate}

It is difficult to systematically determine weights among various datasets suffering from different systematic errors; therefore, a factor multiplied 
by each $\chi^2$ value is subjectively selected in our attempt to find a proper balance between large and small datasets.
We multiply 2 by $\chi^2_{N_{\rm eve}}$ in ${ \tilde{\chi}^2_{\rho_{\rm B}}}$ because our interest is in microlensing study, and we want to have a better agreement on it.
Further, we multiply 2 by $\chi^2_{v_{\rm B}}$ because the number of data points contributed is $\sim 30$ times fewer than ${ \tilde{\chi}^2_{\rho_{\rm B}}}$.
$\chi^2_{\rm pena}$ is multiplied by 5 in steps 2--5 because the $\chi^2_{\rm pena}$ is the sum of the $\chi^2$ penalty set on the nine quantities, as the effect is otherwise easily diminished 
by a small improvement of $\chi^2_{v_{\rm B}}$ or ${ \tilde{\chi}^2_{\rho_{\rm B}}}$, which could be falsely caused due to systematic errors in data.

In step 1, we fit the four parameters for the IMF and the mass normalization factor, $\rho_{\rm 0, B}$, by minimizing ${ \tilde{\chi}^2_{\rm IMF}}$, which concerns 
not only agreement with the $t_{\rm E}$ distribution by $\chi^2_{t_{\rm E}}$, but also agreement with 
star count data via the $\chi^2_{N_{I < 21}}$, $\chi^2_{N_{I < 18}}$, and $\chi^2_{f_{18/21}}$ values. 
This is because the model values for the star count data given by Eq. (\ref{eq-NIc}) are calculated using the luminosity function $L_{{\cal M}_I} ({\cal M}_I)$, and 
$L_{{\cal M}_I} ({\cal M}_I)$ is calculated using a given IMF.
The sum of the three $\chi^2$ values for the OGLE-IV star count data, $\chi^2_{N_{I < 21}} + \chi^2_{N_{I < 18}} + \chi^2_{f_{18/21}}$, 
are multiplied by 0.1 $\times$ 2/3 in ${ \tilde{\chi}^2_{\rm IMF}}$ because the number of data values are 1456 for each of the three datasets 
compared with 40 for the $t_{\rm E}$ distribution, and we desire to place a larger weight on $t_{\rm E}$ because $\chi^2_{N_{I < 21}} + \chi^2_{N_{I < 18}}$ is also included in ${ \tilde{\chi}^2_{\rho_{\rm B}}}$.
The factor 2/3 is multiplied because only two of the three datasets, $N_{I < 21}$, $N_{I < 18}$, and $f_{18/21}$ ($= N_{I < 18}/N_{I < 21}$), are independent data.

As mentioned in Section \ref{sec-Neve}, the $\chi^2$ for the number of microlensing event detections, $\chi^2_{N_{\rm eve}}$, 
is insensitive to a variation of the probability density function for $t_{\rm E}$ in $i$th subfield, $f_i(t_{\rm E})$,
till the models used to calculate $f_i (t_{\rm E})$ show similar values of $\chi^2_{t_{\rm E}}$, the $\chi^2$ for the shape of the $t_{\rm E}$ distribution.
In the fitting procedure, we use the same $f_i(t_{\rm E})$ from step 1 throughout steps 2--6, where $\chi^2_{N_{\rm eve}}$ contributes to $\chi^2_{\rm \rho_{\rm B}}$ in steps 2 and 4.
Using the same $f_i(t_{\rm E})$ can be justified only when $\chi^2_{t_{\rm E}, {\rm ini}}$ is similar to  $\chi^2_{t_{\rm E}, {\rm fin}}$, and thus a similarity between the two values is used as the condition to end the iteration.
The differences among different iteration runs are $f_i(t_{\rm E})$ and the IMF model, and the condition ensures convergences of not only $f_i(t_{\rm E})$, but also the IMF, because 
the IMF parameters are primarily determined by shape of $t_{\rm E}$ distribution when the $\rho_{\rm B}$ and $v_{\rm B}$ models are sufficiently constrained by other datasets.
Note that $\chi^2_{N_{\rm eve}}$ is only used to find the best-fit $\rho_{\rm B}$ models because a model value for the number of microlensing events with $t_{{\rm E}, j}$ in 
$i$th subfield, $N_{{\rm eve}, i}^{\rm mod} (t_{{\rm E}, j})$, given by Eq. (\ref{eq-Neve}) is independent of the $v_{\rm B}$ and IMF models once $f_i(t_{\rm E})$ is fixed.

%Figure 9
\begin{figure}
\centering
\includegraphics[width=18cm]{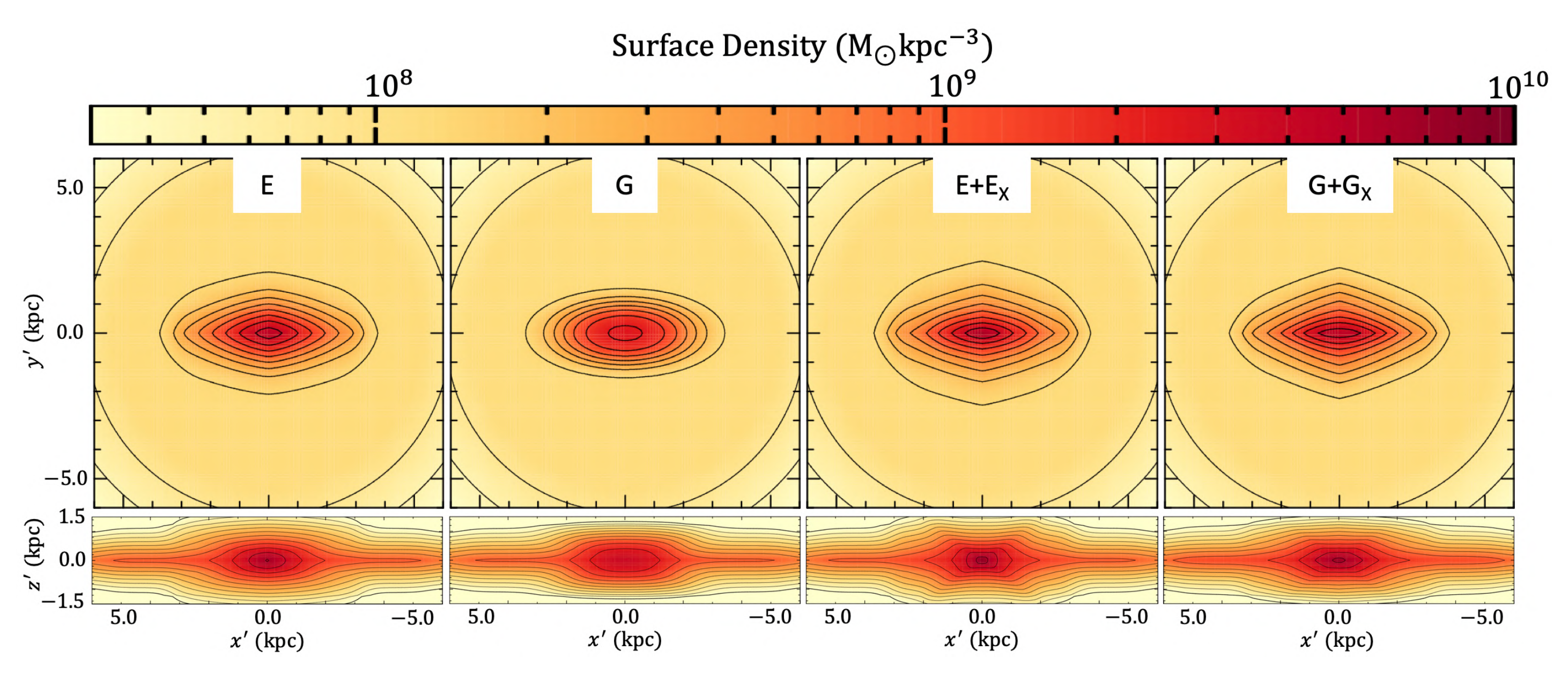}
\caption{Surface density distributions along the bar-axes of the two one-component models (E and G) and two two-components models (E+E$_{\rm X}$ and G+G$_{\rm X}$).
Parameters of each model are listed in Table \ref{tab-params}.
Note that %these models are determined by comparing with the data in Figs. \ref{fig-mod_rhob} and \ref{fig-mod_vb} which lacks information in $|b| \simlt 2$. 
the plotted surface density does not include the contribution from very centered or very thin components such as a nuclear stellar disk \citep{lau02, nis13, por17} or thin bar components \citep{weg15}.
We neither confirm nor confute existence of such components in this paper because we lack data in $|b| \simlt 2^\circ$ or in $|l| > 10^\circ$.
}
\label{fig-best_xyz}
\end{figure}

We use a Monte Carlo simulation of $3 \times 10^6$ microlensing events to calculate the model values for the $t_{\rm E}$ distribution,
$N_o^{\rm mod} (t_{{\rm E}, j})$ ($o = {\rm O_{\rm hi}}, {\rm O_{\rm low}}$), which takes $\sim$20 s to 
calculate a $\chi^2_{t_{\rm E}}$ value under our computational environment.
{ This causes a dispersion of $\chi^2_{t_{\rm E}}$ values $\sim 1.3$ depending on the random seed value.
Thus, we conduct MCMC fits in step 1 with a temperature of $T_{\rm MCMC} \geq 4$ to avoid being stuck at a local minimum created due to the random noise, where 
a proposed parameter set with $\Delta \chi^2 > 0$ is accepted with a probability $\exp[{-0.5 \, \Delta \chi^2/T_{\rm MCMC}}]$ in an MCMC run.}
As a result of this, we use $|\chi^2_{t_{\rm E}, {\rm ini}} - \chi^2_{t_{\rm E}, {\rm fin}}| < 4$ as the condition in step 6.
Note that the $\chi^2_{t_{\rm E}, {\rm ini}}$ and $\chi^2_{t_{\rm E}, {\rm fin}}$ values are calculated with a Monte Carlo simulation of $3 \times 10^7$ events, which is 10 times 
the one during the MCMC calculations.

{ It is important to ensure that a solution from the MCMC fit with the $\chi^2_{t_{\rm E}}$ dispersion of $\sim 1.3$ is not plagued by the random noise.
In Section \ref{sec-fund}, we compare the best-fit model from our fitting procedure with the best-fit model from a grid search 
where $\chi^2_{t_{\rm E}}$ at each grid is calculated with $3 \times 10^7$ events.
We find a consistency between them, and} confirm that the settings for the fitting procedure have successfully determined a converged solution.

In the Monte Carlo simulation of microlensing events, we also consider binary systems for lens objects, which are generated by following 
the binary distribution developed by \citet{kos20b} based on \citet{duc13}.
Because \citet{mro17} and \citet{mro19} did not include binary lens events in the OGLE-IV $t_{\rm E}$ distributions, we reject a detectable binary system 
whose central caustic size $w_{\rm LC}$ is larger than the impact parameter $u_0$, where $u_0$ is randomly generated uniformly from 0 to 1 in each trial of the Monte Carlo simulation.
We use $w_{\rm LC} = 4 q_{\rm LC}/(s_{\rm LC} - s_{\rm LC}^{-1})^2$ \citep{chu05} with the mass ratio, $q_{\rm LC}$, and the
separation in unit of the angular Einstein radius $\theta_{\rm E}$, $s_{\rm LC}$. 
Although the formula for $w_{\rm LC}$ is an approximation for the planetary mass-ratio of $q_{\rm LC} \ll 1$, \citet{kos20b} found that 
the criterion $w_{\rm LC} < u_0$ itself works reasonably well even for a stellar mass-ratio because of the extension of $\theta_{\rm E}$ 
by a factor $\sqrt{1+q_{\rm LC}}$. Section 10.3.1 of \citet{kos20b} presents a more detailed discussion.
Because a tight close binary system with $s_{\rm LC} \ll 1$ can be undetectable under this criterion, the resulting $t_{\rm E}$ distribution includes 
their contribution, as well as single star lens systems.
Such a binary system has a lens mass equal to the total system mass, and their fraction for each $t_{\rm E}$ bin ranges from $\simlt 3\%$ for $t_{\rm E} < 10~{\rm days}$ to 
10--14\% for $t_{\rm E} > 50~{\rm days}$.

\subsubsection{Best-Fit Models} \label{sec-modsel}

% Table 6
{\tabcolsep = 1.2mm
\begin{deluxetable}{lccccccccccccccccccccccccccccccc}
\tabletypesize{\scriptsize}
\tablecaption{Best-fit parameters for each model. \label{tab-params}}
\tablehead{
\colhead{Model} & & \multicolumn{4}{c}{IMF model} & & \multicolumn{9}{c}{$\rho_{\rm B}$ model} \\
\cline{3-7} \cline{9-17}
                    & ${ \tilde{\chi}^2_{\rm sum}}$\tablenotemark{a}  & ${ \tilde{\chi}^2_{\rm IMF}}$\tablenotemark{a} &   $M_{\rm br}$  & $\alpha_{\rm hm}$ & $\alpha_{\rm ms}$ & $\alpha_{\rm bd}$ & & ${ \tilde{\chi}^2_{\rho_{\rm B}}}$\tablenotemark{a} &  $\rho_{\rm 0, B}$     & &   $x_0$        &        $y_0$         &       $z_0$      &     $C_\perp$   & $C_\parallel$   &        $R_{\rm c}$         \\ 
                    &                    &                 &   [$M_{\odot}$]   &                   &                   &                   & &                         & [$M_{\odot}/$pc$^3$]  & &   [kpc]        &        [kpc]         &       [kpc]      &                 &                 &           [kpc]             
}                                                                                                                                                                                                                                        
\startdata                                                                                                                                            
E                   &     80173          &       355       &         0.84      &      2.31         &       1.10        &       0.18        & &        75481            &       9.72            & &    0.67        &         0.28         &       0.24       &       1.4       &     3.3         &          2.8                \\
G                   &     91319          &       391       &         0.90      &      2.40         &       1.18        &       0.17        & &        86105            &       2.43            & &    1.03        &         0.46         &       0.40       &       2.0       &     4.0         &          4.8                \\
E+E$_{\rm X}$       &     76261          &       342       &         0.86      &      2.32         &       1.13        &       0.18        & &        72147            &       4.12            & &    0.93        &         0.37         &       0.24       &       1.2       &     4.1         &          2.6                \\
G+G$_{\rm X}$       &     76500          &       337       &         0.90      &      2.32         &       1.16        &       0.22        & &        72585            &       0.88            & &    1.56        &         0.72         &       0.49       &       1.2       &     3.1         &          2.8                \\\hline\hline
Model               &                                                                          \multicolumn{8}{c}{$\rho_{\rm B}$ model}                                                               & &  \multicolumn{6}{c}{$v_{\rm B}$ model}                              \\
\cline{2-10} \cline{12-17}
                    &    $f_{\rm 0, X}$  &   $b_{\rm X}$   &   $x_{\rm 0, X}$  & $y_{\rm 0, X}$    & $z_{\rm 0, X}$    & $C_{\rm \perp,X}$ & & $C_{\rm \parallel, X}$  & $R_{\rm c, X}$        & & $\chi^2_{v_{\rm B}}$\tablenotemark{a} & $\Omega_{\rm p}$ & $v_0^{\rm str}$ & $y_0^{\rm str}$ & $\sigma_{v_{\rm B}, x', 0}$ & $\sigma_{v_{\rm B}, y', 0}$ \\
                    &                    &                 &    [kpc]          &   [kpc]           &   [kpc]           &                   & &                         &    [kpc]              & &                &     [km/s/kpc]       &   [km/s]         &     [pc]        &     [km/s]      &      [km/s]                 \\\hline
E                   &        --          &       --        &        --         &          --       &         --        &         --        & &           --            &       --              & &       2203     &        49.5          &     49           &     393         &      64         &          75                 \\
G                   &        --          &       --        &        --         &          --       &         --        &         --        & &           --            &       --              & &       2382     &        40.5          &     12           &      20         &      76         &          68                 \\
E+E$_{\rm X}$       &       1.44         &      1.38       &       0.28        &        0.18       &       0.29        &        1.3        & &          2.2            &      1.3              & &       1908     &        47.4          &     43           &     407         &      64         &          76                 \\
G+G$_{\rm X}$       &       3.00         &      1.38       &       0.76        &        0.31       &       0.40        &        1.2        & &          1.3            &      5.2              & &       1831     &        45.9          &     28           &      11         &      64         &          75                 \\\hline\hline
Model               &                                        \multicolumn{15}{c}{$v_{\rm B}$ model}                            \\
\cline{2-17}
                    & $\sigma_{v_{\rm B}, z', 0}$ & $\sigma_{v_{\rm B},x',1}$ & $\sigma_{v_{\rm B},y',1}$ & $\sigma_{v_{\rm B},z',1}$ & $x_{\rm 0,\sigma_R}$ & $y_{\rm 0,\sigma_R}$ & & $z_{\rm 0,\sigma_R}$ & $C_{\rm \perp,\sigma_R}$ & & $C_{\rm \parallel, \sigma_R}$ & $x_{\rm 0,\sigma_{z'}}$ & $y_{\rm 0,\sigma_{z'}}$ & $z_{\rm 0,\sigma_{z'}}$ & $C_{\rm \perp,\sigma_{z'}}$ & $C_{\rm \parallel, \sigma_{z'}}$  \\
                    &         [km/s]              &        [km/s]             &        [km/s]             &        [km/s]             &      [kpc]           &      [kpc]           & &      [kpc]           &                          & &                               &         [kpc]           &         [kpc]           &         [kpc]           &                             &                                   \\\hline  
E                   &           72                &      156                  &          84               &         86                &      0.82            &       9.29           & &      0.86            &              3.8         & &         1.0                   &    0.51                 &      2.90               &         2.19            &         3.0                 &          1.0                      \\
G                   &           75                &      136                  &         109               &        101                &      1.03            &       2.15           & &      0.73            &              4.9         & &         1.0                   &    0.52                 &      1.44               &         1.10            &         2.3                 &          1.0                      \\
E+E$_{\rm X}$       &           71                &      152                  &          78               &         82                &      0.86            &       3.22           & &      0.95            &              4.3         & &         1.0                   &    0.56                 &      2.00               &         3.82            &         3.7                 &          1.1                      \\
G+G$_{\rm X}$       &           70                &      155                  &          78               &         83                &      0.94            &       4.23           & &      0.88            &              4.6         & &         1.0                   &    0.70                 &      1.73               &         2.03            &         4.8                 &          1.0
\enddata
\tablenotetext{a}{ ${ \tilde{\chi}^2_{\rm sum}} \equiv { \tilde{\chi}^2_{\rho_{\rm B}}} + 2 \chi^2_{v_{\rm B}} + 5 \chi^2_{\rm pena}$, ${ \tilde{\chi}^2_{\rm IMF}} \equiv \chi^2_{t_{\rm E}} + 0.2(\chi^2_{N_{I < 21}} + \chi^2_{N_{I < 18}} + \chi^2_{f_{18/21}})/3$, \ \  ${ \tilde{\chi}^2_{\rho_{\rm B}}} \equiv \chi^2_{\rm RC} + \chi^2_{N_{I < 21}} + \chi^2_{N_{I < 18}} + 2\chi^2_{N_{\rm eve}}$, and $\chi^2_{v_{\rm B}} \equiv \chi^2_{\mu} + \chi^2_{\rm BRA}$.}
%\tablecomments{}   
\end{deluxetable}
}

As described in Section \ref{sec-bar_rho}, we consider four different shapes for the $\rho_{\rm B}$ model: two one-component models (E and G models) and 
two two-components models (E+E$_{\rm X}$ and G+G$_{\rm X}$ models).
For each of the E, G, E+E$_{\rm X}$, and G+G$_{\rm X}$ models, we consider four options, corresponding to the four disk models in Table \ref{tab-disk_result}.
We derived the best-fit parameters for each option with each model by the fitting procedure described in Section \ref{sec-fit_proc}.
We define another $\chi^2$ value, ${ \tilde{\chi}^2_{\rm sum}} = { \tilde{\chi}^2_{\rho_{\rm B}}} + 2 \chi^2_{v_{\rm B}} + 5 \chi^2_{\rm pena}$, and use its difference, $\Delta { \tilde{\chi}^2_{\rm sum}}$, to indicate the better model among them.
The $\chi^2$ for the $t_{\rm E}$ distributions, $\chi^2_{t_{\rm E}}$, is not considered in ${ \tilde{\chi}^2_{\rm sum}}$ because the information 
contained in $\chi^2_{t_{\rm E}}$ is included in the $\chi^2$ for the number of microlensing event detections as a function of $t_{\rm E}$, $\chi^2_{N_{\rm eve}}$, in a more statistically proper style.
Although this inclusion is only for the fields covered by \citet{nat13}, we conservatively avoid the partial overlap.
In this subsection, we determine the best-fit model for each of the E, G, E+E$_{\rm X}$, and G+G$_{\rm X}$ models by selecting the best option from the four based on 
comparisons using $\Delta { \tilde{\chi}^2_{\rm sum}}$.

Among the four disk models in Table \ref{tab-disk_result}, we decide to use the all-$z$ $+$ flat $z_{\rm d}^{\rm thin}$ model based on the following two comparisons.
The first comparison is between the all-$z$ and low-$z$ models, where we find $\Delta { \tilde{\chi}^2_{\rm sum}} < 100$ between the two, which is not considered significant.
Thus, we select the all-$z$ models because it is more broadly applicable compared to the low-$z$ models optimized for bulge sky.
The second comparison is between the flat- and linear-scale height models, where we find that the best-fit model with the flat-scale height model is preferable to 
the linear-scale height model by $\Delta { \tilde{\chi}^2_{\rm sum}} > 1000$.
Half of this preference comes from $\chi^2_{N_{I < 21}} + \chi^2_{N_{I < 18}}$ while the other half comes from $2 \chi^2_{v_{\rm B}}$.
These results barely depend on the choice of E, G, E+E$_{\rm X}$, and G+G$_{\rm X}$ models, and we select the all-$z$ $+$ flat $z_{\rm d}^{\rm thin}$ 
model as the fiducial disk model for all of them.

In Appendix \ref{sec-G20}, we consider another option of using a different dataset of the OGLE-IV $t_{\rm E}$ measurements.
In this dataset, the original light curve data are the same as those by \citet{mro17, mro19}, but the Gaussian Process model developed by \citet{gol20} 
is applied for the $t_{\rm E}$ measurements.
Appendix \ref{sec-G20} presents a comparison of $\chi^2$ values between the best-fit models with the original $t_{\rm E}$ distributions by \citet{mro17, mro19} and 
the best-fit models with the $t_{\rm E}$ distributions with the \citet{gol20} model, where we find that the original Mr{\'o}z et al.'s $t_{\rm E}$ distributions are favored with 
respect to the ${ \tilde{\chi}^2_{\rm sum}}$, $\chi^2_{N_{\rm eve}}$, and $\chi^2_{t_{\rm E}}$ values.

Table \ref{tab-params} lists all the best-fit parameters for the fiducial four models.
Figs. \ref{fig-mod_rhob}--\ref{fig-mod_tE} present comparisons between the data and model values using the four models from Table \ref{tab-params}, where we show $\chi^2_p$ values for
each parameter $p$ in each panel of Figs. \ref{fig-mod_rhob},  \ref{fig-mod_vb}, and \ref{fig-mod_tE}, while Figs. \ref{fig-resi_rhob} and \ref{fig-resi_vb} show 
the residuals corresponding to Figs. \ref{fig-mod_rhob} and \ref{fig-mod_vb}, respectively.
Fig. \ref{fig-best_xyz} shows the surface density distributions of each best-fit model.

Among the four models, we find that the two-components models are more favorable than the one-component models by $\Delta { \tilde{\chi}^2_{\rm sum}} > 3600$.
The E model is preferred over the G model by $\Delta { \tilde{\chi}^2_{\rm sum}} \sim 11000$ in the one-component models.
By contrast, the two two-component models {(E+E$_{\rm X}$ vs. G+G$_{\rm X}$)} show very similar ${ \tilde{\chi}^2_{\rm sum}}$ values, as described in Section \ref{sec-bar_rho}.

\subsection{Uncertainty Assessments for Fundamental Parameters} \label{sec-fund}

The far-right column in Table \ref{tab-bar_prior} presents the posterior values for the fundamental parameters on which we applied the prior in the fits.
The fiducial values are from the best-fit G+G$_{\rm X}$ model, while the uncertainties are combinations of statistical and systematic errors for each parameter.

The systematic errors are taken from the variation of each value depending on the model choice listed in Table \ref{tab-params}.
These are the stellar mass within the VVV bulge box ($\pm 2.2 \times \pm 1.4 \times \pm 1.2$ kpc), $M_{\rm VVV} = 1.14^{+0.00}_{-0.11} |_{\rm sys.}  \, \times 10^{10} \, M_{\odot}$, 
mass-weighted velocity dispersions inside the bulge half mass radius along the bar axes, 
$(\VEV{\sigma_{v_{\rm B}, x'}}, \VEV{\sigma_{v_{\rm B}, y'}}, \VEV{\sigma_{v_{\rm B}, z'}}) = (141.0^{+0.0}_{-3.8} |_{\rm sys.}, 113.6^{+3.2}_{-0.0} |_{\rm sys.}, 108.3^{+0.0}_{-1.5} |_{\rm sys.}) \, {\rm km/s}$, 
bar pattern speed, $\Omega_{\rm p} = 45.9^{+3.6}_{-5.4} |_{\rm sys.} \, {\rm km/s/kpc}$, 
break mass in the IMF, $M_{\rm br} = 0.90^{+0.00}_{-0.06}  |_{\rm sys.} \, M_{\odot}$, and IMF slopes for three different mass regions, 
$(\alpha_{\rm hm}, \alpha_{\rm ms}, \alpha_{\rm bd}) = (2.32^{+0.08}_{-0.01} |_{\rm sys.}, 1.16^{+0.02}_{-0.06} |_{\rm sys.}, 0.22^{+0.00}_{-0.05} |_{\rm sys.})$.
The representative values are for the best-fit G+G$_{\rm X}$ model, and the 0 error values indicate that the G+G$_{\rm X}$ model has 
the largest or lowest values among the four models in Table \ref{tab-params}.

Statistical errors for the four velocity parameters are determined using the posterior distribution for the G+G$_{\rm X}$ model 
taken from the MCMC calculation in step 5 in our fitting procedure that is described in Section \ref{sec-fit_proc}.
These are $(\VEV{\sigma_{v_{\rm B}, x'}}, \VEV{\sigma_{v_{\rm B}, y'}}, \VEV{\sigma_{v_{\rm B}, z'}}) = (141.0^{+1.7}_{-2.2} |_{\rm stat.}, 113.6^{+1.0}_{-0.9} |_{\rm stat.}, 108.3^{+0.2}_{-0.9} |_{\rm stat.}) \, {\rm km/s}$ 
and $\Omega_{\rm p} = 45.9^{+0.6}_{-0.2} |_{\rm stat.} \, {\rm km/s/kpc}$.

\begin{figure}
\centering
\includegraphics[width=10cm]{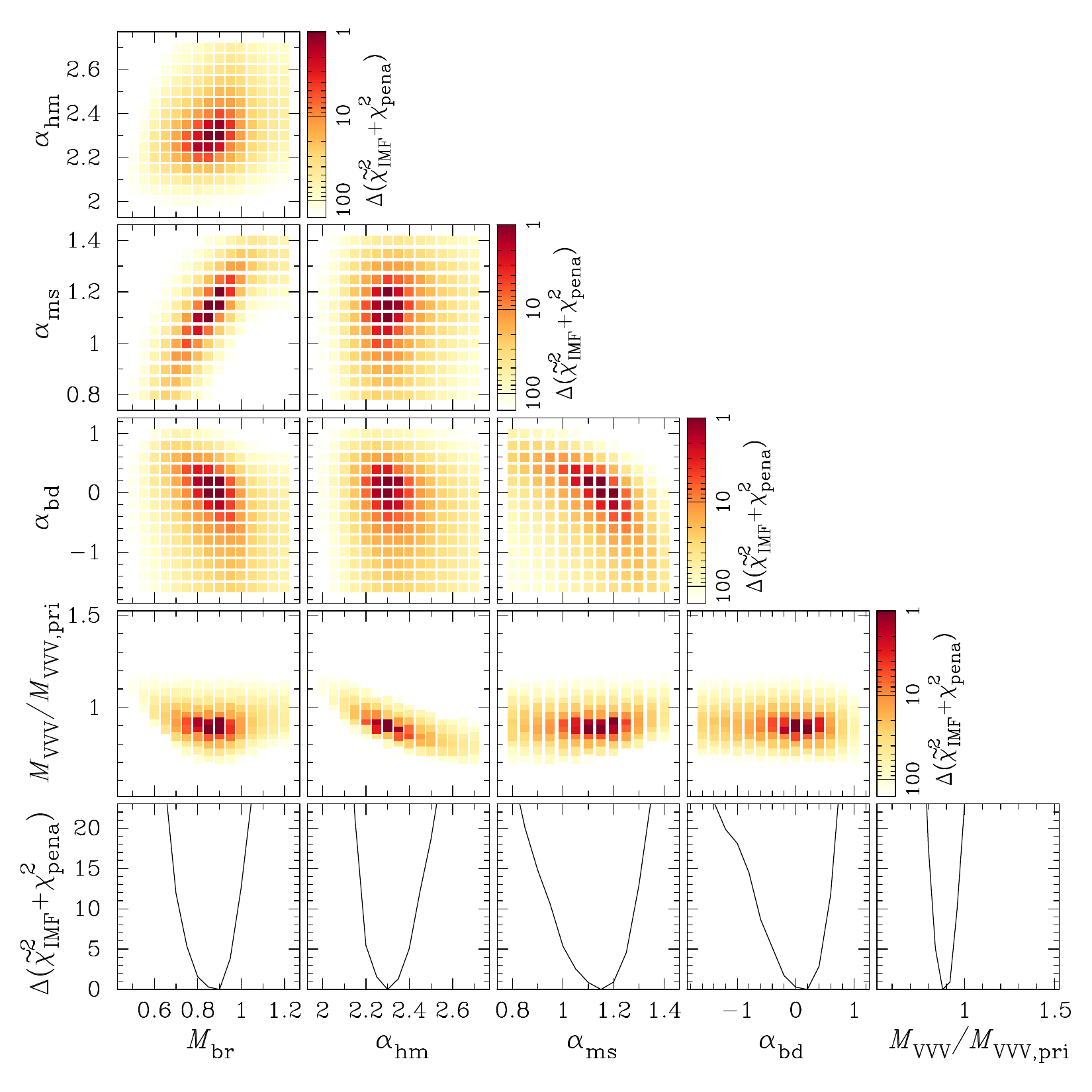}
\caption{$\Delta ({ \tilde{\chi}^2_{\rm IMF}} + \chi^2_{\rm pena})$ map calculated through a grid-search with the best-fit G+G$_{\rm X}$ $\rho_{\rm B}$ and $v_{\rm B}$ models.
$M_{\rm VVV, pri}$ indicates prior value for $M_{\rm VVV}$ of 1.32 $\times 10^{10} \, M_{\odot}$.}
\label{fig-chi2IMFpena}
\end{figure}

Although we have a posterior distribution from the MCMC calculation in step 1 for the other five parameters, 
$M_{\rm VVV}$, $M_{\rm br}$, $\alpha_{\rm hm}$, $\alpha_{\rm ms}$, and $\alpha_{\rm bd}$,
{ the distribution is contaminated by the dispersion of $\chi^2_{t_{\rm E}}$ values of $\sim 1.3$ due to the limited sample size of simulated events ($3 \times 10^6$ events) in 
each step of the MCMC calculation.
Thus, we use the $\Delta \chi^2$ distributions from a grid search with a ten times larger sample size of $3 \times 10^7$ events to determine statistical errors for the five parameters.}
Fig. \ref{fig-chi2IMFpena} shows the resultant $\Delta ({ \tilde{\chi}^2_{\rm IMF}} + \chi^2_{\rm pena})$ maps calculated with the best-fit G+G$_{\rm X}$ $\rho_{\rm B}$ and $v_{\rm B}$ models.
In the grid search, a grid is specified by a combination of $(M_{\rm br}, \alpha_{\rm hm}, \alpha_{\rm ms}, \alpha_{\rm bd})$ distributed uniformly, and 
the mass normalization factor $\rho_{\rm 0, B}$, which is represented by $M_{\rm VVV}$ here, is adjusted such that the ${ \tilde{\chi}^2_{\rm IMF}} + \chi^2_{\rm pena}$ for each grid is minimized.

The best-fit parameters from the grid search are 
$(M_{\rm VVV}, M_{\rm br}, \alpha_{\rm hm}, \alpha_{\rm ms}, \alpha_{\rm bd})$ = ($1.16\times 10^{10} \, M_{\odot}$, $0.90 \, M_{\odot}$, 2.30, 1.15, 0.20), which are 
almost identical to the best-fit G+G$_{\rm X}$ model parameters of ($1.14 \times 10^{10} \, M_{\odot}$, $0.90 \, M_{\odot}$, 2.32, 1.16, 0.22) considering the limited resolution of the grid search.
These two sets of parameters can be, in principle, different because the grid search uses the $\rho_{\rm B}$ and $v_{\rm B}$ models obtained from 
the last iteration of our fitting procedure, whereas the IMF parameters of the best-fit G+G$_{\rm X}$ model are determined using 
the tentative best-fit $\rho_{\rm B}$ and $v_{\rm B}$ models from the last but one iteration.
{ Therefore, the identity of these two sets of parameters confirms a convergence of our iterative fitting procedure.
This also ensures that the number of simulated events used in the MCMC fit, $3 \times 10^6$, was sufficient to find the best-fit parameters.}

We use $\Delta ({ \tilde{\chi}^2_{\rm IMF}} + \chi^2_{\rm pena}) = 4$ as the threshold value to determine the statistical errors for the five parameters 
because the grid search is based on the best-fit $\rho_{\rm B}$ and $v_{\rm B}$ models determined using the IMF parameters from step 1 in which the MCMC temperature of $T_{\rm MCMC} = 4$ is used.
This yields $(M_{\rm VVV}, M_{\rm br}, \alpha_{\rm hm}, \alpha_{\rm ms}, \alpha_{\rm bd})$ = ($1.14^{+0.10}_{-0.02} |_{\rm stat.} \times 10^{10} \, M_{\odot}$, 
$0.90^{+0.05}_{-0.13}  |_{\rm stat.} \, M_{\odot}$, $2.32^{+0.07}_{-0.10} |_{\rm stat.}$, $1.16^{+0.08}_{-0.14} |_{\rm stat.}$, $0.22^{+0.20}_{-0.55} |_{\rm stat.}$).
By combining the statistical and systematic errors above, we derive the posterior values listed in Table \ref{tab-bar_prior}.
Note that these error estimates could be influenced by our non-standard fitting method, an iterative step-by-step procedure, described in Section \ref{sec-fit_proc}.

%Figure 10

\section{Discussion} \label{sec-comp}
In this section, we discuss several fundamental parameters in our Galaxy measured through our modeling process throughout Sections \ref{sec-disk} and \ref{sec-bar} by 
comparison to previous studies.
In Section \ref{sec-disk_comp}, we discuss the 10 best-fit parameters for the $v_{\rm d}$ models in Table \ref{tab-disk_result}.
Section \ref{sec-compIMF} examines how the IMF parameters are constrained in our model framework, and compares these IMF parameters with a local IMF by \citet{kro01} and a bulge IMF by \citet{zoc00}.
In Section \ref{sec-M2L}, we derive the stellar mass-to-light ratio corresponding to our IMF and estimate the dark matter fraction inferred from our results.

\subsection{Disk Velocity Parameters} \label{sec-disk_comp}
A review paper by \citet{bla16} summarizes estimates on the local velocity dispersion values and the scale lengths of the distributions from several previous studies (see their Table 5).
The local velocity dispersion values for the thin disk range from 34 \citep{pif14} to 48 km/s \citep{san15} for $\sigma_{R, \odot}^{\rm thin}$ and 
from 20 \citep{bin12} to 31 km/s \citep{san15} for $\sigma_{z, \odot}^{\rm thin}$, and our best-fit values in Table \ref{tab-disk_result}
are consistent with the estimates for these two parameters in any of the four models considered.
 
\citet{bin12} is the only one who did a fit to the scale length of the velocity dispersion along $R$ for the thin disk, $R_{\sigma_R}^{\rm thin}$, 
in the summary by \citet{bla16}. Their estimate is $R_{\sigma_R}^{\rm thin}=$ 3.3--20 kpc, which includes our estimated values for all models other 
than the all-$z$ $+$ linear $z_{\rm d}^{\rm thin}$ model. 
The value for the all-$z$ $+$ linear $z_{\rm d}^{\rm thin}$ model is $R_{\sigma_R}^{\rm thin} = 21.4$ kpc, and it is only slightly outside the range presented by \citet{bin12}.

No study has presented a fit to the scale length of the velocity dispersion along $z$ for the thin disk,
$R_{\sigma_z}^{\rm thin}$, in the summary by \citet{bla16}, but some studies 
fixed the value at 9.0 \citep{pif14}, 7.8 \citep{san15}, and 7.4 kpc \citep{bin15} in their analysis. 
Our estimates on $R_{\sigma_z}^{\rm thin}$ ranging from 5.9--10.4 kpc are consistent with these values.
The all-$z$ $+$ flat $z_{\rm d}^{\rm thin}$ model, which is our fiducial disk model selected in Section \ref{sec-modsel}, has $R_{\sigma_z}^{\rm thin} = 5.9$ kpc which is 
shorter than the fixed values in the previous papers, though.

Further, our estimates on the slope of age-velocity dispersion relation are $\beta_R = 0.22$--0.34 and $\beta_z = 0.77$--0.82, which 
are fully and fairly consistent with the estimates by \citet{yu18} of $\beta_R = 0.28 \pm 0.08$ and $\beta_z = 0.54 \pm 0.13$, respectively.

By contrast, our estimates for some parameters for the thick disk show inconsistency with previous studies.
In Table 5 of \citet{bla16}, the local velocity dispersion values for the thick disk range from 25 \citep{bin12} to 53 km/s \citep{bin15} for $\sigma_{R, \odot}^{\rm thick}$ and 
from 33 to 65 km/s \citep{bin12} for $\sigma_{z, \odot}^{\rm thick}$, while our estimate for these parameters are $\sigma_{R, \odot}^{\rm thick} =$ 68--75 km/s and 
$\sigma_{z, \odot}^{\rm thick} = 47.8$--61.4 km/s. Therefore, our estimate on $\sigma_{R, \odot}^{\rm thick}$ is significantly higher than the values in previous studies.

Similarly, our estimates on the scale lengths of the velocity dispersion are $R_{\sigma_R}^{\rm thick} = 47.0$--180 kpc and  $R_{\sigma_z}^{\rm thick} = 6.9$--52.0 kpc, which are both 
longer than $R_{\sigma_R}^{\rm thick} = 13$ kpc \citep{pif14} or $11.6$ kpc \citep{bin15} and $R_{\sigma_z}^{\rm thick} = 4.2$ kpc \citep{pif14} or $5.0$ kpc \citep{bin15}, respectively.
These differences probably arise from our rejection of the outer disk ($R > 8440$ pc) region.
In contrast to our preference for the inner disk, most of the previous studies used data primarily from the outer disk rather than the inner disk.
%In fact, Fig. \ref{fig-disk_resi} shows that the $\sigma_R$ values from the models are significantly higher than the data in upper part of the outer disk for the two flat scale height models, which confirms that 
%the estimated values of $\sigma_{R, \odot}^{\rm thick} = 75$ km/s for these models are likely to be too high in the outer disk region.

\subsection{IMF Parameters} \label{sec-compIMF}

%Figure 11
\begin{figure}
\centering
\includegraphics[width=8.7cm]{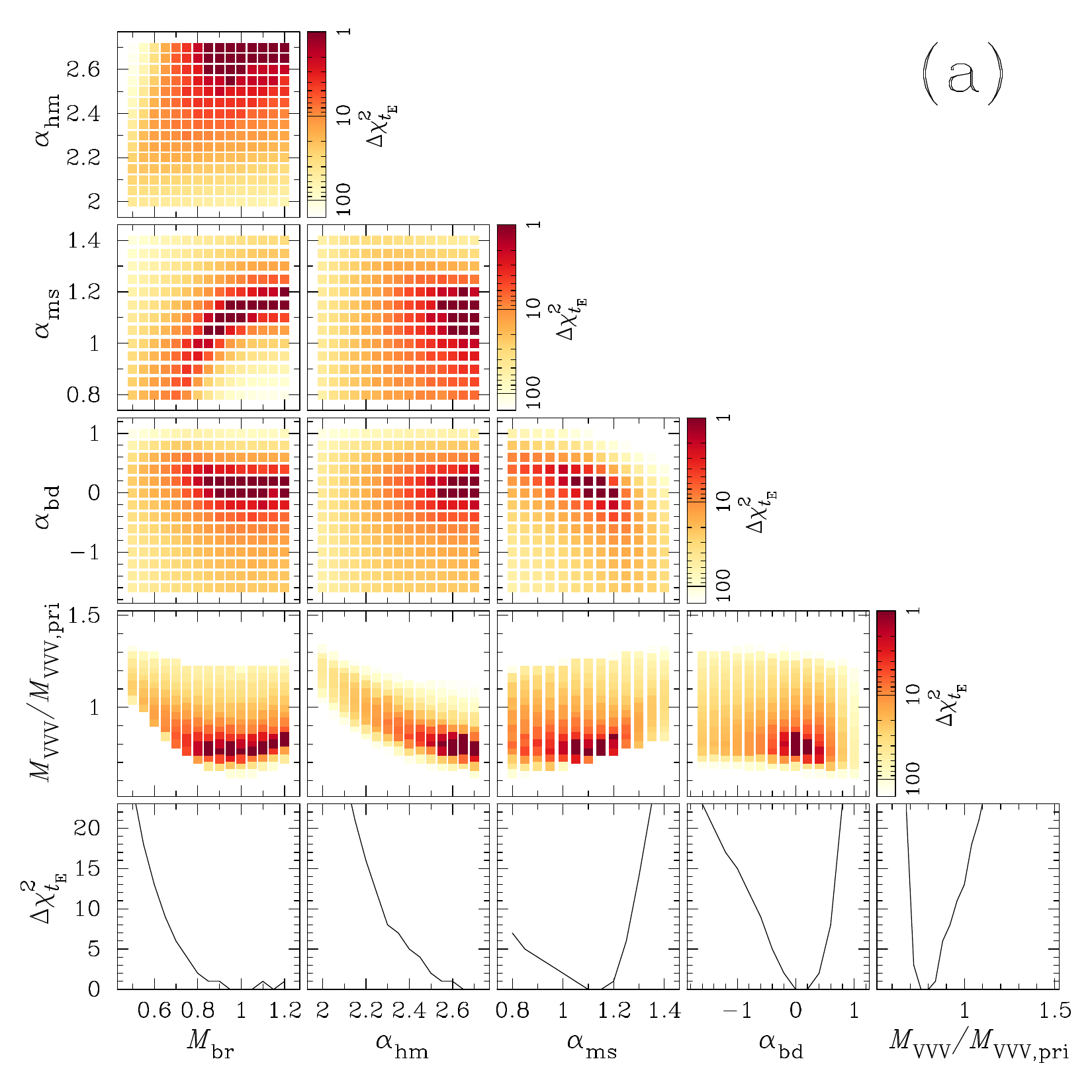}
\includegraphics[width=8.7cm]{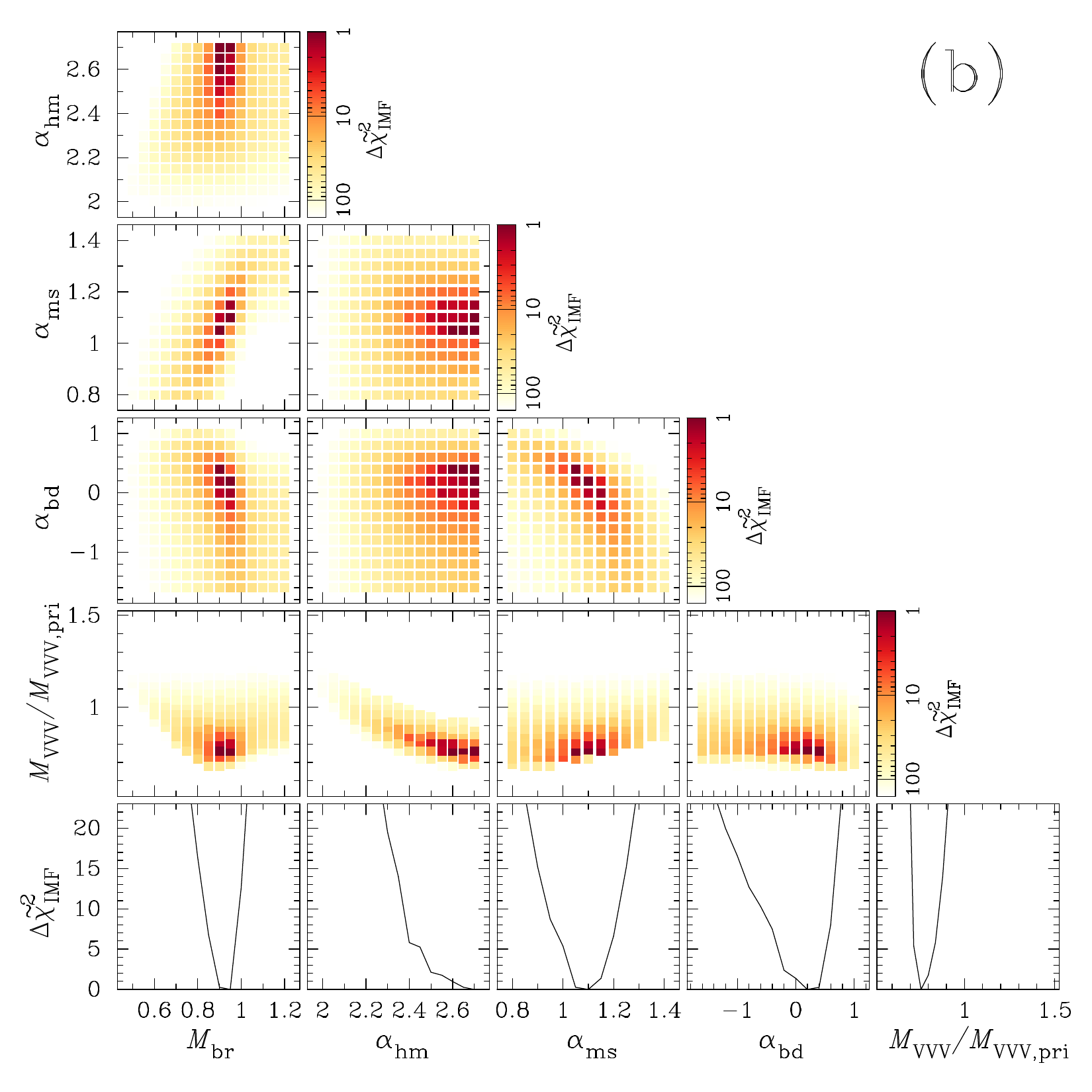}
\caption{
Same as Fig. \ref{fig-chi2IMFpena}, but for (a) $\Delta \chi^2_{t_{\rm E}}$ and (b) $\Delta { \tilde{\chi}^2_{\rm IMF}}$, where 
$\chi^2_{t_{\rm E}}$ is for the OGLE-IV $t_{\rm E}$ distributions and ${ \tilde{\chi}^2_{\rm IMF}} = \chi^2_{t_{\rm E}} + 0.2(\chi^2_{N_{I < 21}} + \chi^2_{N_{I < 18}} + \chi^2_{f_{18/21}})/3$
involves the three $\chi^2$ values for the OGLE-IV star count data, in addition to $\chi^2_{t_{\rm E}}$.
}
\label{fig-chi2IMF_tE}
\end{figure}

Several studies measured the IMF parameters in the bulge field, and two approaches have been attempted so far.
One approach is measuring a very deep luminosity function and deriving the IMF slopes by fitting to it using a mass--luminosity relation.
With this approach, \citet{zoc00} found that a single power--law model of $M^{-1.33}$ had a good agreement in $0.15\, M_{\odot} < M < 1.00\, M_{\odot}$ for 
the luminosity function measured toward $(l, b) = (0.277, -6.167)$ using the {\it HST}.
\citet{cal15} also used the {\it HST} to observe a luminosity function at $(l, b) = (1.25, -2.65)$ and derived $M_{\rm br} \sim 0.56 \, M_{\odot}$ and 
($\alpha_{\rm hm}$, $\alpha_{\rm ms}$) = $(2.41 \pm 0.50, 1.25 \pm 0.20)$ in $0.15\, M_{\odot} < M < 1.00\, M_{\odot}$.
This approach lacks sensitivity to $\alpha_{\rm bd}$ because brown dwarfs are too faint to be observed in a bulge field.

The other approach, in which the $t_{\rm E}$ distributions from microlensing surveys are used, has an advantage with regard to sensitivity to the brown dwarf population, 
although the IMF measured from this method somewhat depends on what Galactic model is used to calculate the $t_{\rm E} / \sqrt{M_{\rm L}}$ ${ = \sqrt{\kappa \pi_{\rm rel}}/\mu_{\rm rel}}$ distribution.
\citet{sum11} measured $\alpha_{\rm bd} = 0.50^{+0.36}_{-0.60}$ by comparing a parametric Galactic model based on \citet{han95} with
the $t_{\rm E}$ distribution of 474 events observed by the MOA-II survey.
\citet{mro17} analyzed the $t_{\rm E}$ distribution of 2617 events from the OGLE-IV survey, where $\alpha_{\rm bd} \sim 0.8$ and
$\alpha_{\rm ms} \sim 1.3$ were measured by comparing with a similar model based on \citet{han95, han03}.
Further, \citet{weg17} measured $\alpha_{\rm bd} = -0.7 \pm 0.9|_{\rm stat.} \pm 0.8|_{\rm sys.}$ and $\alpha_{\rm ms} = 1.31 \pm 0.10|_{\rm stat.} \pm 0.10|_{\rm sys.}$
using a non-parametric dynamical Galactic model developed by \citet{por17} for a comparison with the 3718 $t_{\rm E}$ measurements from the OGLE-III survey \citep{wyr15}.
By contrast, the break mass, $M_{\rm br}$, and the high mass slope, $\alpha_{\rm hm}$, are fixed in these previous studies, where
$M_{\rm br} = 0.5$ or 0.7 $M_{\odot}$ and $\alpha_{\rm hm} = 2.3$ or 2.0 are typically assumed, and no measurements on the two parameters using microlensing are reported in the literature.

Our method is a kind of a hybrid of the above two methods because we used ${ \tilde{\chi}^2_{\rm IMF}} + \chi^2_{\rm pena}$ to determine the IMF model, where 
${ \tilde{\chi}^2_{\rm IMF}}$ is defined by $\chi^2_{t_{\rm E}} + 0.2(\chi^2_{N_{I < 21}} + \chi^2_{N_{I < 18}} + \chi^2_{f_{18/21}})/3$, which is a combination of constraints from 
both the total 8000 $t_{\rm E}$ measurements and the star count data in two different brightness ranges in 1456 lines of sight \citep{mro17, mro19}.
This hybrid method enables us to concurrently measure the three IMF slopes of 
$(\alpha_{\rm hm}, \alpha_{\rm ms}, \alpha_{\rm bd})$ = ($2.32^{+0.14}_{-0.10}$, $1.16^{+0.08}_{-0.15}$, $0.22^{+0.20}_{-0.55}$) over the entire mass range 
defined in Eq. (\ref{eq-MF}), in addition to a break mass $M_{\rm br} = 0.90^{+0.05}_{-0.14} \, M_{\odot}$.

To understand how each constraint contributes to the determination of the IMF parameters, we plot the $\chi^2_{t_{\rm E}}$ and ${ \tilde{\chi}^2_{\rm IMF}}$ maps in Fig. \ref{fig-chi2IMF_tE}.
This is from the same grid search calculation described in Section \ref{sec-fund}.
Fig. \ref{fig-chi2IMF_tE} (a) shows that the $t_{\rm E}$ distribution cannot solely determine upper limits on $M_{\rm br}$ and $\alpha_{\rm hm}$ while 
it sets both upper and lower limits on $\alpha_{\rm ms}$ and $\alpha_{\rm bd}$.
An upper limit on the break mass $M_{\rm br}$ is set when the star count data is added to the restriction, which is shown in Fig. \ref{fig-chi2IMF_tE} (b), although 
that on $\alpha_{\rm hm}$ is not yet set with ${ \tilde{\chi}^2_{\rm IMF}}$.
Because there is a negative correlation between $\alpha_{\rm hm}$ and $M_{\rm VVV}$, an upper limit on $\alpha_{\rm hm}$ is set for the first time by including $\chi^2_{\rm pena}$,  
as shown in Fig. \ref{fig-chi2IMFpena}.
This is because a larger $\alpha_{\rm hm}$ value prefers a lighter bulge mass than the prior of $(1.32 \pm 0.08) \times 10^{10} M_{\odot}$ \citep{por17}.
As indicated in the $M_{\rm VVV}/M_{\rm VVV, pri}$ inset in the bottom right in Fig. \ref{fig-chi2IMF_tE} (b), the used data itself prefers $M_{\rm VVV} \sim 0.75 \, M_{\rm VVV, pri}$, a 
$\sim4.1 \, \sigma$ smaller value than the prior, which results in a large $\chi^2$ penalty value of $5 \times 4.1^2 \sim 84$.

The measured IMF slopes $(\alpha_{\rm hm}, \alpha_{\rm ms}, \alpha_{\rm bd})$ = ($2.32^{+0.14}_{-0.10}$, $1.16^{+0.08}_{-0.15}$, $0.22^{+0.20}_{-0.55}$) seem similar to the 
local values $(\alpha_{\rm hm}, \alpha_{\rm ms}, \alpha_{\rm bd}) = (2.3, 1.3, 0.3)$ by \citet{kro01}.
However, the measured break mass $M_{\rm br} = 0.90^{+0.05}_{-0.14} \, M_{\odot}$ is different from $0.5 \, M_{\rm \odot}$ by \citet{kro01}.
In fact, we find $\Delta \chi^2_{t_{\rm E}} = 496$, $\Delta { \tilde{\chi}^2_{\rm IMF}} = 810$, and $\Delta ({ \tilde{\chi}^2_{\rm IMF}} + \chi^2_{\rm pena}) = 852$ at the grid of 
$(M_{\rm br}, \alpha_{\rm hm}, \alpha_{\rm ms}, \alpha_{\rm bd})$ = ($0.5 \, M_{\rm \odot}$, 2.3, 1.3, 0.3), 
which indicates that the local IMF model by \citet{kro01} is significantly disfavored by all three of the $t_{\rm E}$ data, star count data, and prior on $M_{\rm VVV}$.
Our IMF is similar to \citet{zoc00}, a single power-law of $M^{-1.33 \pm 0.07}$ in $0.15\, M_{\odot} < M < 1.00\, M_{\odot}$, rather than \citet{kro01}.
This is expected because the \citet{zoc00} measurement is for bulge stars.
This similarity becomes clearer when we compare the two using the mass-to-light ratio values in the following section.

\subsection{Mass-to-light Ratio and Dark Matter Fraction} \label{sec-M2L}

\citet{por15} constructed five versions of dynamical models, which are called the M80, M82.5, M85, M87.5 and M90 models, using the made-to-measure method \citep{sye96}.
These models have different degree of maximality ranging from 80 to 90\%, where 
degree of maximality is defined as the proportion of the disk contribution to the total velocity curve, at the radius where the disk velocity curve is maximal.
Each model reproduces the observed distributions toward the Galactic bulge, including
the VVV photometric survey data \citep{sai12}, BRAVA RV data \citep{ric07, kun12}, and OGLE-II proper motion data \citep{rat07a}.
Although the five models have consistent dynamical mass $(1.84 \pm 0.07) \times 10^{10} \, M_{\odot}$ inside the VVV bulge box ($\pm 2.2 \times \pm 1.4 \times \pm 1.2$ kpc), 
the dark matter fractions are different.
This leads to different mass-to-light ratio values for the five models, which allows us to consider the dark matter fraction inferred by our model through it.

% Table 7
\begin{deluxetable}{lllllllllllrrrrrrrrrrrrrc}
\tabletypesize{\small}
\tablecaption{Mass-to-light ratio in $K$-band for indicated models. \label{tab-M2L}}
\tablehead{
\colhead{Model}                 & \colhead{$\Upsilon_K$}                & \colhead{$M_{\rm br}$}  & \colhead{$\alpha_{\rm hm}$} & \colhead{$\alpha_{\rm ms}$}  & \colhead{$\alpha_{\rm bd}$}  \\
\colhead{}                      & \colhead{[$M_{\odot}/L_{K_{\odot}}$]} & \colhead{[$M_{\odot}$]} &                             &                              &
}
\startdata
This work                       & $0.72^{+0.05}_{-0.02}$                & $0.90^{+0.05}_{-0.14}$ & $2.32^{+0.14}_{-0.10}$ & $1.16^{+0.08}_{-0.15}$ & $0.22^{+0.20}_{-0.55}$ \\
\citet{kro01}                   & 1.04                                  & 0.50                   & 2.30                   & 1.30                   &  0.30                  \\
\citet{zoc00}\tablenotemark{a}  & 0.75                                  & 1.00                   & 2.35                   & 1.33                   &  0.30                  \\
% G+G$_{\rm X}$ (G20)           & 0.76                                  & 1.06                   & 2.78                   & 1.54                   & -0.27                  \\
%M80   in \citet{por15}          & $0.83 \pm 0.08$ \tablenotemark{b}     &    --                  &    --                  &   --                   &    --                  \\
%M82.5 in \citet{por15}          & $0.88 \pm 0.08$ \tablenotemark{b}     &    --                  &    --                  &   --                   &    --                  \\
\enddata
\tablenotetext{a}{Modified version of the third one in Table 3 of \citet{zoc00}, where we apply $\alpha_{\rm bd} = 0.30$ because their measurement is insensitive to $M < 0.15 \, M_{\odot}$.}
%\tablenotetext{b}{Visually identified from Figure 15 of \citet{por15}.}
% \tablecomments{}
\end{deluxetable}

Table \ref{tab-M2L} lists the stellar mass-to-light ratio in $K$-band, $\Upsilon_K$, calculated for the three IMF models from this work, \citet{kro01}, and \citet{zoc00}.
The calculations are done with the initial--final mass relationships for the remnants described in Section \ref{sec-MF} and the PARSEC isochrone models for age = 10 Gyr to 
compare with values by \citet{por15}.
To check our calculation, we calculate the $\Upsilon_K$ values for the \citet{kro01} and \citet{zoc00} IMFs. The calculation yields $\Upsilon_K = 1.04 \, M_{\odot}/L_{K_{\odot}}$ and $0.75 \, M_{\odot}/L_{K_{\odot}}$, respectively, which are consistent 
with the values calculated by \citet{por15} for the two IMF models (see their Figure 15).
The $\Upsilon_K$ value calculated for our IMF is $\Upsilon_K = 0.72^{+0.05}_{-0.02} \, M_{\odot}/L_{K_{\odot}}$, and this confirms the similarity with the \citet{zoc00} IMF
described in Section \ref{sec-compIMF}.
The requirement for a mass-to-light ratio lower than the \citet{kro01} IMF value comes mainly from a high ratio of the number of stars with $I < 18$ to 
the number of stars with $I < 21$ in the OGLE-IV data.
The residuals in the bottom two panels of Fig. \ref{fig-resi_rhob} show that our model, which has a 30\% lower $\Upsilon_K$ value than the \citet{kro01} model, still 
slightly underestimates the number of stars with $I < 18$ and slightly overestimates the number of stars with $I < 21$, on average.
%We also find $\Upsilon_K = 0.76 \, M_{\odot}/L_{K_{\odot}}$ even when we use the best-fit IMF parameters in the G+G$_{\rm X}$ model with the G20 option. 
%This is because the requirement of a small $\Upsilon_K$ value mainly comes from the OGLE-IV star count data which show too large $f_{18/21}$ values to be reproduced by the local IMF.

Among the five dynamical models in \citet{por15}, the M80 model has the smallest $\Upsilon_K$ value, $0.83 \pm 0.08\, M_{\odot}/L_{K_{\odot}}$, and the largest dark matter mass inside the VVV bulge box, $M_{\cal DM} = 0.33 \times 10^{10} \, M_{\odot}$\footnote{The nuclear stellar disk mass based on \citet{por17}, $0.20 \times 10^{10} \, M_{\odot}$, is subtracted from the fiducial dark matter mass of the M80 model, $0.53 \times 10^{10} \, M_{\odot}$, because the nuclear stellar disk mass was not considered in the dark matter mass estimation of \citet{por15}.}.
Because our $\Upsilon_K$ value is even smaller than the M80 model's value, we linearly fit the relation between $M_{\cal DM}$ and 
$\Upsilon_K$ values among the five models in \citet{por15} and extrapolate it to derive 
$M_{\cal DM} = 0.45^{+0.02}_{-0.06} \times 10^{10} \, M_{\odot}$ at $\Upsilon_K = 0.72^{+0.05}_{-0.02} \, M_{\odot}/L_{K_{\odot}}$ for our model.
Note that the error of $M_{\cal DM}$ does not include the uncertainties of the relation, which is expected to be dominant, but it is sufficient to assess 
the reliability of the following independent estimate by a different approach.

Another approach to estimate $M_{\cal DM} $ is using $M_{\rm VVV}$, the model integrated mass within the VVV bulge box.
By simply subtracting the sum of $M_{\rm VVV} = 1.14^{+0.10}_{-0.11} \times 10^{10} \, M_{\odot}$ and the central nuclear stellar disk mass of $0.2 \times 10^{10} \, M_{\odot}$ suggested by \citet{por17} from the 
well-constrained dynamical mass $(1.85 \pm 0.05) \times 10^{10}$ \citep{por17}, we have $M_{\cal DM}  = 0.51^{+0.12}_{-0.11} \times 10^{10} \, M_{\odot}$, which can be larger than 
$M_{\cal DM}  = (0.32 \pm 0.05) \times 10^{10} \, M_{\odot}$ by \citet{por17}.
A drawback of this approach is an implicit assumption that no additional stellar mass other than the central nuclear stellar disk is missed inside the VVV bulge box in our density model that is fitted to data in $|b| \simgt 2^{\circ}$.
However, because the $M_{\cal DM}$ estimation with this approach is consistent with $M_{\cal DM} = 0.45^{+0.02}_{-0.06} \times 10^{10} \, M_{\odot}$ estimated by the
independent approach using $\Upsilon_K$ above, the implicit assumption seems reasonable. Thus, we consider $M_{\cal DM}  = 0.51^{+0.12}_{-0.11} \times 10^{10} \, M_{\odot}$ as our fiducial value.
This corresponds to a dark matter fraction in the VVV bulge box of $28 \pm 7 \%$, which could be larger than a previous estimate of $17 \pm 2$\% by \citet{por17}.

\section{Application to Microlensing Analysis} \label{sec-app}

In microlensing studies, a Galactic model is most commonly used as a prior probability distribution to calculate the posterior 
probability distributions of the lens mass and distance for individual events \citep{alc95, bea06, kos14, ben14}.
This is done when no or only one quantity that provides a mass--distance relation is measured, and such a Bayesian analysis usually yields a loosely constrained posterior distribution.
Because of their large uncertainties, results of the Bayesian estimates for individual events are not very sensitive to the choice of Galactic models \citep{yan20}, and thus 
we do not apply our Galactic model to such an analysis in this paper.

Using a refined Galactic model is important when it is applied to a statistical study with many events because small differences for individual events are combined and become significant.
In Section \ref{sec-stat} below, we repeat part of the analysis by \citet{kos20} using the new Galactic models and compare the result with the 
one using the \citet{ben14} model, which is a Galactic model frequently used for microlensing studies.

Another useful way to use a Galactic model is to distinguish between degenerate solutions, which indicate different lens system physical 
quantities for the analysis of an individual event.
We apply our Galactic model to calculate the prior probability distribution for OGLE-2011-BLG-0950 \citep{cho12, suz16} in Section \ref{sec-dege}, 
in which degenerate models indicate different lens-source relative proper motion $\mu_{\rm rel}$ values with each other and very different mass-ratios.
This is the only ambiguous event suffering from a degeneracy between planetary and binary solutions out of the 29 events in 
the \citet{suz16} combined sample, and we show that the stellar binary solution is preferred by our Galactic model.

\subsection{Statistical Study Using a Galactic Model} \label{sec-stat}

%Figure 12
\begin{figure}
\centering
\includegraphics[width=16cm]{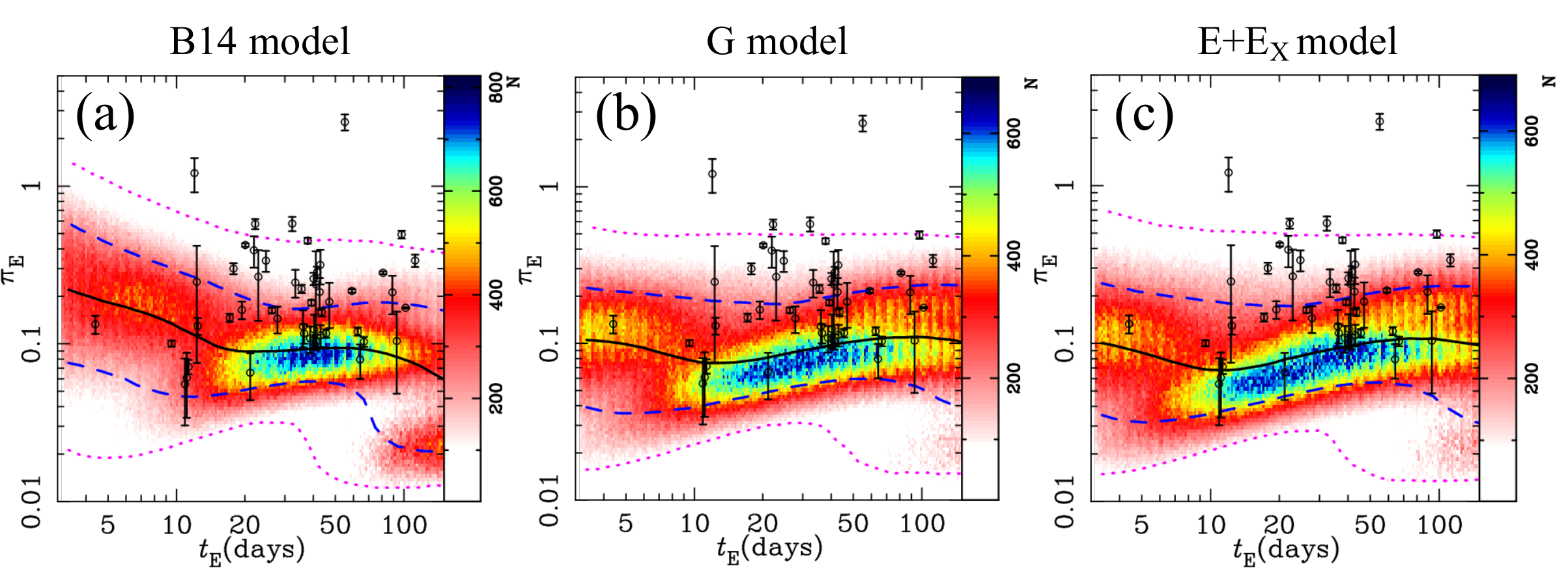}
\caption{Reproductions of Figure 1 of \citet{kos20} using the Galactic models developed here, where the probability of microlens parallax, $\pi_{\rm E}$, as a 
function of fixed $t_{\rm E}$, $\Gamma_{\rm Gal} (\pi_{\rm E} | t_{\rm E})$, 
is plotted using (a) the B14 model from \citet{kos20}, (b) the G model, and (c) the E+E$_{\rm X}$ model. Some updates, such as inclusion of 
neutron stars and black holes, are applied compared to the original one in \citet{kos20}.
The black dots are measured values for 50 events in the raw sample of \citet{zhu17}. 
The solid black, dashed blue, and dotted magenta lines indicate the median, $1~\sigma$, and $2~\sigma$ for $\Gamma_{\rm Gal} (\pi_{\rm E} \, | \,  t_{\rm E})$, respectively.
}
\label{fig-tEpiE}
\end{figure}

\citet{kos20} compared 50 microlens parallax, $\pi_{\rm E}$, measurements from the 2015 {\it Spitzer} campaign \citep{zhu17} to three 
different Galactic models \citep{sum11,ben14,zhu17}, commonly used in microlensing analyses.
They found that $\geq 37$ events have $\pi_{\rm E}$ values higher than the medians predicted by the Galactic models, and concluded that the difference is mainly attributed to 
systematic errors in the {\it Spitzer} microlens parallax measurements.
They considered that part of the inconsistency might originate from some simplistic features in the Galactic models, such as constant velocity dispersion 
regardless of Galactic distance, which is one of the main motivations for this work.
Because our Galactic model includes the dependency of the velocity dispersion on the Galactic location, as well as further updates to match recent observations, we test their claim using the updated models.

Following \citet{kos20}, we calculate $\Gamma_{\rm Gal} (\pi_{\rm E} | t_{\rm E})$, the probability distribution of $\pi_{\rm E}$ as a function of a given $t_{\rm E}$ value, which 
can be directly compared with observed $\pi_{\rm E}$ values without a detection efficiency correction.
Color maps in Fig. \ref{fig-tEpiE} show the distributions toward a typical sky direction of the 50 {\it Spitzer} events of $(l, b) = (1.0^{\circ}, -2.2^{\circ})$, 
where a total of $10^5$ artificial events { contribute} to each bin of $t_{\rm E}$ with width 0.05 dex in $\log t_{\rm E}$.
Fig. \ref{fig-tEpiE} (a) shows the result with the B14 model from \citet{kos20} for comparison, while the (b) and (c) panels show the results with the best-fit G and E+E$_{\rm X}$ models, respectively.
We selected the G and E+E$_{\rm X}$ models here to show the largest variation caused by one's choice from the E, G, E+E$_{\rm X}$, and G+G$_{\rm X}$ models, 
because these two models have the largest $\chi^2_{\rm fit}$ difference from each other ($\Delta \chi^2_{\rm fit} \sim 15000$) among the four models.
In the calculation, we also included the neutron star and black hole populations, which were ignored in \citet{kos20} as negligible possibilities.
The distinct population in the bottom right for each color map primarily consists of black hole lenses, which confirms the same feature predicted by the \citet{lam20} simulation.
In contrast to our previous expectation, this population changes the distribution in $t_{\rm E} \simgt 60~{\rm days}$, seen as undulations of the median, $1~\sigma$, and $2~\sigma$ lines in the figure.

There is a clear difference in the distributions between the B14 model and our two new models; compared with the G and E+E$_{\rm X}$ models, 
the B14 model tends to have higher $\pi_{\rm E}$ values with short $t_{\rm E}$ values, while it tends to have lower $\pi_{\rm E}$ values with long $t_{\rm E}$ values.
By contrast, there is no major difference between the G and E+E$_{\rm X}$ models.
We calculated the same distributions with the E and G+G$_{\rm X}$ models, and found that they were almost identical to the E+E$_{\rm X}$ model's distribution.

Black dots in each panel are the reported $\pi_{\rm E}$ values in the raw-sample of \citet{zhu17}, in which the Galactic prior is not applied unlike Figure 1 of \citet{kos20}.
This results in 43, 43, and 44 events whose $\pi_{\rm E}$ values above the median of $\Gamma_{\rm Gal} (\pi_{\rm E} | t_{\rm E})$ with the B14, G, and E+E$_{\rm X}$ models, respectively.
Even though there is a clear visual difference in the distributions between the old and new models, statistical quantities inferred from this analysis happen to be very similar to the old models,
which implies no change to the \citet{kos20} conclusions.
Nevertheless, Fig. \ref{fig-tEpiE} shows that it could be very different depending on the observed distributions, indicating that the choice of the models is important 
for a statistical analysis using a Galactic model.

\subsection{Prior Calculation to Distinguish Degenerate Models} \label{sec-dege}

%Figure 13
\begin{figure}
\centering
\includegraphics[width=5cm]{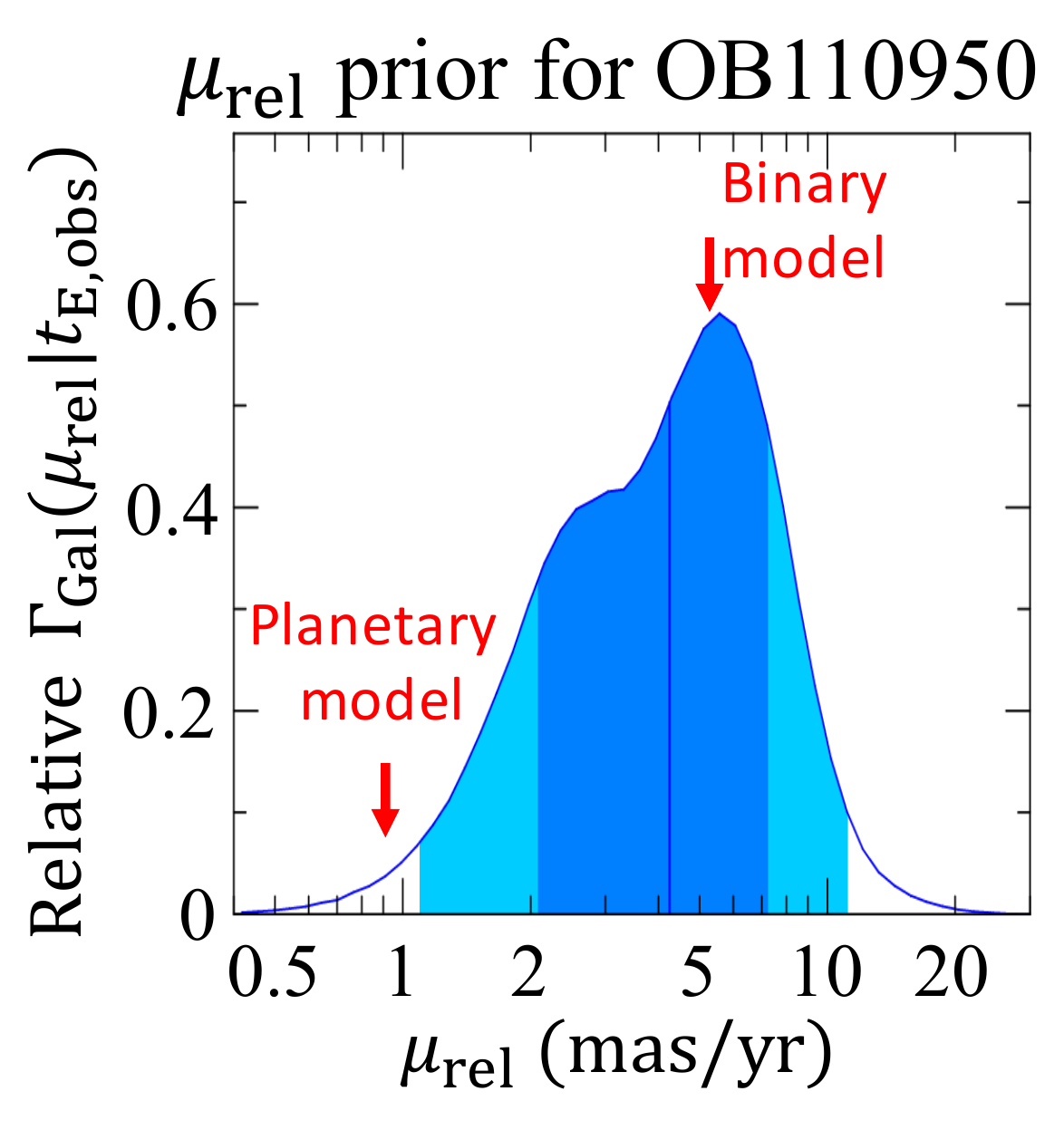}
\caption{Prior probability distribution of the lens-source relative proper motion $\mu_{\rm rel}$ for OGLE-2011-BLG-0950, calculated using the best-fit G+G$_{\rm X}$ model.
The indicated two $\mu_{\rm rel}$ values are calculated using the parameters in Table 3 of \citet{suz16}.}
\label{fig-murel_pri}
\end{figure}

\citet{suz16} has analyzed the largest statistical sample of 29 planetary events till now and found a peak in the mass-ratio function for the first time.
Out of the 29 planetary events, there is one ambiguous event, OGLE-2011-BLG-0950, which has degenerate planetary and stellar binary mass-ratio models with $\Delta \chi^2 \sim 20$ \citep{cho12, suz16}.
The preferred planetary model has mass-ratio $q \sim 6 \times 10^{-4}$ while the binary model has $q \sim 0.5$ \citep{suz16}.
Although our Galactic model cannot assess the relative prior probability for these two mass-ratio values because 
we do not know the relative frequency of systems with mass ratios with $q \sim 6 \times 10^{-4}$ compared to systems with $q \sim 0.5$,
these two models have very different lens-source proper motion $\mu_{\rm rel}$ values, which can be assessed using our model.

Fig. \ref{fig-murel_pri} shows the calculated $\mu_{\rm rel}$ prior, $\Gamma_{\rm Gal} (\mu_{\rm rel} | t_{\rm E, obs})$, using the best-fit G+G$_{\rm X}$ model, where $t_{\rm E, obs}$ is 
the observed $t_{\rm E}$ value for this event and we applied $t_{\rm E, obs} = 68 \pm 3$ days \citep{suz16}.
This figure shows that our Galactic model significantly prefers the $\mu_{\rm rel}$ value for the stellar binary model over that for the planetary model.
The preference of the stellar binary model is also suggested by a high angular resolution follow-up imaging by Keck (Terry et al., in preparation).

%\subsection{Estimate of physical quantity for individual events}
%Application to OGLE3-ULENS-PAR-02  (BH candidate from Wyrzykowski et al. 2016, 2020)

\section{Summary and Conclusion} \label{sec-con}
In this study, we developed parametric Galactic models using constraints from the spatial distributions of the median velocity and velocity dispersion from the Gaia DR2 \citep{kat18}, 
OGLE-III RC star count data \citep{nat13}, VIRAC proper motion data \citep{smi18, cla19}, BRAVA radial velocity data \citep{ric07}, and OGLE-IV star count 
and microlensing event data \citep{mro17, mro19}.
Our modeling indicated the following (we note that the error estimates could be influenced by our non-standard fitting method of an iterative step-by-step procedure.):

\begin{enumerate}
\item Local velocity dispersions for the thin and thick disks of 
$(\sigma_{R, \odot}^{\rm thin}, \, \sigma_{z, \odot}^{\rm thin},\, \sigma_{R, \odot}^{\rm thick},\, \sigma_{z, \odot}^{\rm thick}) =$ 
(35.2--44.0, 22.2--25.4, 68--75, 47.8--61.4) km/s, slopes of age-velocity dispersion relation for the thin disk of $(\beta_R, \beta_z) =$ (0.22--0.34, 0.77--0.82), and 
scale lengths for the velocity dispersion distributions of ($R_{\sigma_R}^{\rm thin}$, $R_{\sigma_z}^{\rm thin}$, $R_{\sigma_R}^{\rm thick}$, $R_{\sigma_z}^{\rm thick}$) = 
(9.5--21.4, 5.9--10.4, 47--180, 6.9--52.0) kpc by a grid search with the modified Shu DF model \citep{sha14} over the Gaia data in 3340 pc $< R <$ 8440 pc.
The value ranges correspond to variations due to our choice for the scale height model (linear or flat) and for the fitting range ($|z| < 3400$ pc or $|z| < 900$ pc).
\item Although a linear scale height disk model is preferred with the Gaia data, a flat scale height model is preferred with $\Delta \chi^2_{\rm fit} > 1000$ with 
the other datasets for bulge stars when combined with the bulge model, which might indicate a best choice somewhere in the middle.
\item A bar pattern speed of $\Omega_{\rm p} = 45.9^{+4.0}_{-5.4}~{\rm km/s/kpc}$, which could be slightly larger than but consistent with recent measurements of $39.0 \pm 3.5~{\rm km/s/kpc}$ \citep{por17} and 
$41 \pm 3~{\rm km/s/kpc}$ \citep{san19b}.
\item Stellar mass inside the VVV bulge box, defined as the central region inside $(x', y', z') = (\pm 2.2, \pm 1.4, \pm 1.2)$ kpc \citep{weg13}, 
of $M_{\rm VVV} = 1.14^{+0.10}_{-0.11} \times 10^{10} \, M_{\odot}$. 
This results in the dark matter mass inside the box of $M_{\cal DM}  = 0.51^{+0.12}_{-0.11} \times 10^{10} \, M_{\odot}$ when we assume 
the dynamical mass of $(1.85 \pm 0.05) \times 10^{10}$ and the central nuclear stellar disk of $0.2 \times 10^{10} \, M_{\odot}$ suggested by \citet{por17}, 
which could be larger than a previous estimate of $M_{\cal DM}  = (0.32 \pm 0.05) \times 10^{10} \, M_{\odot}$ \citep{por17}.
\item An IMF with a break mass at $M_{\rm br} = 0.90^{+0.05}_{-0.14} \, M_{\odot}$ and slopes for different mass ranges of 
$(\alpha_{\rm hm}, \alpha_{\rm ms}, \alpha_{\rm bd})$ = ($2.32^{+0.14}_{-0.10}$, $1.16^{+0.08}_{-0.15}$, $0.22^{+0.20}_{-0.55}$), 
which is different from the \citet{kro01} local IMF but similar to the bulge IMF measured by \citet{zoc00}.
\item The IMF indicates stellar mass-to-light ratio in $K$-band, $\Upsilon_K  = 0.72^{+0.05}_{-0.02} \, M_{\odot}/L_{K_{\odot}}$.
A comparison with the $\Upsilon_K$ values of a series of five dynamical models by \citet{por15} yields $M_{\cal DM} = 0.45^{+0.02}_{-0.06} \times 10^{10} M_{\odot}$, 
which is consistent with the above value implied by $M_{\rm VVV}$.
\end{enumerate}

We used our new Galactic models to test the result of \citet{kos20} in which the existence of systematic errors in {\it Spitzer} microlens parallax measurements is claimed based on 
older, simpler Galactic models.
We saw a significant difference in the predicted microlens parallax distributions between our model and a model used in 
the \citet{kos20} analysis, but this had no effect on the conclusions of \citet{kos20} that the {\it Spitzer} microlensing parallax measurements were 
contaminated by systematic errors.

We also applied the new model to calculate a prior probability distribution of the lens-source relative proper motion $\mu_{\rm rel}$ for OGLE-2011-BLG-0950, the only ambiguous event in the 
\citet{suz16} planet sample.
Our calculation shows that the $\mu_{\rm rel}$ value for the stellar binary solution is significantly preferred over that for the planetary solution.

Although the influence of model choice for individual event analysis is expected to be small, it becomes significant for statistical studies using multiple events \citep{yan20}.
With many recent microlensing event discoveries, a refined Galactic model like the one in this work is beneficial to study populations of various objects ranging from planets to black holes.
Our model is purely parametric, which makes it easy to implement and reproduce.
The demand is expected to increase in the era of the {\it Nancy Grace Roman Space Telescope}, previously known as {\it WFIRST}, which is expected to find 54,000 microlensing events 
during the primary 5 years survey toward the Galactic center \citep{gau19}. 
Because our current models lack constraints from data in $|b| < 2^{\circ}$, where the {\it Roman} fields are likely to be located, a further update should be expected in the future.
In such future updates, survey data toward the Galactic center expected to be collected by coming missions like 
Japan Astrometry Satellite Mission for INfrared Exploration
\citep[JASMINE;][]{Gouda2012}\footnote{
\url{http://jasmine.nao.ac.jp/index-en.html}}
or PRime-focus Infrared Microlensing Experiment (PRIME\footnote{\url{http://www-ir.ess.sci.osaka-u.ac.jp/prime/index.html}}) will be very useful.

\acknowledgments
We are grateful to Przemek Mr{\'o}z, Nathan Golovich, and David Katz who provided us with valuable data from their leading studies, which significantly helped us do this work.
We would like to thank Shogo Nishiyama for his helpful suggestion for our model.
NK was supported by JSPS KAKENHI Grant Number JP18J00897 and the JSPS overseas research fellowship.
JB acknowledges the support by JSPS KAKENHI Grant Numbers 18K03711, 18H01248, 19H01933, 21H00054 and 21K03633.
DPB and NK  were supported by NASA through grant NASA-80NSSC18K0274 and award number 80GSFC17M0002.

\appendix

\section{Comparing to another set of the Einstein radius crossing time data} \label{sec-G20}

% Table 5
% based on sortfiles_ref/v1-2/sortchi2_iloop2_ZFIN.out, sortfiles_ref/v1-2/sortchi2_X5_iloop2_ZFIN.out, sortfiles_ref/v1-2/sortchi2_X6_iloop2_ZFIN.out
\begin{deluxetable}{lrrrrrrrrrrrrrrrrrrrrrrrrrrc}
\tabletypesize{\small}
\tablecaption{${ \tilde{\chi}^2_{\rm sum}}$, $\chi^2_{N_{\rm eve}}$, and $\chi^2_{t_{\rm E}}$ values for each model with and without G20 option. \label{tab-chi2op}}
\tablehead{
\colhead{Model}       & \multicolumn{3}{c}{w/o G20} &  & \multicolumn{3}{c}{w/ G20}  \\
\cline{2-4} \cline{6-8}
\colhead{}            & \colhead{${ \tilde{\chi}^2_{\rm sum}}$\tablenotemark{a}} & \colhead{$\chi^2_{N_{\rm eve}}$} & \colhead{$\chi^2_{t_{\rm E}}$} & & \colhead{${ \tilde{\chi}^2_{\rm sum}}$} & \colhead{$\chi^2_{N_{\rm eve}}$} & \colhead{$\chi^2_{t_{\rm E}}$}
}
\startdata
E                     &        80173    &      21887  &      78   &  &      81411 & 22018 &     115  \\      
G                     &        91319    &      22022  &      89   &  &      92261 & 22283 &     133  \\      
E+E$_{\rm X}$         &        76261    &      21857  &      67   &  &      77583 & 22058 &     113  \\      
G+G$_{\rm X}$         &        76500    &      21852  &      68   &  &      77797 & 22041 &     109  \\
\enddata
\tablenotetext{a}{ ${ \tilde{\chi}^2_{\rm sum}} \equiv { \tilde{\chi}^2_{\rho_{\rm B}}} + 2 \chi^2_{v_{\rm B}} + 5 \chi^2_{\rm pena}$.}
\tablecomments{G20 option is disfavored in any model with any $\chi^2$ value.}
\end{deluxetable}

Measurements of $t_{\rm E}$ for a microlensing event can be occasionally erroneous due to systematic errors in light curve data.
\citet{gol20} recently reanalyzed all the 5788 events of \citet{mro19} using their newly developed model to simultaneously handle 
microlens parallax due to Earth's motion, systematic instrumental effects, and unlensed stellar variability with a Gaussian Process model.
Consequently, they found fewer long $t_{\rm E}$ events and more short $t_{\rm E}$ events compared to the \citet{mro19} $t_{\rm E}$ distribution.
To consider this possible systematic effect in the $t_{\rm E}$ distribution, we attempt another option, hereafter called the G20 option, 
using the $t_{\rm E}$ distribution from \citet{gol20} instead of the original Mr{\'o}z et al.'s distribution.

\citet{gol20} conducted their modeling on the 5788 events in the low-cadence fields \citep{mro19}, but not on the 2212 events in the high-cadence fields \citep{mro17}.
To approximately apply the systematic effect correction by \citet{gol20} to the 2212 events, we derived the following factor 
\begin{align}
g (t_{\rm E}) = \frac{N_{\rm G20}^{\rm obs} (t_{\rm E})}{N_{\rm O_{\rm low}}^{\rm obs} (t_{\rm E})}
\end{align}
by comparing the two $t_{\rm E}$ distributions of the 5788 events in the low-cadence fields by \citet{mro19} and \citet{gol20} (Golovich, private communication).
We multiply $g (t_{\rm E})$ by the $t_{\rm E}$ distribution of the 2212 events in the high-cadence fields, where the resultant number is rounded to the nearest whole number and 
use as $N_{\rm O_{\rm hi}}^{\rm obs} (t_{\rm E})$ in the G20 option.
The $t_{\rm E}$ data with and without the G20 option are plotted in Fig. \ref{fig-mod_tE}.

With the G20 option, we should also consider modifying the number of detected events as a function of $t_{\rm E}$ for each $i$th subfield, $N_{{\rm eve}, i}^{\rm obs} (t_{{\rm E}, j})$.
Whereas the numbers by \citet{gol20} are used for $i \in {\rm O_{\rm low}}$, we apply $g (t_{\rm E})$ for $i \in {\rm O_{\rm hi}}$.
Multiplying $g (t_{\rm E})$ by $N_{{\rm eve}, i}^{\rm obs} (t_{{\rm E}, j})$ is not feasible because $N_{{\rm eve}, i}^{\rm obs} (t_{{\rm E}, j})$ is mostly 0, and its typical non-zero value is 1 or 2, which is largely affected by 
the round-off process needed to calculate the Poisson probability in Eq. (\ref{eq-chi2Neve}).
Therefore, the compared model value of $N_{{\rm eve}, i}^{\rm mod} (t_{{\rm E}, j})$ is instead divided by $g (t_{{\rm E}, j})$, and thereafter, $\chi^2_{N_{\rm eve}}$ is calculated.
Applying $g (t_{\rm E})$ to the events in the high-cadence fields is a crude attempt for the \citet{gol20} correction because the factor should depend on both the $t_{\rm E}$ value, as well as 
characteristics of individual events or fields, such as extinction, event brightness, impact factor, and light curve coverage.
For analyzing the impact qualitatively, this is sufficient, though.

We compare the best-fit ${ \tilde{\chi}^2_{\rm sum}}$, $\chi^2_{N_{\rm eve}}$, and $\chi^2_{t_{\rm E}}$ values for each model, with and without the G20 option in Table \ref{tab-chi2op}.
We find that all the three $\chi^2$ values increase when the \citet{gol20} correction based on a Gaussian Process model is applied to the $t_{\rm E}$ data regardless of the model choice.
As discussed above, the G20 application to the $t_{\rm E}$ distribution in the high-cadence fields is a crude approximation.
However, as shown in Fig. \ref{fig-mod_tE}, the breakdown of $\Delta \chi^2_{t_{\rm E}}$ is 7 from $\chi^2_{t_{\rm E}, {\rm O_{\rm hi}}}$ and 34 
from $\chi^2_{t_{\rm E}, {\rm O_{\rm low}}}$; therefore, the worse $\chi^2_{t_{\rm E}}$ with the G20 option is attributed to the the $t_{\rm E}$ distribution in the low-cadence fields, 
to which the G20 correction is accurately applied, rather than the $t_{\rm E}$ distribution in the high-cadence fields.
Although Fig. \ref{fig-mod_tE} is for the G+G$_{\rm X}$ model, this is same for the other models.
%; it seems that the G20 correction makes the number of long $t_{\rm E}$ events too small to be reproduced within our model framework.
Thus, we conclude that with our model framework, the original Mr{\'o}z et al.'s $t_{\rm E}$ distributions are favored to 
the Golovich et al's one and use models without the G20 option as the fiducial best-fit models in this study.

\end{document}